\documentclass[aps,pre,twocolumn,superscriptaddress,floatfix,amsmath]{revtex4}

\usepackage{ae}
\usepackage{graphicx}
\usepackage{psfrag}
\usepackage{bm}
\usepackage{wasysym}
\usepackage{amssymb}

\begin{document}

\title{Domain growth morphology in curvature driven 
two dimensional coarsening}
\vskip 10pt
\author{Alberto Sicilia}
\affiliation{Universit\'e Pierre et Marie Curie -- Paris VI, LPTHE UMR 7589,
4 Place Jussieu,  75252 Paris Cedex 05, France}
\author{Jeferson J. Arenzon}
\affiliation{Instituto de F\'\i sica, Universidade Federal do 
Rio Grande do Sul, CP 15051, 91501-970 Porto Alegre RS, Brazil} 
\author{Alan J. Bray}
\affiliation{School of Physics and Astronomy, University of Manchester, 
Manchester M13 9PL, UK}
\author{Leticia F. Cugliandolo}
\affiliation{Universit\'e Pierre et Marie Curie -- Paris VI, LPTHE UMR 7589,
4 Place Jussieu,  75252 Paris Cedex 05, France}

\begin{abstract}
We study the distribution of domain areas, areas enclosed by domain
boundaries (``hulls''), and perimeters for curvature-driven
two-dimensional coarsening, employing a combination of exact analysis
and numerical studies, for various initial conditions. We show that
the number of hulls per unit area, $n_h(A,t)\,dA$, with enclosed area
in the interval $(A,A+dA)$, is described, for a disordered initial
condition, by the scaling function $n_h(A,t) = 2c_h/(A + \lambda_h
t)^2$, where $c_h=1/8\pi\sqrt{3} \approx 0.023$ is a universal
constant and $\lambda_h$ is a material parameter. For a critical
initial condition, the same form is obtained, with the same
$\lambda_h$ but with $c_h$ replaced by $c_h/2$. For the distribution
of domain areas, we argue that the corresponding scaling function has,
for random initial conditions, the form $n_d(A,t) = 2c_d (\lambda_d
t)^{\tau'-2}/(A + \lambda_d t)^{\tau'}$, where $c_d$ and $\lambda_d$
are numerically very close to $c_h$ and $\lambda_h$ respectively,
and $\tau' = 187/91 \approx 2.055$. For
critical initial conditions, one replaces $c_d$ by $c_d/2$ 
and the exponent is $\tau = 379/187 \approx
2.027$. These results are extended to describe the number density of
the length of hulls and domain walls surrounding connected clusters of
aligned spins.  These predictions are supported by extensive numerical
simulations. We also study numerically the geometric properties of the
boundaries and areas.

\end{abstract}

\maketitle

\section{Introduction}

\begin{figure}[h]
\includegraphics[width=4.cm]{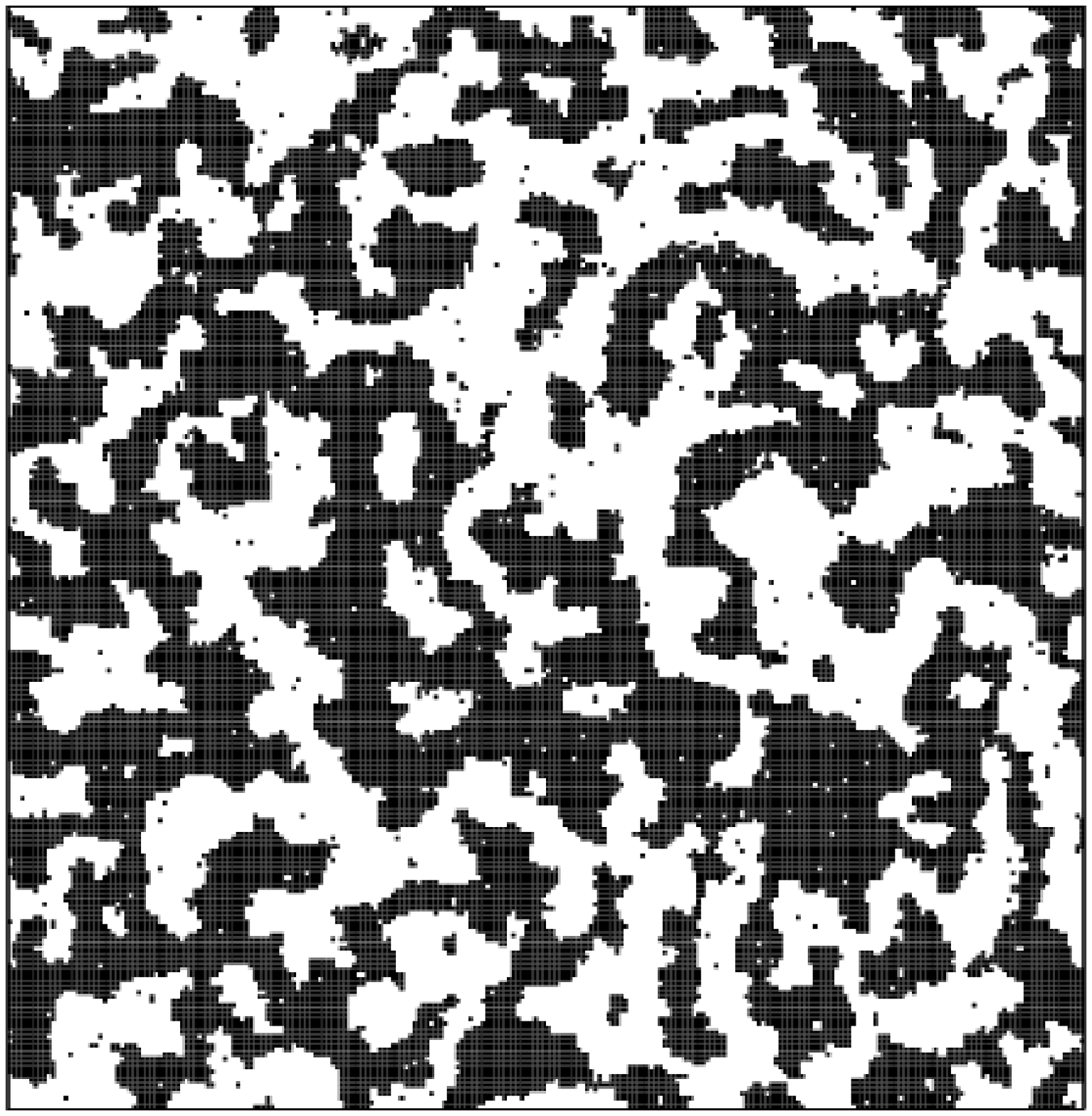}
\includegraphics[width=4.cm]{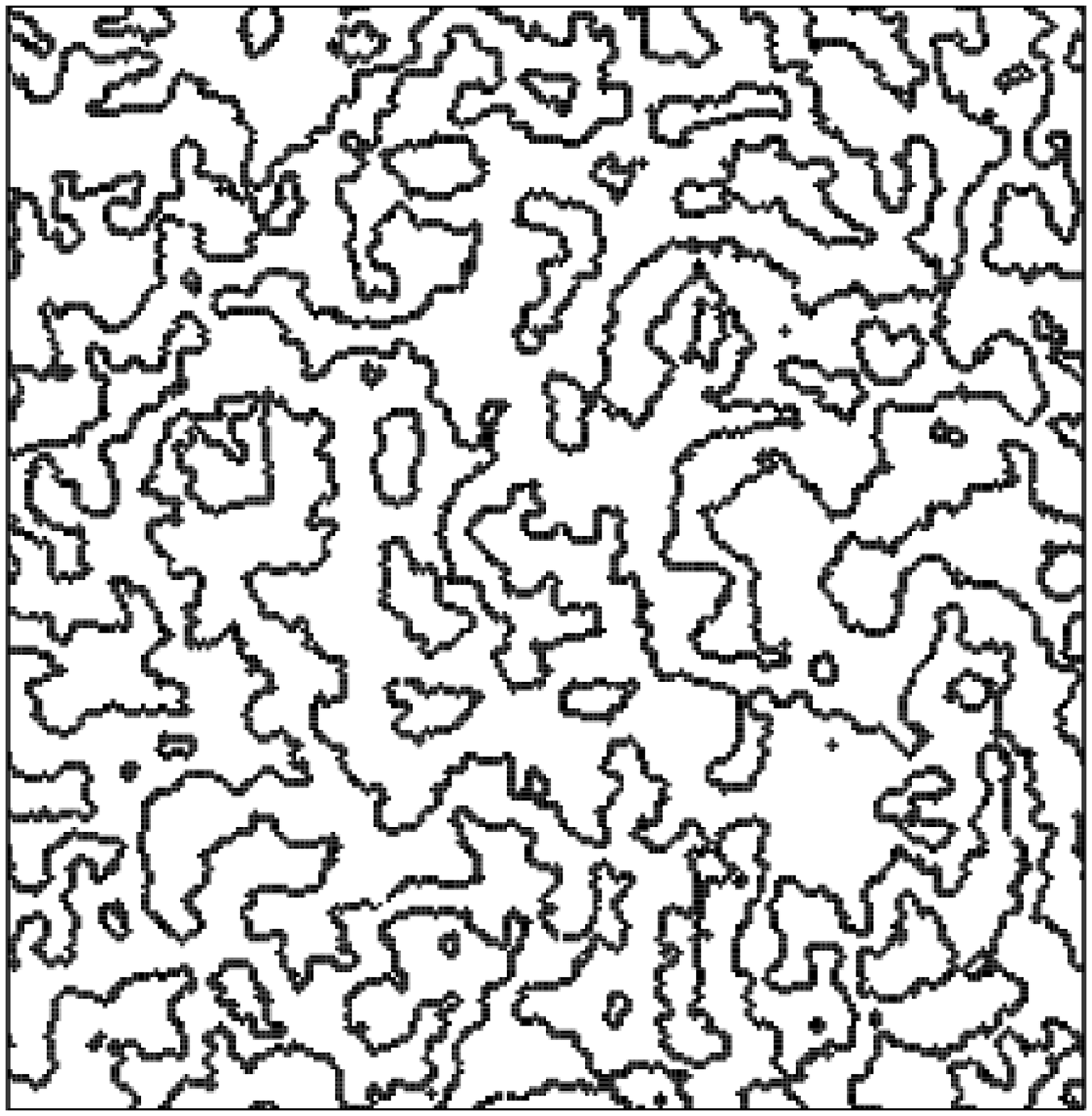}
\caption{Snapshots of the $2d$ Ising model at time
$t=32$ MCs after a quench from infinite temperature, $T_0\to\infty$,
to the working temperature $T=1.5\simeq 0.66T_c$. In the left panel we
show the raw data, where the domain structure as well as the thermal
fluctuations within the domains are visible.  In the right panel 
we show the boundaries between regions of opposite sign in the configuration
to the left, using a variant of the algorithm in
Ref.~\cite{Derrida} to make the domain structure clearer.}
\label{fig:snapshots0}
\end{figure}

In this work we obtain some exact and new results for the coarsening
dynamics~\cite{BrayReview} of a non-conserved scalar field in two
dimensions, demonstrating, {\em en passant}, the validity of the {\em
dynamical scaling hypothesis} for this system. We study the morphology
of the domain structure, which is illustrated in
Fig.~\ref{fig:snapshots0}
for the coarsening
of the two-dimensional Ising model ($2d$IM) on a square lattice
quenched from an equilibrium state at $T_0>T_c$.

A {\it domain} is a region of connected aligned spins. Each domain has
one external perimeter which is called the {\it hull}. The {\it hull
enclosed area} is the total area within this perimeter, {\it i.e.} the
domain area plus the area of any internal sub-domain.  The {\it domain
perimeter} is the total length of the interface between the chosen
domain and the neighbouring ones -- including the hull and internal
borders.  See
Fig.~\ref{fig:sketch-domains-hulls} for a sketch explaining these
definitions.

The paper deals primarily with the distributions of two characteristic
areas, the domain area and the hull enclosed area, and their 
associated lengths, the domain wall perimeter and the hull length.  

Naively, one may imagine that coarsening is basically due to the
coalescence of small domains that form larger ones. However, in two
dimensional curvature driven coarsening coalescence processes are
quite unimportant as shown by the Allen-Cahn result. All the
interfaces move with a local velocity that is proportional to the
local curvature and points in the direction of decreasing the
curvature; therefore, interfaces tend to disappear independently of
one another. This is the reason why we first focused on the statistics
of hull enclosed areas, quantities that depend on the motion of a
single and connected interface, and not on the statistics of the more
natural domain areas. Next we expressed the statistics of the domain
areas in terms of the simpler and more clear statistics of hull
enclosed areas.

Hull enclosed and domain areas have distributions that, at late times
after the quench into the ordered phase, exhibit, according to the
scaling hypothesis, the scaling form $n(A,t) = t^{-2} f(A/t)$, where
$n(A,t)dA$ is the number of hulls (domains) per unit area with area in
the range $(A, A+dA)$. The argument of the scaling function arises
from the fact that the characteristic length scale is known to grow as
$t^{1/2}$, so the characteristic area (of hulls and domains) grows as
$t$. The scaling function $f(x)$ will be different for domains and
hulls. The prefactor $t^{-2}$ follows from the fact that there is of
order one hull (or domain) per scale area.
Our analytical result is an elegant application of the Gauss-Bonnet
theorem and, it should be emphasized, its simplicity relies on the
dimensionality of the system being 2. Indeed, in three dimensions, the time
variation of the hull enclosed volume depends on a characteristic
size of the domain. A similar dependence on the dimension
is also observed in the von Neumann's law for cellular systems, whose
simple form, independent of any linear size of the system, is also
only observed in two dimensions \cite{Glazier, Mac}.

In this paper we derive these scaling forms from first principles (i.e. 
without recourse to the scaling hypothesis), and determine explicitly 
the scaling functions. Some of our results have appeared earlier 
in Letter form \cite{us}. 

Hulls and domain boundary lengths are themselves distributed
quantities related in a non-trivial manner to their corresponding
areas. In this paper we examine the geometry of these
structures and we derive the number density of hull and domain
wall lengths showing that these distributions also satisfy
scaling.

The organisation of the paper is as follows. In
Sect.~\ref{sec:equilibrium} we recall known results about the
equilibrium distribution of hull enclosed and domain areas at critical
percolation and critical Ising initial conditions. We also summarize known
results about the equilibrium distribution of domain walls and hulls
and their geometrical relation to their associated areas.  In
Sect.~\ref{sec:generic} we derive some generic results that stem from
a number of sum rules and the use of the scaling hypothesis.  In
Sect.~\ref{sec:analytic} we explain the analytic derivation of the
time-dependent hull enclosed and domain area distributions. These
arguments do not rely on any scaling hypothesis but rather demonstrate its
validity. In Sect.~\ref{sec:numeric} we show our numerical results for
the statistics of areas in the $2d$IM evolving with Monte Carlo
dynamics.  Section~\ref{sec:domain-walls} is devoted to the analysis,
both analytical and numerical, of the geometry of hulls and domain
walls during the dynamics and their relation to their corresponding
areas.  Finally, in the Conclusion we discuss future studies along
these lines. We also add two appendices in which we describe the
algorithm used to identify and count the hull enclosed areas, and for
the sake of comparison we present the distribution of domain lengths
in the one-dimensional Ising model.

\section{Equilibrium distributions}
\label{sec:equilibrium}

\begin{figure}[h]
\begin{center}
\psfrag{r1}{$R_1$}
\psfrag{r2}{$R_2$}
\includegraphics[width=3cm]{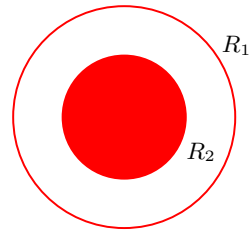}
\end{center}
\caption{(Colour online.) A sketch of a configuration with two
concentric and circular interfaces with radius $R_1$ and $R_2$ is
shown to illustrate the definition of hull enclosed and domain areas
as well as hull and domain-wall perimeters.  This configuration has
two hull enclosed areas $A_h^{(1)}=\pi R_1^2$ and $A_h^{(2)}=\pi
R_2^2$ and two domains with areas $A_d^{(1)}=\pi (R_1^2-R_2^2)$ and
$A_d^{(2)}=\pi R_2^2$. There are two hulls with length 
$p_h^{(1)}=2\pi R_1$ and $p_h^{(2)}=2\pi R_2$, and two domain walls
with length $p_d^{(1)}=2\pi (R_1+R_2)$ and $p_d^{(2)}=2\pi R_2$.
}
\label{fig:sketch-domains-hulls}
\end{figure}

In this Section we summarize known results on the {\it equilibrium} hull
enclosed and domain area distributions as well as the number density
of perimeters, both  at critical percolation and critical Ising conditions in
two-dimensions, see Fig.~\ref{fig:sketch-domains-hulls}.
These will act as initial conditions for the coarsening dynamics.

\subsection{Area distributions}

A {\it hull enclosed area} is defined as the full interior of a domain
boundary -- irrespective of there being other interfaces and thus
regions of the opposite phase within.  The equilibrium hull enclosed
area distributions 
at percolating criticality and Ising criticality, have been computed by Cardy and 
Ziff~\cite{Cardy} in two dimensions:

\begin{eqnarray}
n_h(A,0) \sim \left\{
\begin{tabular}{ll}
 $2c_h/A^2 \;$ , & \qquad \mbox{critical percolation}, \\
 $c_h/A^2 \;$ , & \qquad \mbox{critical Ising}.
\end{tabular}
\right.
\label{eq:nhpercolation}
\end{eqnarray}
These results are valid for $A_0 \ll A \ll L^2$, with $A_0$ a
microscopic area and $L^2$ the system size. Note also that we are
taking an extra factor 2 arising from the fact that there are
two types of hull enclosed areas, corresponding to the two phases,
while the Cardy-Ziff results accounts only for clusters of occupied
sites (and not clusters of unoccupied sites). 
$n_h(A,0)\,dA$ is the
number density of hulls per unit area with enclosed area in the
interval $(A,A+dA)$ (we keep the notation to be used later and set
$t=0$).  The adimensional constant $c_h$ is a universal quantity that
takes a very small value: $c_h = 1/(8\pi\sqrt{3})\approx
0.022972$. The smallness of $c_h$ plays an important role in the
analysis of Sect.~IV.

The distribution of {\it domain} areas -- we recall that domains are
clusters of connected aligned spins -- at critical percolation is
given by~\cite{Stauffer}
\begin{equation}
n_d(A,0) \sim \frac{2c_d A_0^{\tau'-2}}{A^{\tau'}}
\; , 
\qquad \mbox{with}\ \ \tau' = \frac{187}{91} \approx 2.055
\; . 
\label{eq:domequilTinfty}
\end{equation}
Of course, the constants $c_d$ and $A_0$ are not separately defined by 
this relationship, only the ratio $c_d A_0^{\tau'-2}$. In practice 
it is convenient to choose $c_d$ to be the value appearing in the 
domain area distribution at general times -- see Eq.\ (\ref{eq:guess2-bis}). 
The quantity  $A_0$ can be interpreted as a microscopic area, and will 
be introduced at various points as a small-area cut-off. The quantity 
$A_0^{\tau'-2}$ in Eq. (\ref{eq:domequilTinfty}) sets the units in such 
a way that $[n_d]=A_0^{-2}$. This result is also valid in the limit 
$A_0\ll A \ll L^2$. 

In equilibrium at $T_c$, Stella and Vanderzande~\cite{Stella} computed
the number density of domains with area $A$
\begin{equation}
n_d(A,0) \sim \frac{c_d A_0^{\tau-2}}{A^\tau}
\; , \qquad \mbox{with} \ \ \tau=\frac{379}{187} \approx 2.027
\label{eq:Janke}
\end{equation}
in the large $A$ limit.
Janke and Schakel~\cite{Janke} confirmed this claim numerically
finding $\tau\approx 2.0269$. Motivated by the Cardy-Ziff result for 
hull enclosed areas, we conjecture that the prefactor $c_d$ in 
(\ref{eq:Janke}) is the same $c_d$ (up to terms of order $c_h^2$) 
as that appearing in the prefactor $2c_d$ for critical percolation. 
We discuss this point in detail at the beginning of section IV.B, and 
we check this hypothesis numerically in Sect.~\ref{sec:numeric}.

We find it useful to include the small-area cut-off, $A_0$, in these number
densities, transforming the denominators to $(A+A_0)^2$ 
or $(A+A_0)^{\tau,\tau'}$ for hull enclosed and domain areas,
respectively. 

\subsection{Perimeter distributions}

One can equally study the length of {\it hulls} (external perimeters) and 
{\it domain walls} (including external and internal boundaries). 

The number density $n_h(p,0)$ of hulls with length $p$, defined as 
the number of spins on the hull, at the critical point 
of the $2d$IM was computed by Vanderzande and Stella~\cite{Vander}
\begin{equation}
n_h(p,0) \sim \frac{c_{p_h} \; p_0^{\zeta_h-3}}{(p+p_0)^{\zeta_h}} \; , 
\;\; \mbox{with} \;\; \zeta_h=\frac{27}{11}
\approx 2.454
\; . 
\end{equation}
The value of the constant $c_{p_h}$ was not estimated.  As far as we
know there is no analytic prediction for the number density of domain
walls at critical Ising conditions that should have the same
functional form though with a possibly different exponent $\zeta_d$
and a different constant $c_{p_d}$. $p_0$ is a microscopic length that 
we define as $p_0^2=A_0$.

For critical percolation, the number density of hulls with given perimeter 
was obtained by Saleur and Duplantier~\cite{Saleur-Duplantier}
\begin{equation}
n_h(p,0) \sim 
\frac{c'_{p_h} {p'_0}^{\zeta'_h-3}}{(p+p_0)^{\zeta'_h}} \; , 
\;\; \mbox{with} \;\; 
\zeta'_h=\frac{15}{7}
\approx 2.143 \; , 
\label{perimeters}
\end{equation}
compatible with the numerical study~\cite{Ziff-perc}.
Again, the numerical value of the constant $c'_{p_h}$ is not
known.  As far as we know, the analog of (\ref{perimeters}) for domain 
walls at critical percolation is not known, though we expect the same 
functional form with different constant $c'_{p_d}$ and exponent $\zeta'_{d}$.

In Sect.~\ref{sec:domain-walls} we show numerical results for the
equilibrium perimeter length number densities of critical Ising 
and infinite temperature configurations.
 
\subsection{Fractal properties}
\label{sec:Cambier}

Several authors studied the fractal properties of areas and perimeters
in critical Ising and critical percolation equilibrium conditions
using different analytic methods that include conformal invariance and
renormalization group and Coulomb gas
techniques~\cite{Stella-Vanderzande,Vander,Stella,Saleur-Duplantier,Duplantier-Saleur}.
Many numerical studies~\cite{Janke,Ziff-perc,Aharony,Voss,Grassberger} 
confirmed and complemented the results in these analytic works.
These works focused on the fractal dimensions of the domain area, $D_d$, 
and of the hull length, $D_h$. In two-dimensions 
these exponents are linked to the distribution 
exponents $\tau$ and $\zeta_h$ as~\cite{Stauffer}
\begin{equation}
D_d = \frac{2}{\tau-1} \; , \qquad 
D_h = \frac{2}{\zeta_h-1} \; , 
\end{equation}
and similarly for the primed quantities.

We concentrate on the 
fractal properties of geometric structures by 
comparing the area of the clusters to their associated 
perimeter. This approach was used by 
Cambier and Nauenberg~\cite{Cambier} who studied domain
walls in the $2d$IM -- with internal and external border -- at
equilibrium below, but near $T_c$, 
and found 
\begin{equation}
A \sim p^{1.43}\;.
\label{eq:equil-Ap}
\end{equation}
The proportionality constant is not given. In
Sect.~\ref{sec:domain-walls} we revisit the geometric 
properties of the clusters at $T_c$ and the
corresponding one at very high temperature as well as their zero
temperature evolution.

\section{Some general results}
\label{sec:generic}

A number of general properties of hull enclosed and domain areas as
well as hull and domain wall perimeters can be easily derived just by
using the scaling hypothesis and two sum rules.  We summarize them
here.

\subsection{Scaling}
\label{sec:scaling}

At long times, and irrespective of the initial condition, the total
 number of domain and hull enclosed areas per unit area, 
$N_{d,h}(t) = \int_0^\infty dA \; n_{d,h}(A,t)$, should
 scale as $R^{-d}(t)$ in $d$ dimensions, with $R(t)$ a 
characteristic length scale usually associated to 
the `typical' 
domain radius. For pure ferromagnetic coarsening, 
$R(t) \sim (\lambda_d t)^{1/2}$ from which it follows that 
$N_{d,h}(t) \sim t^{-1}$ in $d=2$. Since characteristic areas scale as 
$R^2(t) \sim t$, the scaling hypothesis implies that the domain and hull 
enclosed area distributions have the forms $n_{d,h}(A,t) = 
t^{-2}f_{d,h}(A/t)$. 

In Sect.~IV we present arguments that the initial distributions 
of hull enclosed and domain areas determine the forms of these distributions 
at late times. In particular, the initial forms are modified in a 
rather simple way at later times, such that the scaling forms are already 
suggested by the initial conditions. In the following we outline the 
consequences of this line of argument and defer detailed analysis to 
Sect.~IV.   

Let us first discuss domains with critical Ising initial conditions.  
Retaining the initial form~(\ref{eq:Janke}), but including an extra 
time-dependent factor to satisfy the desired scaling 
form at large times, the expression for $n_d(A,t)$ is,

\begin{equation}
n_d(A,t) \sim  
\frac{c_d \; [\lambda_d(t+t_0)]^{\tau-2}}{[A+\lambda_d (t+t_0)]^\tau}
\label{eq:guess1}
\end{equation}
at long times. Here, $t_0$ is a microscopic time, and $\lambda_d$ is a 
phenomenological parameter with the dimensions of a diffusion constant. 
Setting $t=0$ one recovers
Eq.~(\ref{eq:Janke}) provided $A_0 = \lambda_d t_0$ and $A\gg
A_0$. Indeed, $t_0$ is defined through this requirement while $A_0$ is
the by now usual microscopic area. For large $t$, $t_0$ can be neglected 
and the conventional scaling form is recovered.  

For infinite temperature initial conditions we propose
\begin{equation}
n_d(A,t) \sim  
\frac{2c_d \; [\lambda_d(t+t_0)]^{\tau'-2}}{[A+\lambda_d (t+t_0)]^{\tau'}}
\label{eq:guess2-bis}
\end{equation}
that also satisfies the scaling form 
asymptotically. 

One can easily check that the obvious generalization of 
Eq.~(\ref{eq:nhpercolation}) including time
\begin{equation}
n_h(A,t) \sim \frac{(2) c_h}{(A+\lambda_h t)^2} 
\label{eq:guess3}
\end{equation}
also has the desired scaling form. The constant $\lambda_h$ is another
phenomenological parameter. Its value will turn out to be very close to
$\lambda_d$. The factor 2 in the brackets in Eq.~(\ref{eq:guess3})
takes into account the two types of initial condition.

We remark that in Eq.~(\ref{eq:guess3}) we do not explicitly include
a short-time cut-off $t_0$, while it is necessary in Eqs.~(\ref{eq:guess1}) 
and (\ref{eq:guess2-bis}) to connect smoothly to the initial condition. 

Scaling can also be used to predict the time-dependence of the 
number density of hull and domain wall lengths. It yields
\begin{equation}
n_{d,h}(p,t) \sim
\frac{c_{p_d,p_h}[(\lambda_{d,h} t)^{1/2}+p_0]^{\zeta_{d,h}-3}}
{[p+(\lambda_{d,h} t)^{1/2}+p_0]^{\zeta_{d,h}}} \; .
\label{eq:scaling-np}
\end{equation}
with the constant and exponent values depending on the 
initial condition and whether we are studying hulls or domain
walls. This form is based on the assumption that the characteristic
perimeters evolve in time as $(\lambda_{d,h} t)^{1/2}$.

\subsection{Sum rules}

We now present two exact sum rules which provide useful input for the 
analysis of $n_{d,h}(A,t)$. These sum rules apply at all times $t\geq 0$, 
for any initial condition, and for any working temperature.

The first sum rule follows from the fact that the total domain area,
per unit area of the system, is unity since each space point (or
lattice site) belongs to one and only one domain. This gives
\begin{equation}
\int_0^\infty dA \; A \; n_d(A,t) = 1 
\; . 
\label{eq:totalarea}
\end{equation}

The second sum rule follows from the fact that the total number of hull 
enclosed areas, 
$N_h(t)$, is equal to the total number of domains, $N_d(t)$, since each 
domain can be associated with a unique hull, namely the hull that forms 
its external boundary. This yields
\begin{eqnarray} 
N_d(t) &\equiv& 
\int_0^\infty dA \; n_d(A,t) 
\nonumber\\
&=&\int_0^\infty dA \; n_h(A,t) 
\equiv N_h(t)
\; . 
\label{eq:hulls-equal-domains}
\end{eqnarray}

Equations~(\ref{eq:nhpercolation})-(\ref{eq:domequilTinfty}) and their
generalization describing the time-dependence, have been shown to hold
for large areas ($A\gg A_0$) only and the number density can take a
different form at small values of $A$. From the constraints
(\ref{eq:totalarea}) and (\ref{eq:hulls-equal-domains}), using
Eqs.~(\ref{eq:guess1}), (\ref{eq:guess2-bis}) and (\ref{eq:guess3}),
we shall derive {\it approximate} relations between the constants
$c_d$, $c_h$, $\lambda_d$, $\lambda_h$, $\tau$ and $\tau'$ expected to
hold at any working temperature $T$. These relations are exact to
first order in the small quantity $c_h$.

For the number densities of hulls or domain walls with given perimeter 
we have only one sum rule. The total number density must equal the number 
density of domains and hull enclosed areas:
\begin{equation}
N_{p_d,p_h}(t) \equiv 
\int_0^\infty dp \; n_{d,h}(p,t) = N_d(t) =N_h(t)
\; . 
\end{equation}

\subsubsection{Critical Ising initial conditions.}$\;$

The constraint on the total area (\ref{eq:totalarea}) 
using Eq.~(\ref{eq:guess1}) for the number density of domain areas
yields 
\begin{equation}
c_d = (\tau-2)(\tau-1) \approx 0.02745
\label{eq:relation1}
\end{equation}
where the numerical value was obtained for $\tau=379/187$. 
If one takes into account that the minimal area is $A_0$ (and not $0$)
the result is slightly different and it varies, though very weakly,
with $t$. Indeed, the value of $c_d$ decreases from $c_d=(\tau-2)(\tau-1)
2^{\tau-1}/\tau$ at $t=0$ to $c_d=(\tau-2)(\tau-1)$ at $t\gg t_0 =
A_0/\lambda_d$. Inserting $\tau=379/187$ gives 
\begin{equation}
0.02745
\leq c_d \leq 0.02760,
\end{equation}
a rather narrow interval. 
Note that $c_d$ computed using (\ref{eq:relation1}) is quite close to the 
analytical result for $c_h$, namely $c_h^{\rm an} = 1/8\pi\sqrt{3} 
\sim 0.02297$. Indeed, $c_d - c_h^{\rm an} \approx 0.00448$  
(which is order $c^{\rm an}_h/5$) 
where we used the minimum value of $c_d$ and the analytical value of $c_h$. 

We now consider the sum rule~(\ref{eq:hulls-equal-domains}), at times
$t=0$ and $t \gg A_0/\lambda_h$, distinguishing the integral over the
full interval $[0,\infty)$ from the one that takes into account the
finite minimal area $[A_0,\infty)$.  We denote by $A_1$ the lower
limit of the integration interval.  We also include a small area
cut-off $A_0$ in the hull enclosed area number density, $n_h(A,0) \sim
c_h/(A+A_0)^2$.

At time $t=0$ we find
\begin{eqnarray}
c_d &=&
c_h\; (\tau-1) \; 
\frac{(A_1+A_0)^{\tau-2}}{A_0^{\tau-2}}
\; . 
\label{eq:relation1a}
\end{eqnarray}
Using now the expressions for $c_d$ derived above 
we  relate $c_h$ to $\tau$:
\begin{eqnarray}
c_h &=& 
\left\{
\begin{array}{lcll}
(\tau-2) & \approx & 0.02674
\; , \qquad &A_1=0
\; , 
\\
\frac{2 (\tau-2)}{\tau}  & \approx & 0.02639
\; , \qquad & A_1= A_0
\; . 
\end{array}
\right.
\end{eqnarray}
These values are quite close to the analytic one
$c^{\rm an}_h \approx 0.02297$. The relative difference is
$(c_h-c_h^{\rm an})/c_h^{an} \approx 0.164$ ($A_1=0$) and
$(c_h-c_h^{\rm an})/c_h^{\rm an} \approx 0.149$ ($A_1=A_0$). 
Note that using the sum rules we over-estimate the value of $c_h$. 
We can then expect to have over-estimated the value of $c_d$ too. 
This remark will be important when comparing to numerical data.

Finally, we can evaluate condition (\ref{eq:hulls-equal-domains}) at
late times, $t \gg A_0/\lambda_h$, using the conjecture
(\ref{eq:guess1}) for $n_d$ and the result $n_h(A,t) \sim
c_h/(A+\lambda_h t)^2$ deduced from scaling in Sect.~\ref{sec:scaling}
and to be shown analytically in Sect.~\ref{sec:hulls}. In this case we find,
independently of $A_1$:
\begin{equation}
\frac{\lambda_d}{\lambda_h} = \frac{c_d}{c_h} \frac1{(\tau-1)}
\; . 
\label{eq:relation2}
\end{equation}
This equation can be used to relate the two factors $\lambda_d$ and 
$\lambda_h$. Indeed, 
replacing $c_d$ and $c_h$ by their expressions as functions of 
$\tau$ one finds 
\begin{eqnarray}
\frac{\lambda_d}{\lambda_h}
&=&
\left\{
\begin{array}{lll} 
 1 \;, \qquad & A_1=0 
\; , 
\\
 2^{\tau-2} \approx 1.019 \;, \qquad & A_1=A_0 
\; .
\end{array}
\right.
\end{eqnarray}
If, instead, we use $c_d$ as derived above and the analytic $c_h$,
$c_h^{\rm an}=0.022972$ we
find 
\begin{eqnarray}
1.164 \leq  \frac{\lambda_d}{\lambda_h} \leq 1.170
\; . 
\label{eq:rel3}
\end{eqnarray}
Since we derived Eq.~(\ref{eq:relation2}) using the sum rules, it seems
more appropriate to use the values of $c_d$ and $c_h$ obtained
from the same relations. The values of $\lambda_d$ and $\lambda_h$ are
then very close and consistent with the relation that we shall find in
the next Section using the approximate equation for the evolution of
domain areas [while Eq.~(\ref{eq:rel3}) yields a too large value 
for $\lambda_d$].

The condition $N_{p_h}(t)=N_{p_d}(t)=N_h(t)=N_d(t)$ implies
\begin{eqnarray}
c_{p_h} &=& -c_h \, (1-\zeta) 
\; , 
\nonumber\\
c_{p_d} &=& c_d \; \frac{1-\zeta}{1-\tau}
\; , 
\end{eqnarray}
for critical Ising initial conditions.

\subsubsection{Infinite-temperature initial conditions.}$\;$ 

The infinite-temperature initial conditions turn out to be, after just a few
time steps, equivalent to critical percolation ones -- see the numerical 
evidence in Sect.~\ref{sec:infTinit}. The sum rules yield, in this case, 
\begin{eqnarray}
c_d &=& \frac{(\tau'-2) (\tau'-1)}{2} \approx  0.029\; , 
\\
c_h &=& \frac{c_d}{(\tau'-1)} \approx 0.027 \; , 
\\
\frac{\lambda_d}{\lambda_h} &=& \frac{c_d}{c_h \, (\tau'-1)}
=1 
\; , 
\end{eqnarray}
where, for simplicity, we present results obtained with 
$A_1=0$ only. 

The conditions $N_{p_h}(t)=N_{p_d}(t)=N_h(t)=N_d(t)$ imply
\begin{equation}
c'_{p_h} = -2c_h \, (1-\zeta'_h) 
\; , 
\qquad
c'_{p_d} = 2c_d \; \frac{1-\zeta'_d}{1-\tau'}
\; , 
\end{equation}
for critical percolation.

\section{Statistics of areas: analytic results}
\label{sec:analytic}

Our analytic results are obtained using a continuum description of
domain growth in which the non conserved order parameter is a scalar field,
$\phi(\vec x,t)$, defined on a $d$-dimensional space. For a
review in this problem, see \cite{BrayReview}.
Its evolution is determined by the time-dependent
Ginzburg-Landau equation or model A dynamics:
\begin{eqnarray}
\gamma \frac{\partial \phi(\vec x,t)}{\partial t} = \nabla^2 \phi(\vec x,t)
- \frac{\delta V(\phi)}{\delta \phi(\vec x,t)} + \xi(\vec x,t)
\; . 
\label{eq:langevin}
\end{eqnarray}
The potential $V$ is a symmetric double well, with $V(\phi\to\pm
\infty) = \infty$ and two minima at $\pm\phi_0$.  $\xi$ is a Gaussian
distributed random scalar field with zero mean and correlation
\begin{equation}
\langle \, \xi(\vec x,t) \xi(\vec x', t') \, \rangle 
= 2 k_{\scriptscriptstyle\rm B} T \gamma \; \delta^d(\vec x-\vec x') \, 
\delta (t-t')
\; . 
\end{equation}
This white noise introduces thermal agitation.  $T$ is the temperature
of the thermal bath, $k_{\scriptscriptstyle\rm B}$ is the Boltzmann 
constant and $\gamma$ is
the friction coefficient. From now on we set the units in such a way
that $k_{\scriptscriptstyle\rm B}=\gamma=1$.   The
low-temperature ordering dynamics from a disordered initial condition
corresponds to the growth of ordered domains of the two equilibrium states,
$\phi(\vec x,t) =\pm \phi_0$, separated by
interfaces.  Using the evolution equation (\ref{eq:langevin}) at
{\it zero temperature}, Allen and Cahn showed that in any dimension $d$
the velocity, $v$, of each element of a domain boundary is
proportional to the local interfacial mean curvature,
$\kappa$~\cite{AC,BrayReview},
\begin{equation}
v = - \frac{\lambda_h}{2\pi} \; \kappa
\; .
\label{allen-cahn}
\end{equation}
$\lambda_h$ is  a  material  constant with  the  dimensions of  a
diffusion constant, and the  factor $1/2\pi$ is for later convenience.
The velocity is normal to the interface and 
points in the direction of reducing the curvature.
The dynamics is then purely curvature driven at zero temperature.

Temperature fluctuations have a two-fold
effect. On the one hand they generate {\it equilibrium} thermal
domains that are not related to the coarsening process. On the other
hand they {\it roughen} the domain walls thus opposing the curvature driven
growth and slowing it down. Then equation (\ref{allen-cahn})  
is no longer valid. However, it has been conjectured and verified 
numerically that, at least for averaged dynamic quantities 
well-described by the scaling hypothesis, all temperature 
effects are captured by introducing a $T$ dependent $\lambda_h$ 
parameter. We shall use this working hypothesis in the 
analytic part of our paper and we shall put it to the 
test numerically. 

The number density of hull enclosed or domain areas at time $t$ as a
function of their initial distribution is
\begin{equation}
n(A,t) = \int_0^\infty dA_i \; \delta(A-A(t,A_i)) \, n(A_i,t_i) 
\;
\label{eq:integration} 
\end{equation}
with $A_i$ the initial area and $n(A_i,t_i)$ their number 
distribution at the initial time $t_i$. $A(t,A_i)$ is the
hull enclosed/domain area, at time $t$, having started from
an area $A_i$ at time $t_i$.

\subsection{Hull enclosed areas}
\label{sec:hulls}

In two dimensions~\cite{Comment} we can immediately deduce the
time-dependence of the area contained within any finite hull by
integrating the velocity around the hull:
\begin{equation}
\frac{dA}{dt}  = \oint v\,dl = -\frac{\lambda_h}{2\pi} \oint
\kappa\,dl = -\lambda_h
\end{equation}
In the second equality we used the zero-temperature 
Allen-Cahn Eq.\  (\ref{allen-cahn}),
and in the final one we used the Gauss-Bonnet theorem.
Integrating over time, with initial time $t_i$, we find 
\begin{equation}
A_h(t,A_i) = A_i-\lambda_h (t-t_i)
\; .
\end{equation}
Therefore
\begin{eqnarray}
n_h(A,t) &=& 
\int_0^\infty dA_i \; \delta(A-A_i+\lambda_h (t-t_i)) \, n_h(A_i,t_i)
\nonumber\\
&=& n_h(A+\lambda_h(t-t_i),t_i) 
\; . 
\label{eq:advection}
\end{eqnarray}
In deriving this result we have implicitly assumed that a single domain 
cannot split into two, and that two domains cannot coalesce. A little 
thought shows that neither process is possible for two-dimensional 
curvature-driven growth since both processes require that two parts 
of a single domain boundary (for splitting) or parts of two different 
domain boundaries (for coalescence) come together and touch. But it is 
clear that the curvature driven dynamics always acts to prevent this 
happening, since the velocities of the domain boundaries at the incipient 
contact point are in opposite directions. 

The initial distributions, $n_h(A_i,t_i)$, are given by the Cardy-Ziff 
results displayed in Eq.\ (\ref{eq:nhpercolation}) -- assuming, in the case 
of a quench from infinite temperature, that the system rapidly sets into 
the critical percolation condition (see Sect.~\ref{sec:infTinit}). 
For $t\gg t_i$ one immediately recovers the results in~\cite{us}:
\begin{eqnarray}
n_h(A,t) & = & \frac{2c_h}{(A+\lambda_h t)^2}
\; , \qquad T_0\to\infty
\; , 
\label{eq:analytic-nh-Tinf}
\\
n_h(A,t) & = & \frac{c_h}{(A+\lambda_h t)^2}
\; , \qquad T_0=T_c
\; , 
\label{eq:analytic-nh-Tc}
\end{eqnarray}
in the limit $A_0 \ll A \ll L^2$, i.e.\ for hull enclosed areas much 
larger than microscopic areas but much smaller than the area of the 
system. 

Equations~(\ref{eq:analytic-nh-Tinf})   and  (\ref{eq:analytic-nh-Tc})
have   the  expected   scaling  forms   $n_h(A,t)  =   t^{-2}  f(A/t)$
corresponding to a system with characteristic area proportional to $t$
or characteristic  length scale $R(t)  \propto t^{1/2}$, which  is the
known  result if  scaling is  {\em assumed}  \cite{BrayReview}.  Here,
however, we do not {\em assume} scaling -- rather, it emerges from the
calculation.  Furthermore,  the conventional scaling  phenomenology is
restricted to the `scaling limit': $A \to \infty$, $t \to \infty$ with
$A/t$       fixed.      Equations~(\ref{eq:analytic-nh-Tinf})      and
(\ref{eq:analytic-nh-Tc}),  by  contrast, are  valid  whenever $t$  is
sufficiently large and  $A \gg A_0$. This follows  from the fact that,
for   large    $t$,   the   forms    (\ref{eq:analytic-nh-Tinf})   and
(\ref{eq:analytic-nh-Tc}) probe, for any  $A \gg A_0$, the tail (i.e.\
the large-$A$  regime) of  the Cardy-Ziff results,  which is  just the
regime in  which the latter is  valid. The restriction $A  \gg A_0$ is
needed  to justify the  use of  Eq.\ (\ref{allen-cahn}),  which breaks
down when the reciprocal of  the curvature becomes comparable with the
width of a domain wall.

The averaged area enclosed by a hull is then given by
\begin{eqnarray}
\langle  A\rangle(t)   &=&  \frac{\int  dA'   A'  n_h(A',t)}{\int  dA'
n_h(A',t)} \\ &\propto& \lambda_h t
\end{eqnarray}
with a  time-independent prefactor that  behaves as $(A_0^2  \ln L^2)$
for large system sizes. The  reason for the divergent prefactor in the
infinite size limit is that a site can belong to several hulls.

\subsection{Domains}
\label{sec:domains}

For the domains we need to  write an evolution equation and derive, at
least approximately, the  area at time $t$, $A_d(t,A_i)$,  of a domain
with initial area $A_i$. We  shall show that the time-dependent number
density   of   domain   areas   is   indeed  given   by   our   guess,
Eqs.~(\ref{eq:guess1}) and (\ref{eq:guess2-bis}),  for the two classes
of initial conditions.

Our strategy is to exploit the smallness of the parameter $c_h \approx
0.023$.   Although  $c_h$ is  a  constant,  we  can exploit  a  formal
expansion in $c_h$  in the following sense. Since  the total number of
hulls per unit  area is proportional to $c_h$,  the number of interior
hulls within  a given hull is  also proportional to $c_h$,  and so on.
This means  that, in  dealing with domains  we need consider  only the
first generation of interior hulls, since the number of ``hulls within
hulls'' is  smaller by a factor  $c_h$.  With this  approach, only one
approximation  --  a   kind  of  mean-field  one  on   the  number  of
first-generation  hulls  within  a  parent  hull  (see  below)  --  is
necessary.

The same  line of  reasoning shows that,  in a hypothetical  theory in
which $c_h$ can be treated  as variable, the distinction between hulls
and domains  will disappear in the  limit $c_h \to 0$.  In this limit,
therefore,  the exponents $\tau$  and $\tau'$  must both  approach the
value 2, i.e.\ we can formally write $\tau = 2 +O(c_h)$ and $\tau' = 2
+ O(c_h)$. Furthermore,  due to the factor 2 that  appears in 
(\ref{eq:analytic-nh-Tinf}) but not in (\ref{eq:analytic-nh-Tc}), 
the ratio  $(\tau'-2)/(\tau-2)$ must approach the value 2
in the limit $c_h \to 0$. The  actual value of this ratio is $187/91 =
\tau'= 2.055$, not very far from  2. Indeed the difference is of order
$c_h$ as expected.

We can use the same line of argument to discuss $c_d$, $c_h$, $\lambda_d$ 
and $\lambda_h$. Since in the (hypothetical) limit $c_h \to 0$, hulls 
and domains become identical, it follows that in this limit one must have 
$c_d \to c_h$ and $\lambda_d \to \lambda_h$, i.e.\ $c_d = c_h + O(c_h^2)$, 
and $\lambda_d = \lambda_h[1 + O(c_h)]$. All of these results are 
consistent with the relations (\ref{eq:relation1}) and 
(\ref{eq:relation2}) derived from the sum rules (\ref{eq:totalarea}) 
and (\ref{eq:hulls-equal-domains}).

\subsubsection{The evolution of domain areas.}$\;$

\begin{figure}[h]
\psfrag{r1}{$R_1$}
\psfrag{r2}{$R_2$}
\psfrag{r3}{$R_3$}
\psfrag{r4}{$R_4$}
\begin{center}
\includegraphics[width=4cm]{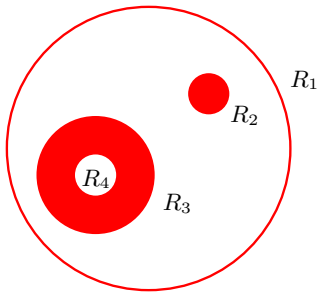}
\end{center}
\caption{(Colour online.) Sketch of a configuration with four circular
hulls and domains. The parent hull has radius $R_1$. There are two
first generation hulls with radius $R_2$ and $R_3$ and one second
generation hull with radius $R_4$. $\nu=2$ in this example.  The
interior border of the external domain is disconnected and has two
components.}
\label{fig:snapshot-Ising}
\end{figure}

Take a hull with enclosed area $A_h$ at time $t$.  This hull is also the
external border of a domain, which may itself contain one or more `first 
level' sub-domains whose external borders form the internal border (which 
may be disconnected) of the original domain. These external borders of 
the first level sub-domains are themselves `first generation' hulls lying 
within the parent hull. These interior hulls can themselves have 
interfaces in their bulk separating domains of the reversed phase (higher 
generation hulls), see Fig.~\ref{fig:snapshot-Ising} where we show a 
sketch with this structure.

Let us call $\nu(t)$ the number of first-generation hulls within the
parent one.  It is clear that $\nu(t)$ is semi-positive definite,
monotonically decreasing as a function of time and reaching zero at a
given instant $t_{max}$, when all interior hulls disappear and
$A_d=A_h$ thereafter. One can estimate $t_{max}$ from
$0=A^{int}_h(t_{max})=A^{int}_h(t_i)-\lambda_h (t_{max}-t_i)$, which 
yields $t_{max} -t_i= A^{int}_h(t_i)/\lambda_h$ where the index $int$
indicates that we are studying here the first generation hull with
maximal initial area (all others having already
disappeared). It is clear that $t_{max}-t_i$ is smaller but of the order 
of $A_h(t_i)/\lambda_h$ where we replaced $A_h^{int}(t_i)$ by the initial
area of the parent hull:
\begin{equation}
(t_{max}-t_i) \stackrel{<}{\sim} \frac{A_h(t_i)}{\lambda_h}
\; . 
\label{eq:tmax}
\end{equation}

We wish to write a differential equation for the time-evolution 
of the parent domain area. It is clear that, at first order in $dt$:
\begin{equation}
A_d(t+dt) = A_d(t) - \lambda_h dt + \nu(t) \lambda_h dt  
\end{equation}
where the second term in the right-hand-side represents the 
loss in area due to the inward motion of the external domain-wall 
while the last term is the gain in area due to the outward
motion of the first-generation internal domain-walls. This gives
\begin{equation}
\frac{dA_d(t)}{dt} = - \lambda_h \; [1- \nu(t)]
\; .
\label{eq:diff-eq-domains} 
\end{equation}

Differently from hull enclosed areas, that always decrease in size
as time passes, domain can either diminish ($\nu = 0$), increase
($\nu>1$) or conserve ($\nu=1$) their area in time. 

\subsubsection{The number of first generation interior hulls.}$\;$
\label{subsubsec:nu}

We cannot, of course, know the exact number of first generation 
hulls falling within a selected hull with enclosed area $A_h$. 
We can, however, estimate it with an 
upper bound obtained by counting all interior hulls
and  averaging over all parent hulls 
using $n_h(A,t)$ derived in Sect.~\ref{sec:hulls}.
Thus, we expect  
\begin{eqnarray}
\qquad\qquad\nu(t) &<& \langle \, \nu(t) \, \rangle_{A_h(t)}  
\nonumber\\
\langle \, \nu(t) \, \rangle_{A_h(t)} 
&\sim& 
A_h(t) \int_0^{A_h(t)} dA \; n_h(A,t) 
\nonumber\\
&=& 
\frac{c_h A^2_h(t)[\lambda_h (t-t_i)+A_0]^{-1}} 
{[A_h(t) + \lambda_h (t-t_i) +A_0]}
\; , 
\label{eq:numedia}
\end{eqnarray}
where we include a small area cut-off, $A_0$, in the denominator
of $n_h$ and, for concreteness, we use the hull enclosed
area distribution for critical Ising initial conditions.
This equation can be further simplified if one uses
that at time $t$ the hull enclosed area we are interested in is given by 
\begin{equation}
A_h(t) = A_h(t_i)-\lambda_h (t-t_i)
\; . 
\end{equation}
(We call here $A_h(t_i)$ the initial area of the hull.)   Then
\begin{eqnarray}
\langle \, \nu(t) \, \rangle_{A_h(t)}  = 
\frac{c_h [A_h(t_i)-\lambda_h (t-t_i)]^2}{[\lambda_h (t-t_i) +A_0]\; 
[A_h(t_i) +A_0]}
\; . 
\label{eq:nuaveraged}
\end{eqnarray}
Note that, although we over-counted the interior hulls by including 
second-generation, third-generation, etc.\ hulls, the number of these 
is of order $c_h^2$, $c_h^3,\ldots$ respectively, so this treatment is 
exact to leading order in $c_h$ except for the replacement of $\nu(t)$ 
by its average over all first-generation hulls of the same area.

The most interesting cases are such that $A_h(t_i) \gg A_0$, otherwise
the hull and domain areas are just identical or very similar. In these
cases $\langle \, \nu(t_i) \, \rangle_{A_h(t_i)} \sim c_h A_h(t_i)/A_0$.
Expression (\ref{eq:nuaveraged}) has the following limiting values
\begin{eqnarray*}
\langle \, \nu(t) \, \rangle_{A_h(t)} \sim 
\left\{
\begin{array}{ll}
\displaystyle
\frac{c_h\ A_h(t_i)}{\lambda_h (t-t_i)+A_0}, 
& A_h(t_i) \gg \lambda_h (t-t_i)
\; , 
\\ & \\ & \\
a c_h, 
&  A_h(t_i) \sim \lambda_h (t-t_i)
\; ,
\end{array}
\right.
\label{eq:limits}
\end{eqnarray*}
we used $A_h(t_i) \gg A_0$ in the last case, and $a$ is 
a numerical constant of the order of $A_h(t_i)$. 
The result is a very small quantity, of the order of $c_h$, in both
cases.  The remaining mathematical possibility, $A_h(t_i) < \lambda_h
(t-t_i)$ is not realized because $A_h(t)$ cannot be negative.   

While $\nu(t)$ vanishes at $t_{max}$, see
Eq.~(\ref{eq:tmax}), $\langle \, \nu(t) \, \rangle_{A_h(t)}$ is different
from zero at all times. Thus,
Eq.~(\ref{eq:nuaveraged}) cannot be used beyond the limit $t_{max}$
when all internal hulls have already disappeared and it is no longer
correct to replace $\nu(t)$ by $\langle \, \nu(t) \,
\rangle_{A_h(t)}$.

The analysis of infinite temperature initial conditions is identical 
to the one above with $c_h$ replaced by $2c_h$. Thus, $\langle \, \nu(t) \,
\rangle_{A_h(t)}$ is expected to take twice the value it takes for 
critical Ising initial configurations. 

\begin{figure}[h]
\centerline{
\psfrag{x}{\large\hspace{-10mm} $A_h(t)/\lambda_h t$}
\psfrag{y}{\large $\nu$}
\psfrag{a}{$T_0=T_c$}
\psfrag{b}{$T_0=\infty$}
\includegraphics[width=8cm]{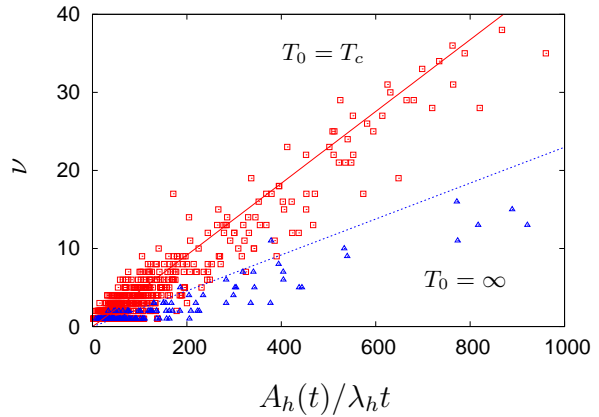}
}
\caption{(Colour online.) Comparison between $\nu(t)$ and $\langle \, \nu(t) \,
\rangle_{A_h(t)}$ for the $T=0$ evolution of the $2d$IM with
$T_0\to\infty$ and $T_0=T_c$ initial conditions. The measuring time is
$t=64$ MCs. The curves are given by Eq.~(\ref{eq:numedia}) in the
limit $t\gg t_i$ and $A_h(t) \gg A_0$, leading to the functional form
$\langle \, \nu(t) \, \rangle_{A_h(t)} = c_h x^2/(1+x) \sim c_h x$
when $x\gg 1$, with $x=A_h(t)/\lambda_h t$ and $\lambda_h=2.1$, see
Sect.~\ref{sec:numeric} for an explanation of the choice of this
value.}
\label{fig:nu}
\end{figure}

We have checked the accuracy of this approximation numerically by
counting the number of first generation internal hulls falling within
each parent hull at different times. 
Figure~\ref{fig:nu} shows the results for the zero
temperature evolution of the $2d$IM starting from $T_0\to\infty$ and
$T_0=T_c$ initial conditions.  While at very short times one sees
deviations between the numerical data and analytic prediction, the
agreement between the two becomes very satisfactory for times of the
order of $t=64$ MCs and longer, as shown in the figure. 

\subsubsection{The instantaneous domain area.}$\;$

If we now replace $\nu(t)$ by $\langle \, \nu(t) \, \rangle_{A_h(t)}$
given in Eq.~(\ref{eq:nuaveraged}), 
it is quite simple to integrate the differential
equation~(\ref{eq:diff-eq-domains}).  One finds
\begin{eqnarray*}
A_d(t) &=& A_d(t_i) - \lambda_h (1+2c_h) (t-t_i) 
\nonumber\\
&&
+ \,  
\frac{c_h A_0^2}{2[A_h(t_i)+A_0]} 
\left\{\left[1+\frac{\lambda_h (t-t_i)}{A_0}\right]^2-1 \right\}
\nonumber\\
&&
+ \,  
c_h [A_h(t_i)+A_0] \ln \left[1 +\frac{\lambda_h (t-t_i)}{A_0} \right]
\; . 
\end{eqnarray*}
Setting $t=t_i$ one recovers $A_d(t)=A_d(t_i)$ as required. 
In the natural cases in which $A_h(t_i) \gg A_0$ and 
 for long times such that $\lambda_h(t-t_i) \gg A_0$ 
this expression can be rewritten as
\begin{eqnarray}
A_d(t) &=& 
A_d(t_i) - \lambda_h 
\left[1+2c_h-\frac{c_h}{2} \frac{\lambda_h (t-t_i)}{A_h(t_i)} \right] 
(t-t_i) 
\nonumber\\
&&
+ \,
c_h A_h(t_i) \ln \left[1+\frac{\lambda_h (t-t_i)}{A_0} \right]
\; . 
\label{Ad(t)}
\end{eqnarray}
The factor in the second term 
\begin{eqnarray}
\lambda_d(t) \equiv 
\lambda_h \left[1+2c_h-\frac{c_h}{2} \frac{\lambda_h (t-t_i)}{A_h(t_i)}\right]
\; 
\end{eqnarray}
is a very weakly time-dependent function. Since
$t$ can take values between the initial time, $t=t_i$, and the maximum
time before the first generation hull itself disappears,
$t_{max}=t_i+A_h(t_i)/\lambda_h$, $\lambda_d(t)$ varies within the interval:
\begin{eqnarray}
\lambda_h \left(1+\frac{3c_h}{2} \right) \leq \lambda_d(t) \leq 
\lambda_h (1 + 2c_h)
\; . 
\end{eqnarray} 
These bounds are indeed very close. As expected from the analysis 
of the sum rules, see Sect.~\ref{sec:generic}, 
$\lambda_d$ takes a slightly higher value than 
$\lambda_h$; it equals
$\lambda_h$ plus a small correction of order $c_h$ (in practice, $1.035
\, \lambda_h \leq \lambda_d \leq 1.046 \, \lambda_h$ using the analytic
value for $c_h$).

The coefficient in front of the logarithm, $c_h A_h(t_i)$, is ${\cal
O}(c_h)$.  The sum rules, imply $c_h =c_d +{\cal O}(c_h^2)$.
Neglecting the higher order correction we can then replace $c_h$ by
$c_d$. The same applies to $A_h(t_i)$, which equals $A_d(t_i)$ plus a
term ${\cal O}(c_h)$ that we can equally neglect. Thus $A_h(t_i)
\approx A_d(t_i) \equiv A_i$. In this way we obtain
\begin{displaymath}
A_d(t,A_i) \simeq A_i - \lambda_d (t-t_i) +  c_d A_i 
\; 
\ln \left[1+\frac{\lambda_h(t-t_i)}{A_0} \right] 
\; .
\label{eq:expectation} 
\end{displaymath}
Inserting this result into Eq.~(\ref{eq:integration}), 
including the microscopic area $A_0$ as a small-area cut-off in the 
denominator of Eq.~(\ref{eq:nhpercolation}), then gives 
\begin{eqnarray}
n_d(A,t) & \simeq  & c_d A_0^{\tau-2}
\left\{1+ c_d 
\ln\left[ 1 + \frac{\lambda_d(t-t_i)}{A_0}\right] \right\}^{\tau-1}
\nonumber \\
& & \times [A + \lambda_d (t-t_i) + A_0]^{-\tau}
\label{eq:replace}
\end{eqnarray}
where we have replaced $\lambda_h$ by $\lambda_d$ inside the logarithm, 
which is correct to leading order in $c_h$.  
Using the fact that $c_d$ is very small and of the order of 
$(\tau-2)(\tau-1) = (\tau-2) + {\cal O}(c_h^2)$, as implied by the sum rules,
we can now exponentiate, correct to leading order in $c_d$, the logarithm 
in the curly brackets to obtain 
\begin{equation}
n_d(A,t) \simeq \frac{c_d \, [A_0 + \lambda_d(t-t_i)]^{\tau-2}}
{[A + A_0 + \lambda_d(t-t_i)]^\tau}
\; . 
\end{equation}
Finally we set the initial time, $t_i$, to zero and write the 
microscopic area, $A_0$, as $\lambda_d t_0$ to obtain the expected form
(\ref{eq:guess1}),
\begin{equation}
n_d(A,t) \simeq  
\frac{c_d \; [\lambda_d(t+t_0)]^{\tau-2}}{[A+\lambda_d (t+t_0)]^\tau}
\; , 
\label{guess1confirmed}
\end{equation}
 for the time-dependent number density of domain areas.

The same sequence of steps for infinite-temperature initial
conditions leads to the same form but with $c_d$ replaced by 
$2c_d$ and $\tau$ replaced by $\tau'$. The effects of temperature
are expected to appear only through the parameters $\lambda_d$ and
$\lambda_h$ once thermal fluctuations are extracted from the analysis.

The averaged domain area is then given by 
\begin{eqnarray}
\langle A\rangle(t) 
&=& 
\frac{\int dA' A' n_d(A',t)}{\int dA' n_d(A',t)} = \frac{1}{N_d(t)}
\\
&\propto& \lambda_d t
\; .  
\end{eqnarray}

\section{Statistics of areas: numerical tests}
\label{sec:numeric}

To test our analytic results we carried out numerical simulations on the
$2d$ square-lattice Ising model ($2d$IM) with periodic boundary
conditions using a heat-bath algorithm with random sequential
updates. All data have been obtained using systems with size
$L^2=10^3\times 10^3$ and $2\times 10^3$ runs using independent
initial conditions.  

Domain areas are identified with the Hoshen-Kopelman 
algorithm~\cite{Hoshen} while hull-enclosed ones are measured by
performing a directed walk along the interfaces, in analogy with the
algorithm in~\cite{Aharony}. A detailed description of our
algorithm is given in Appendix~\ref{app:algorithm}.

The equilibrium critical Ising initial conditions have
one spanning cluster (since the system is at the limit
of the percolation threshold), that grows during evolution.  No other
spanning cluster is later formed. By contrast, equilibrium infinite
temperature initial conditions are below the critical random percolation
point in $d=2$ but often after 2 MC steps two spanning
clusters appear that also grow during evolution.  After 20 MCs roughly
50\% of the spins lie typically on these clusters. This
implies that we need to simulate a large number of independent samples
to obtain a good statistics. 

It is important to note that the dynamics of the discrete model
includes processes that are not taken into account in the continuous
model, as given in Eq.~(\ref{allen-cahn}), for which we derived our
analytical results. Some of these processes are the fission of a big
domain into two smaller ones (that usually occurs by cutting a thin
bottle neck that joined them), or the coalescence of two domains to 
form a bigger one.  However, we shall prove that these processes are not
important and the dynamics of the discrete model is well described by
the analytic results.

\subsection{Initial conditions.}
\label{sec:infTinit}

We used three types of initial conditions: equilibrium at infinite
temperature, $T_0\to\infty$; equilibrium at the critical point,
$T_0=T_c$; equilibrium within the high temperature phase,
$T_0=T_c+\Delta T$ with $\Delta T>0$.

\begin{figure}[h]
\centerline{
\psfrag{x}{\large\hspace{-5mm} $A$}
\psfrag{y}{\large\hspace{-5mm} $n_h(A)$}
\includegraphics[width=8cm]{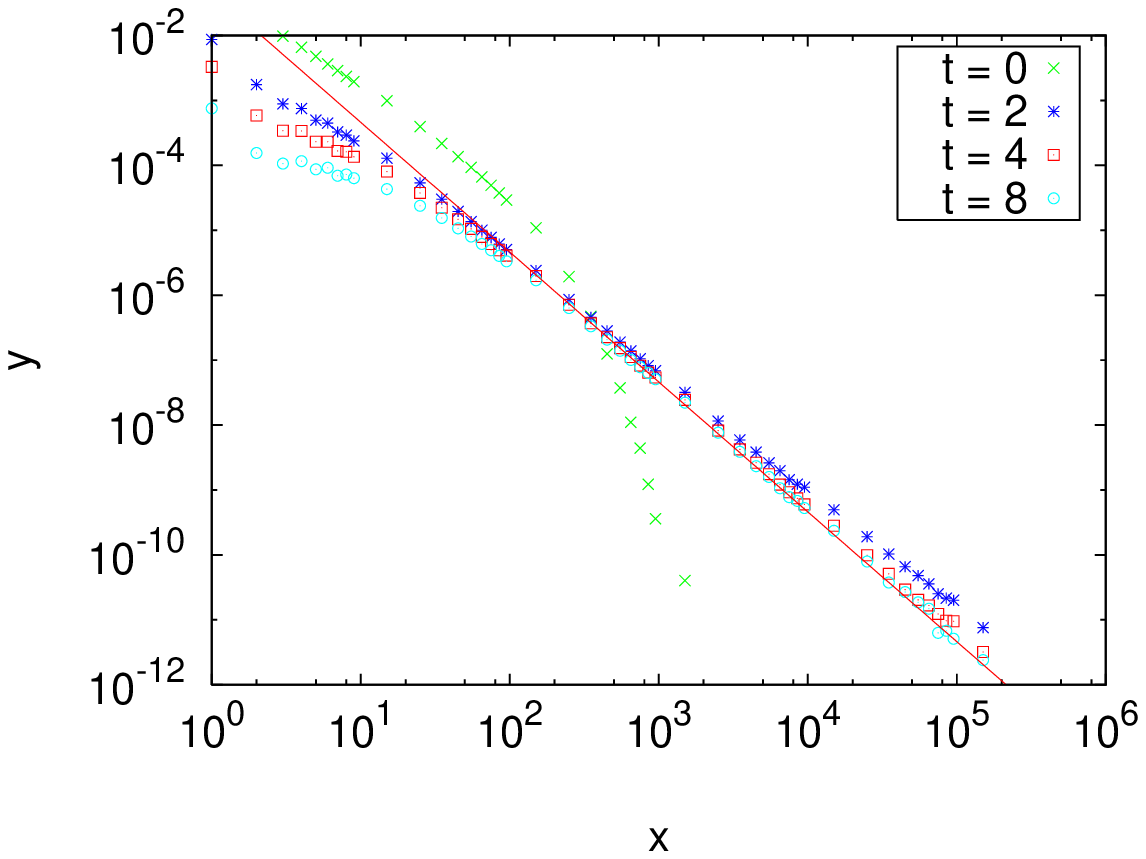}
}
\centerline{
\psfrag{x}{\large\hspace{-5mm} $A$}
\psfrag{y}{\large\hspace{-5mm} $n_d(A)$}
\includegraphics[width=8cm]{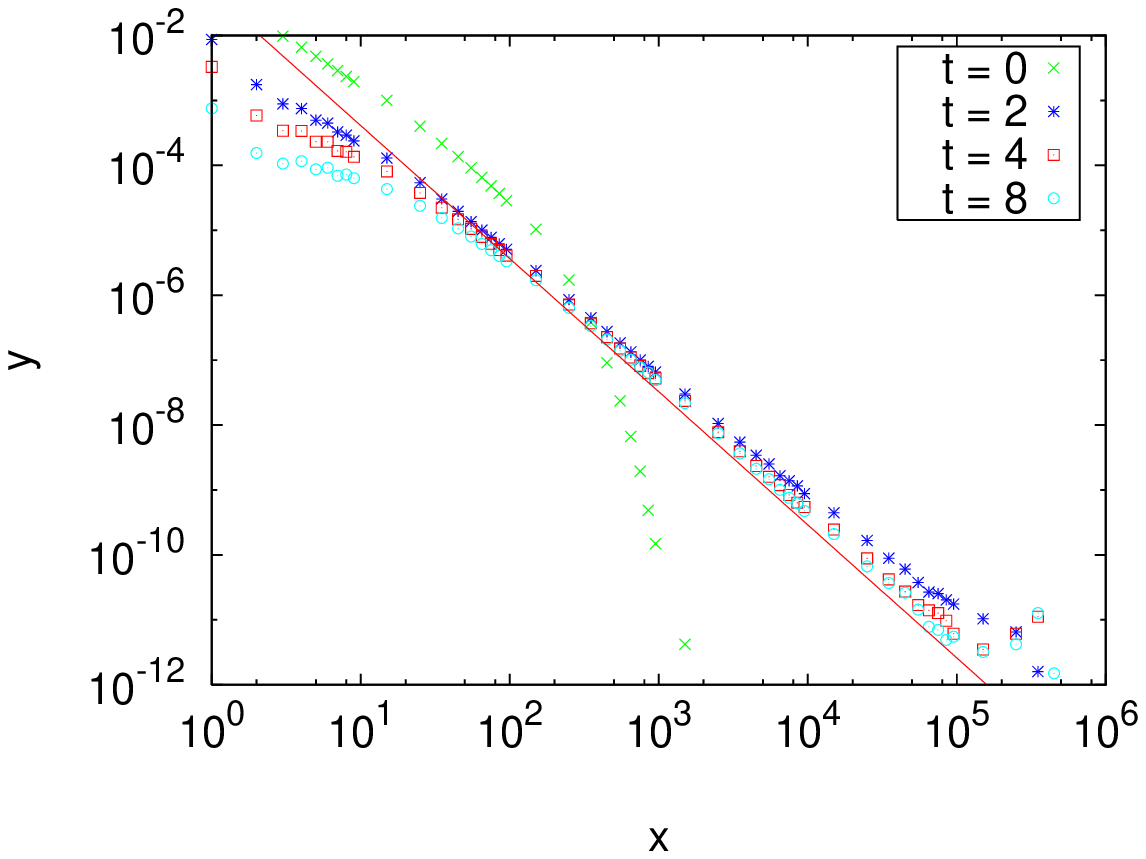}
}
\caption{(Colour online.) Early evolution of the infinite temperature
initial condition.  Top panel: in a few MC steps the hull enclosed area
distribution reaches the one of critical percolation
(\ref{eq:nhpercolation}), whose slope, -2, is shown by the straight
line~\cite{Cardy}.  Bottom panel: the domain area distribution. The
straight line is the power-law decay $A^{-2.055}$~\cite{Vander}.}
\label{fig:evolinitcondT05fromTinfhull}
\end{figure}

We mimicked an instantaneous quench from $T_0\to\infty$ by using
random initial states with spins pointing up or down with probability
$1/2$.  Assigning site occupation to up spins and vacant sites to down
spins the infinite temperature initial condition can be interpreted as
a percolation problem at $p=0.5$ and thus below the random percolation
transition $p_c=0.5927$ in a square bi-dimensional lattice.  Even if
initially away from criticality, in a few MC steps the hull enclosed
area distribution becomes the one in Eq.~(\ref{eq:nhpercolation}), as
shown in the top panel in 
Fig.~\ref{fig:evolinitcondT05fromTinfhull}. The initial
distribution lacks large areas, there being almost none with $A>10^3$,
and the tail of $n_h$ falls off too quickly well below the critical
percolation curve. In a few time steps large structures appear and the
tail of the distribution approaches the expected form at critical
percolation.  Simultaneously, the weight at small areas diminishes and
the curve progressively gets flatter. This effect can also be seen in
Fig.~\ref{fig:pdf-perim-equil-Tinf} where we display data for
perimeter lengths.  In the bottom panel in 
Fig.~\ref{fig:evolinitcondT05fromTinfhull}
we display our numerical results for the domain area distribution,
which are compatible with the form (\ref{eq:domequilTinfty}). It is
intuitively clear why this must be so.  If the system is
coarse-grained on the domain typical scale, $R(t)$, it will look
completely disordered.  When $R(t)$ is large compared to the lattice
spacing, the disorder will be that of continuum percolation, for which
the critical density is one half by symmetry in two
dimensions~\cite{continuous-perc,continuous-perc2}.  
It follows that the coarsening
system will be asymptotically at percolative criticality, i.e.\ the
dynamics self-tunes the system to percolative criticality in two
dimensions [provided $R(t)$ remains much smaller than the system
size]. The data show that, as far as the hull and domain area
distributions are concerned, this only takes a few Monte Carlo steps
in practice. During these few steps many small domains coalesce to
form larger ones meaning that the dynamics is dominated by processes
that are not taken into account by Eq.~(\ref{allen-cahn}). 
This argument also shows that the domain distribution in the scaling limit
indeed has the predicted $A^{-\tau'}$ tail. It is interesting to
remark that the system approaches the percolative critical state not
by increasing $p$ (indeed, the magnetization during the coarsening
process initially remains close to zero), but by decreasing the value
of $p_c$, from 0.5927 to 0.5, as the correlation between spins
increases.

We can look at this from another perspective in the context of the
continuum model. Consider a random field $\phi(\vec x)$, symetrically
distributed with respect to zero, with bounded variance and two-point
correlator $C(r)=\langle\, \phi(\vec x)\phi(\vec x + \vec r) \,
\rangle$ with $r=|\vec r|$.  The zero contour lines of this field can
be imagined to divide the plane into regions of black and white with
each contour line forming a boundary between black and white
regions. Provided that $C(r)$ falls off faster than $r^{-3/4}$ for
large $r$, this problem is known to belong to the standard percolation
universality class~\cite{continuous-perc2}.  If we
now identify $\phi(\vec x)$ with the order parameter field when
well-defined domain walls (the zero contour lines) have formed, we see
that the resulting domain structure corresponds to critical
percolation.

In the plots we use a  double logarithmic scale that serves as a first
check of the  power-law decay of the probability  distributions but it
is not accurate enough to examine the value of the constants $c_h$ and
$c_d$.  We delay  the presentation  of a  very precise  test  of these
parameters  to Sect.~\ref{subsec:coarseningT0}  where  we analyse  the
time-evolution of the distribution functions.

We obtained the initial states for the coarsening dynamics at the critical 
temperature, $T_0=T_c$, and at $T_0=T_c+\Delta T$, after running $10^3$ 
Swendsen-Wang algorithm steps. We checked that the systems are well 
equilibrated after these runs. The distribution of hull enclosed and domain 
areas at $T_c$ are consistent with the analytic forms -- not shown.

\begin{figure}[h]
\begin{center}
\psfrag{x}{\large $A$}
\psfrag{y}{\large\hspace{-5mm} $n_h(A,t)$}
\includegraphics[width=8cm]{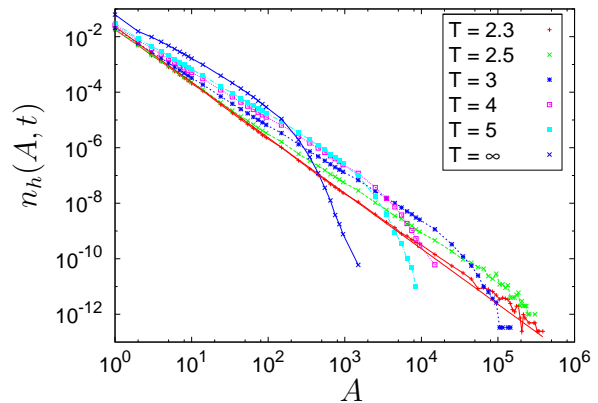}
\end{center}
\caption{(Colour online.) Equilibrium number density of hulls per unit
area for the high temperature phase of the $2d$IM. The distributions
seem power law for small areas, with a temperature dependent exponent
that approaches 2 when $T\to T_c$. Indeed, data for $T=2.3\simeq 1.01
\; T_c$ is almost coincident with Eq.~(\ref{eq:analytic-nh-Tc}).}
\label{fig:tmaiortc}
\end{figure}

\subsection{Coarsening at zero temperature}
\label{subsec:coarseningT0}

\subsubsection{Hull enclosed areas.}

\begin{figure}[h]
\begin{center}
\psfrag{x}{\large $A$}
\psfrag{y}{\large $n_h(A,t)$}
\includegraphics[width=8cm]{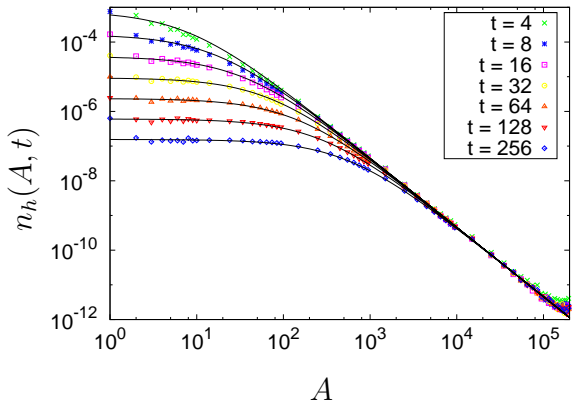}
\end{center}
\caption{(Colour online.) Number density of hull enclosed areas per
unit system area for the zero-temperature dynamics of the $2d$IM at
seven times evolving from an infinite temperature initial
condition. The lines represent Eq.~(\ref{eq:analytic-nh-Tinf})
with $c_h=1/8\pi\sqrt{3}$.}
\label{fig:hulls}
\end{figure}

In Fig.~\ref{fig:hulls} we show the time-dependent hull enclosed area
distribution in double logarithmic scale, at seven different times, 
following a quench from $T_0\to\infty$. The figure shows a strong
time dependence at small areas and a very weak one on the tail, which
is clearly very close to a power law. The curves at small areas move
downwards and the breaking point from the asymptotic power law decay
moves towards larger values of $A$ for increasing $t$.

\begin{figure}[h]
\begin{center}
\psfrag{x}{\large $A/\lambda_ht$}
\psfrag{y}{\large $(\lambda_ht)^2 \,  n_h(A,t)$}
\includegraphics[width=8cm]{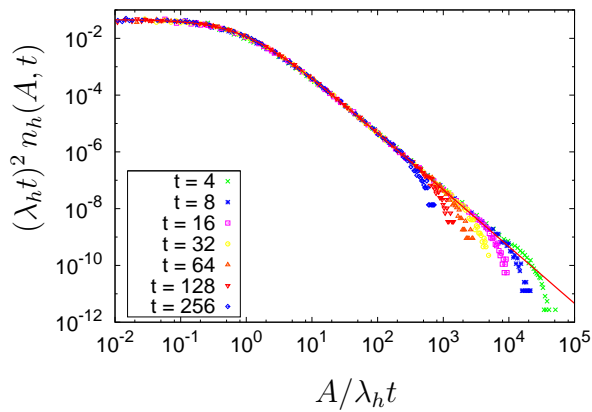}
\end{center}
\caption{(Colour online.) 
Number density of hull enclosed areas per unit system area for the 
zero temperature dynamics of the $2d$IM
evolving from an infinite temperature initial condition. The full 
line is the prediction (\ref{eq:analytic-nh-Tinf}) with $c_h=1/8\pi\sqrt{3}$ 
and $\lambda_h=2.1$.}
\label{fig:scaling-hulls-prl}
\end{figure}

\begin{figure}[h]
\begin{center}
\psfrag{x}{\large $A/\lambda_ht$}
\psfrag{y}{\large $(\lambda_ht)^2 \, n_h(A,t)$}
\psfrag{1000}{1000}
\psfrag{500}{500}
\psfrag{200}{200}
\psfrag{L=100}{$L=100$}
\includegraphics[width=8cm]{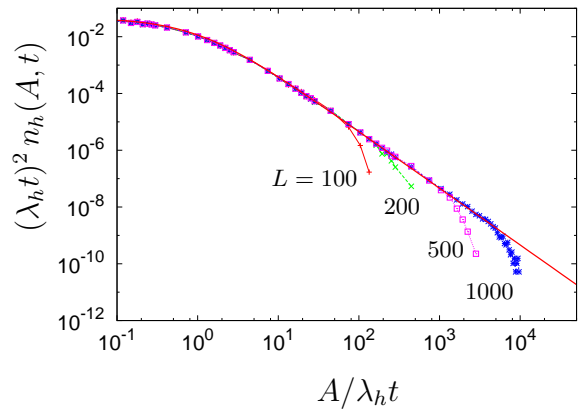}
\end{center}
\caption{(Colour online.) 
Finite size effects at $t=16$ MCs;  
four linear sizes of the sample are used and indicated by the 
data-points. The value of $A/t$ at which the data
separate from the master curve grows very fast with $L$ with
an exponent close to 2.}
\label{fig:finite-size}
\end{figure}

\begin{figure}[h]
\begin{center}
\psfrag{x}{\large $A/\lambda_ht$}
\psfrag{y}{\large $(\lambda_h t)^2 \, n_h(A,t)$}
\includegraphics[width=8cm]{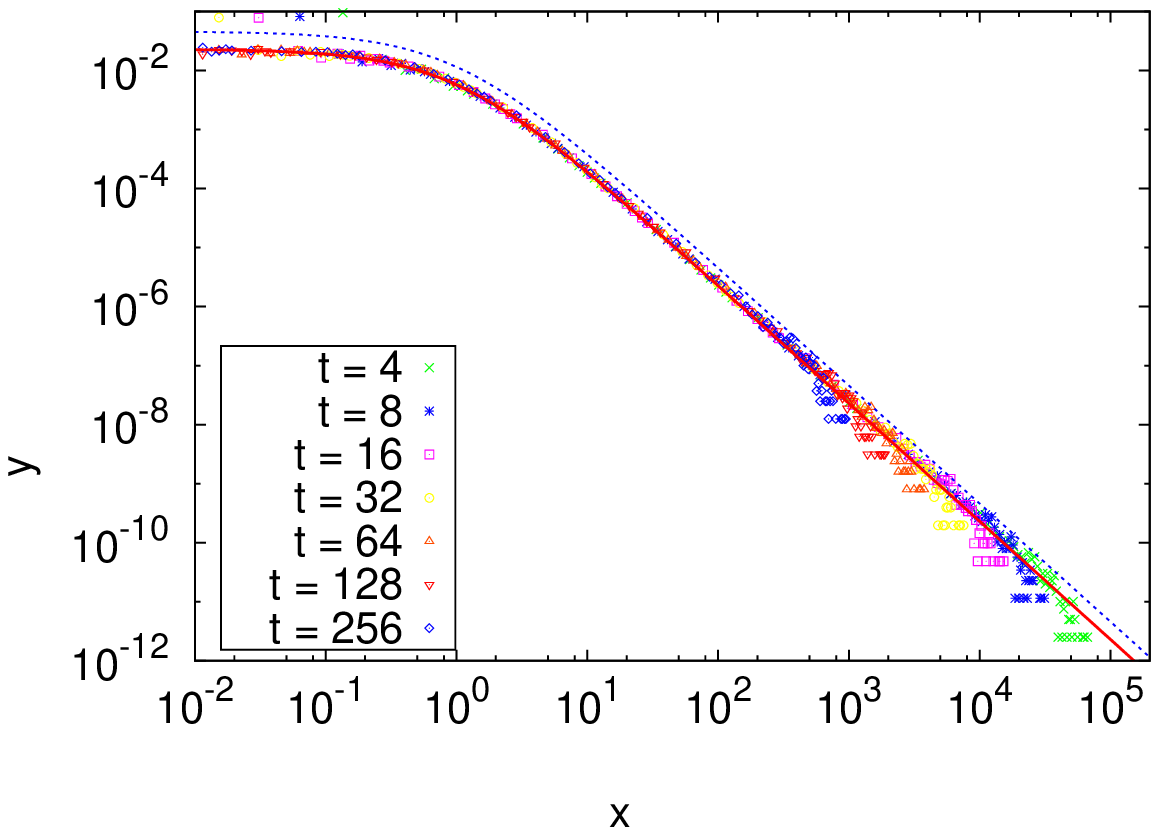}
\includegraphics[width=8cm]{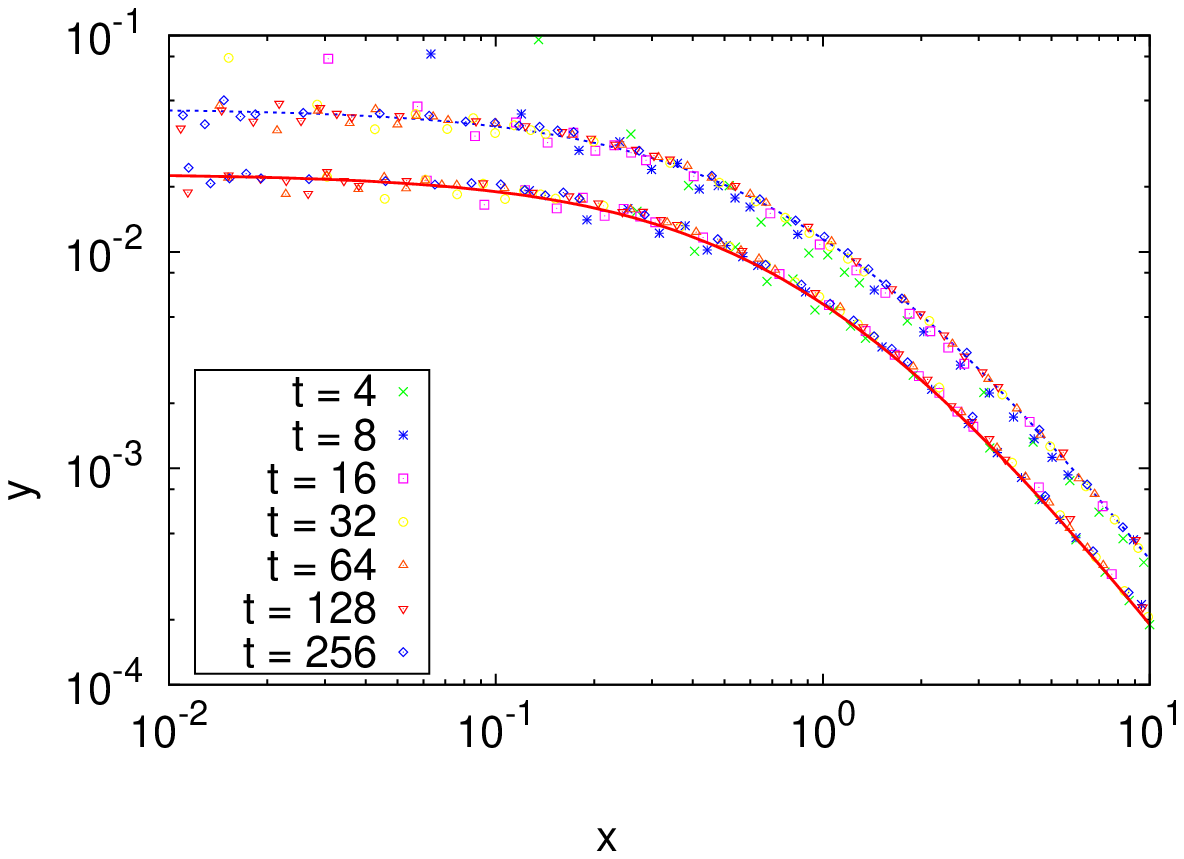}
\end{center}
\caption{(Colour online.) Number density of hulls per unit area for
the zero-temperature $2d$IM evolving from critical initial conditions.
The initial states are obtained after running $10^3$ Swendsen-Wang
algorithm steps. Top panel: the full (red) line is
Eq.~(\ref{eq:analytic-nh-Tc}) with $\lambda_h = 2.1$ which again
yields the best fit of the data at small areas.  For comparison we
include with a dotted (blue) line the analytic prediction for an
infinite temperature initial condition,
i.e. Eq.~(\ref{eq:analytic-nh-Tinf}).  Bottom panel: more details on
the influence of the initial conditions. The two data sets correspond
to configurations taken at several times after a quench from
$T_0\to\infty$ and $T_0=T_c$. The solid lines are the analytic
predictions (\ref{eq:analytic-nh-Tinf}), blue line, and
(\ref{eq:analytic-nh-Tc}), red line.}
\label{fig:critic-init-cond}
\end{figure}

In Fig.~\ref{fig:critic-init-cond} (bottom) we zoom on the small area
region ($A<10^3$) where the time-dependence is clearer and we scale
the data by plotting $(\lambda_h t)^2 n_h(A,t)$ against $A/\lambda_h
t$ with $\lambda_h=2.1$. We tried other time-dependent factors but
$\lambda_h t$ with this particular value of $\lambda_h$ is the one
yielding the best collapse of data at small areas, $A_0 \ll A <
\lambda_h t$. For $A$ larger than the `typical' value $\lambda_h t$
the time and $\lambda_h$ dependence becomes less and less important.
In Fig.~\ref{fig:scaling-hulls-prl} we show the data in their full
range of variation in log-log form to test the prediction $n_h(A,t)
\propto A^{-2}$ for large $A$. The data are in remarkably good
agreement with the prediction~(\ref{eq:analytic-nh-Tinf}) -- shown as
a continuous curve in the figure -- over the whole range of $A$ and
$t$.  The downward deviations from the scaling curve are due to
finite-size effects. The latter are shown in more detail in
Fig.~\ref{fig:finite-size}, where we display the $t=16$ MCs results
for several linear sizes. Finite size effects appear only when the
weight of the distribution has fallen by many orders of magnitude (7
for a system with $L=10^3$) and are thus quite irrelevant.  In the
tail of the probability distribution function (pdf) the numerical
error is smaller than the size of the data points.  The nearly perfect
agreement between the analytical theory and the data is all the more
impressive given that the curvature-driven growth underlying the
prediction (\ref{eq:analytic-nh-Tinf}) only holds in a statistical
sense for the lattice Ising model~\cite{Anisotropy}. Indeed, even at
small values of $A/\lambda_h t$, where the lattice and continuous
descriptions are expected to differ most, the difference is only a few
percent, as we shall show below.

It is clear that the evolution of the hull-enclosed area distribution
follows the same `advection law' (\ref{eq:advection}), with the same
value of $\lambda_h$, for other initial conditions. The evolution from
a critical temperature initial condition is shown in the top panel of
Fig.~\ref{fig:critic-init-cond}. A fit of the data at small areas
yields the value of the parameter $\lambda_h$ that, consistently with
the analytic prediction, takes the same value $\lambda_h=2.1$.  In the
bottom panel, we compare the time-dependent hull enclosed area
distributions for the initial conditions $T_0\to\infty$ and $T_0=T_c$
and we zoom on the behaviour of $n_h(A,t)$ at small areas,
$A/\lambda_h t \leq 10$. The two solid lines correspond to the
numerator in $n_h$ being equal to $2c_h$ for infinite temperature
initial conditions and $c_h$ for critical Ising initial
conditions. The difference between the numerical data for the two
initial states is clear and it goes in the direction of the analytic
prediction (a factor 2 difference in the constant).  Finally, while
the log-log plot in Fig.~\ref{fig:scaling-hulls-prl} suggests that the
data are compatible with $c_h \approx 0.023$ this way of presenting
the data is not precise enough to let us quantify the accuracy with
which we match the analytic prediction.  We test the numerical values
of the constant $c_h$ in detail in Sect.~\ref{sec:numeric} where the
numerical error is also estimated.

\begin{figure}[h]
\begin{center}
\psfrag{x}{\large\hspace{-5mm} $A/\lambda_ht$}
\psfrag{y}{\large\hspace{-1cm} $(\lambda_ht)^2 \, n_h(A,t)$}
\psfrag{a}{$T_0=T_c$ [Eq.~(\ref{eq:analytic-nh-Tc})]}
\psfrag{b}{$T_0=\infty$ [Eq.~(\ref{eq:analytic-nh-Tinf})]}
\includegraphics[width=8cm]{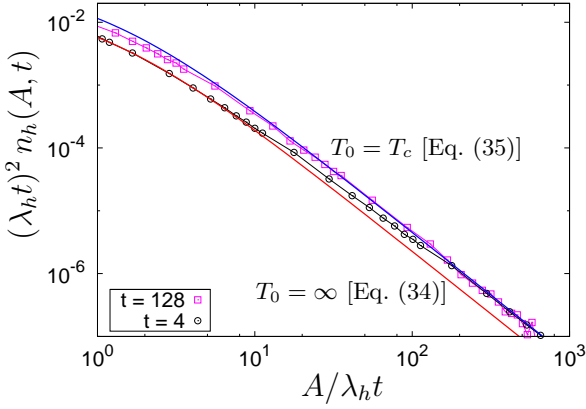}
\end{center}
\caption{(Colour online.) Effect of the initial condition temperature
on the hull enclosed area distribution. The continuous lines are the analytic
results for equilibrium initial conditions at $T_0=\infty$ and
$T_0=T_c$, bottom (red) and top (blue) lines, respectively. In
between, we present numerical data for two different times given 
in the key after the
quench from an initial state equilibrated at $T_0=2.5$.}
\label{fig:intermTihulls}
\end{figure}

Moreover, Eq.~(\ref{eq:analytic-nh-Tinf}) applies to {\it any}
$T_0>T_c$ equilibrium initial condition asymptotically. Equilibrium
initial conditions at different $T_0>T_c$ show only a different
transient behaviour: the closer they are from $T_c$, the longer it
takes to reach the asymptotic law, Eq.~(\ref{eq:analytic-nh-Tinf}).
Equilibrium initial distributions, for $T_c<T_0<\infty$, are shown in
Fig.~\ref{fig:tmaiortc}, while Fig.~\ref{fig:intermTihulls} shows an
example of their subsequent evolution. In the latter, both analytic
predictions, for $T_0=T_c$ and $T_0=\infty$, are shown as solid lines
along with data for increasing times after a zero-temperature quench
from $T_0=2.5$. In the first steps, the curve follows the one for
critical initial conditions at small $A/\lambda_ht$ and then departs
to reach the one for infinite temperature initial conditions at large
$A/\lambda_ht$. At longer times, the deviation from the critical
initial condition line occurs at a smaller value of $A/\lambda_ht$.
Initially the system has a finite, though relatively small,
correlation length $\xi(T)$. Thermal fluctuations with linear size of
the order of $\xi$ and also significantly larger than $\xi$ exist (see
the discussion on the effect of thermal fluctuations in
Sect.~\ref{sec:effectoftemperature}).  Notice that $\xi(T)$ does not
correspond exactly to the size of geometric domains: thermal
fluctuations are not perfectly described by domains of aligned spins,
since not all of them are correlated.  At any given temperature above
$T_c$, fluctuations smaller than $\xi(T)$ have the same statistics
than those occurring at $T_c$ and are thus described by
Eq.~(\ref{eq:analytic-nh-Tc}), while domains larger than $\xi(T)$ are
not made of correlated spins and thus are described by the infinite
temperature distribution, Eq.~(\ref{eq:analytic-nh-Tinf}). As time
increases, the system loses memory of the finite-size fluctuations and
the asymptotic state does not differ from the infinite temperature
one. Only when fluctuations exist over all spatial scales does the
asymptotic state differ.

This behaviour can be interpreted as follows. At
fixed $A/t$, shorter times correspond to small areas while longer
times are related to larger areas.  Very small areas correspond to
short linear sizes, of the order of the domains in the initial
configurations, and thus reminiscent of critical ones. Instead large
areas correspond to long linear sizes that are much longer than the
correlation length and closer to the ones reached from the infinite
temperature initial condition.

\subsubsection{Domain areas.}

We now study the domain areas and perform the same tests
as above though focusing on the analytic predictions  
(\ref{eq:guess1}) and (\ref{eq:guess2-bis}). 

\begin{figure}[h]
\psfrag{x}{\large\hspace{-5mm} $A/\lambda_dt$}
\psfrag{y}{\large\hspace{-1cm} $(\lambda_dt)^2 \, n_d(A,t)$}
\includegraphics[width=225pt]{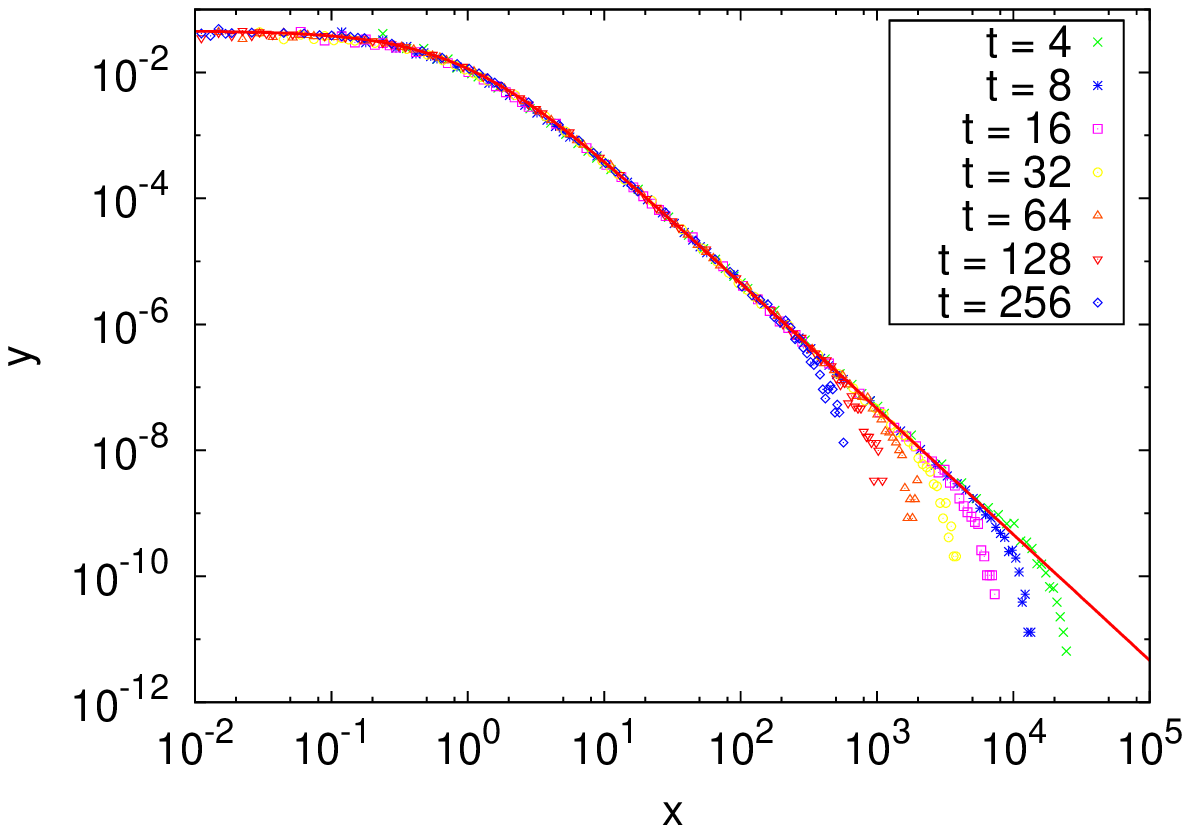}
\includegraphics[width=225pt]{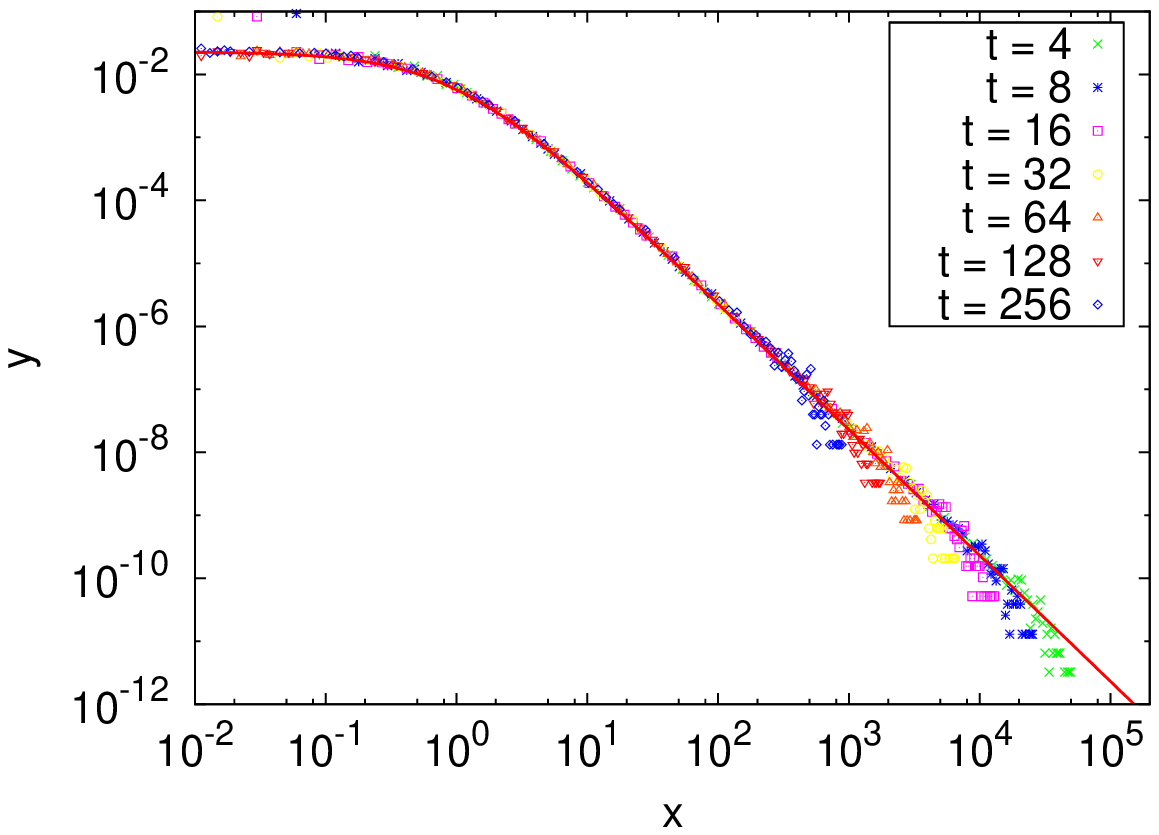}
\caption{(Colour online.) Number density of domains per unit area
for the zero-temperature $2d$IM evolving from $T_0\to\infty$ (top) and
$T_0=T_c$ (bottom) initial conditions. In both figures the spanning
clusters have been extracted from the analysis (compare with 
Fig.~\ref{fig:domains-complete} where we include them).
The full (red) line represents Eq.~(\ref{guess1confirmed}), with
$c_d=0.025$ and $\tau'=2.055$, $c_d\to c_d/2$ and $\tau=2.027$ (bottom panel),
and $\lambda_d=2.1$ in both cases.}
\label{fig:domains}
\end{figure}

\begin{figure}[h]
\psfrag{x}{\large\hspace{-5mm} $A/\lambda_dt$}
\psfrag{y}{\large\hspace{-1cm} $(\lambda_dt)^2 \, n_d(A,t)$}
\includegraphics[width=225pt]{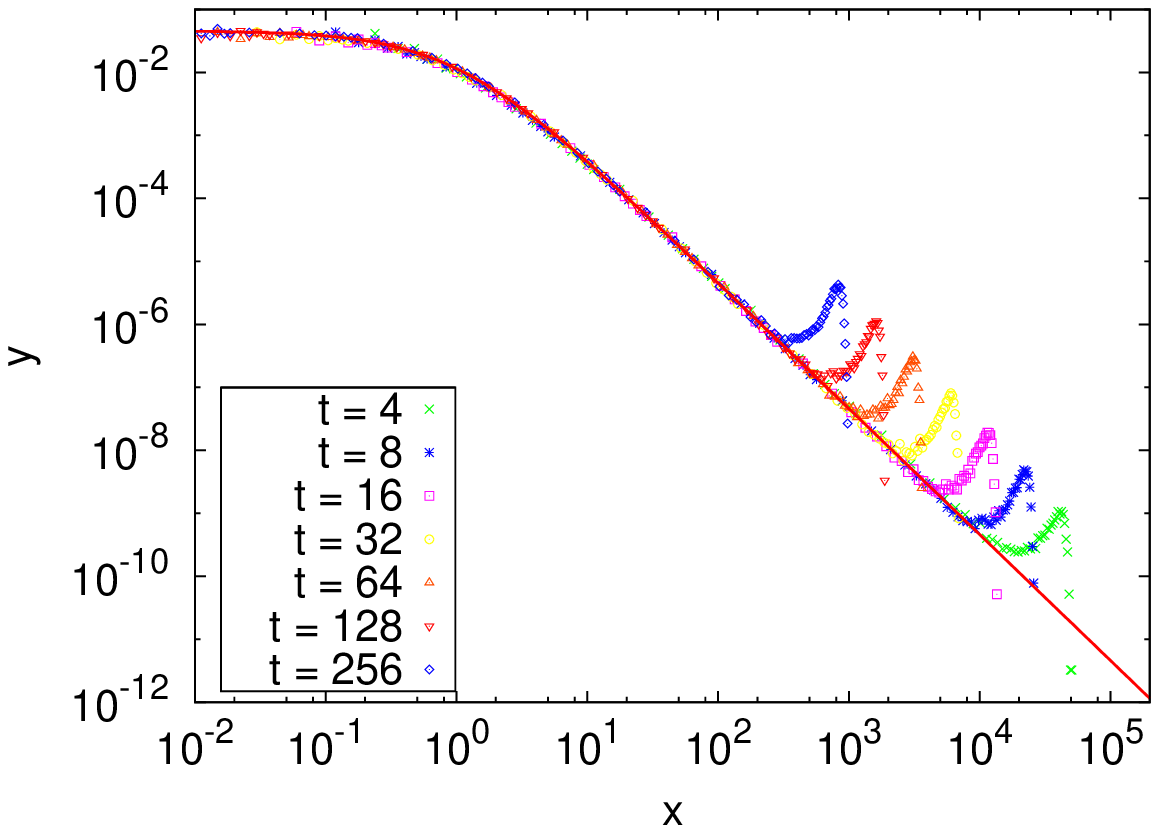}
\includegraphics[width=225pt]{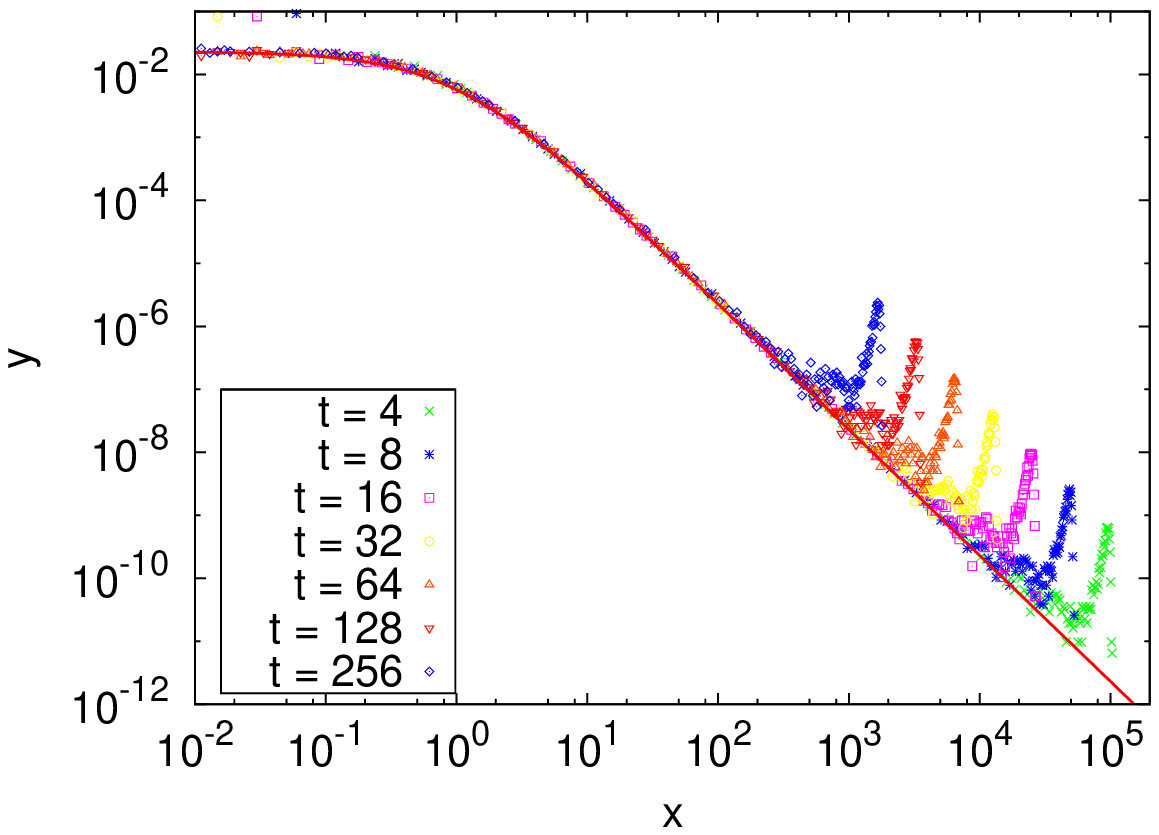}
\caption{(Colour online.) The same as Fig.~\ref{fig:domains}, but with
the spanning domains included in the statistics. Notice that due to
the finiteness of the lattice, the actual area of these spanning
domains are smaller than the value they would have on an infinite
system, generating the overshoot of the distribution for values of $A$
close to $L^2$. The larger is the system, the more to the right these
peaks are localized.}
\label{fig:domains-complete}
\end{figure}

In Fig.~\ref{fig:domains} we display the number density of domain
areas in the scaled form for two initial conditions, $T_0\to\infty$
and $T_0=T_c$, after removing any spanning domain from the statistics. 
For comparison, in Fig.~\ref{fig:domains-complete}, the same distributions
with the spanning domains are shown. As done for the hull enclosed areas 
we fit the parameter $\lambda_d$ by analysing the behaviour at small areas,
$A<\lambda_d t$, and we find, once again that $\lambda_d=2.1$ yields
the best collapse of data (see the discussion in
Sect.~\ref{sec:domains}).  We use the extrapolated value $c_d=0.025$
obtained with the numerical analysis described in full detail in
Sect.~\ref{subsub:constants}. Note that we expect the difference between  
$c_d$ and $c_h$ to be of order $c_h^2$, and thus rather hard to observe 
numerically.

Both sets of figures, \ref{fig:domains} and
\ref{fig:domains-complete}, exhibit finite size effects in the tail of
the distributions, where the number of domain areas has already
decreased by several orders of magnitude. As for the hull enclosed
areas, the point where these finite size effects cause the deviation
from the collapsed curve moves towards the right as the system size
increases, becoming less and less relevant.  In
Fig.~\ref{fig:domains}, large domain areas (violating the limiting
condition $A\ll L^2$) that would nonetheless be accounted for in an
infinite system are here removed since they span the system in one of
the directions, leading to the downward bending of the
distribution. On the other hand, in Fig.~\ref{fig:domains-complete},
when counting these domains, they are chopped by the system
boundaries, thus contributing to the distribution in a region shifted
to the left, accounting for the bumps seen in the figure.

\subsubsection{Study of the constants $c_h$ and $c_d$.}$\;$
\label{subsub:constants}

In order to improve the data analysis we followed the procedure used
by Cardy and Ziff, who studied the `finite area' scaling of the
cumulative distribution between $A$ and $2A$~\cite{Cardy}. The method
is as follows. For hull enclosed and domain areas the total number of
areas between $A$ and $2A$ is $N_{h,d}(A,2A)= \int_{A}^{2A} dA' \;
n_{h,d}(A')$.  Using the analytic prediction for $n_h$ one finds $2A
N_{h}(A,2A)\sim c_h$ for large areas at critical Ising conditions and  
without the factor two at critical percolation. Following Ref.~\cite{Cardy},
we assume that there are power-law finite area corrections and add
a term like  $a A^{-b}$ to the above expressions. From this relation one
extracts the value of $c_h$. Similarly, for domains one can use
$(1-\tau) N_{d}(A,2A)/[(2A)^{1-\tau}-A^{1-\tau}]= c_d + a A^{-b}$
at critical Ising initial and its modified form at critical percolation
initial conditions.

\begin{figure}[h]
\centerline{
\psfrag{x}{\large\hspace{-5mm} $A^{-0.875}$}
\psfrag{y}{\large $c_h$}
\includegraphics[width=8cm]{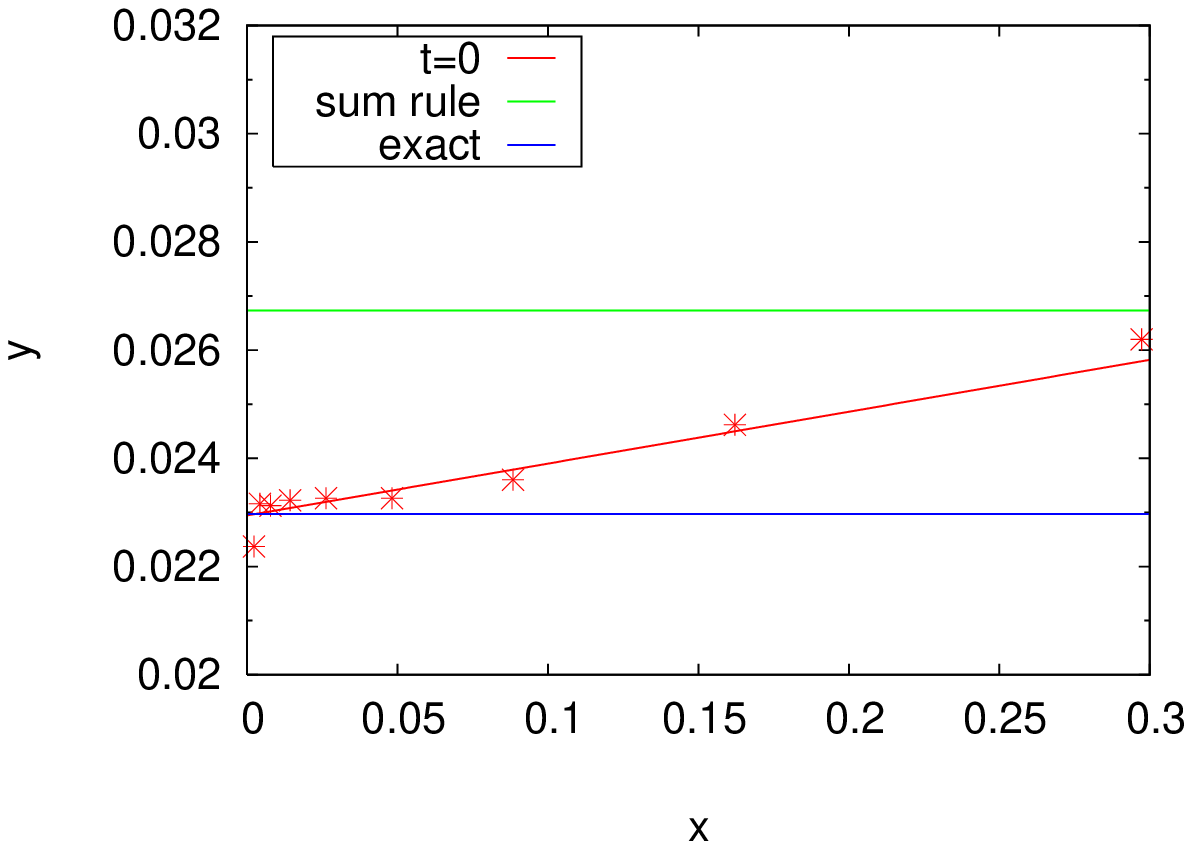}
}
\centerline{
\psfrag{x}{\large\hspace{-7mm} {$ {(A/\lambda_dt)}^{-0.875}$} }
\psfrag{y}{\large $c_h$}
\includegraphics[width=8cm]{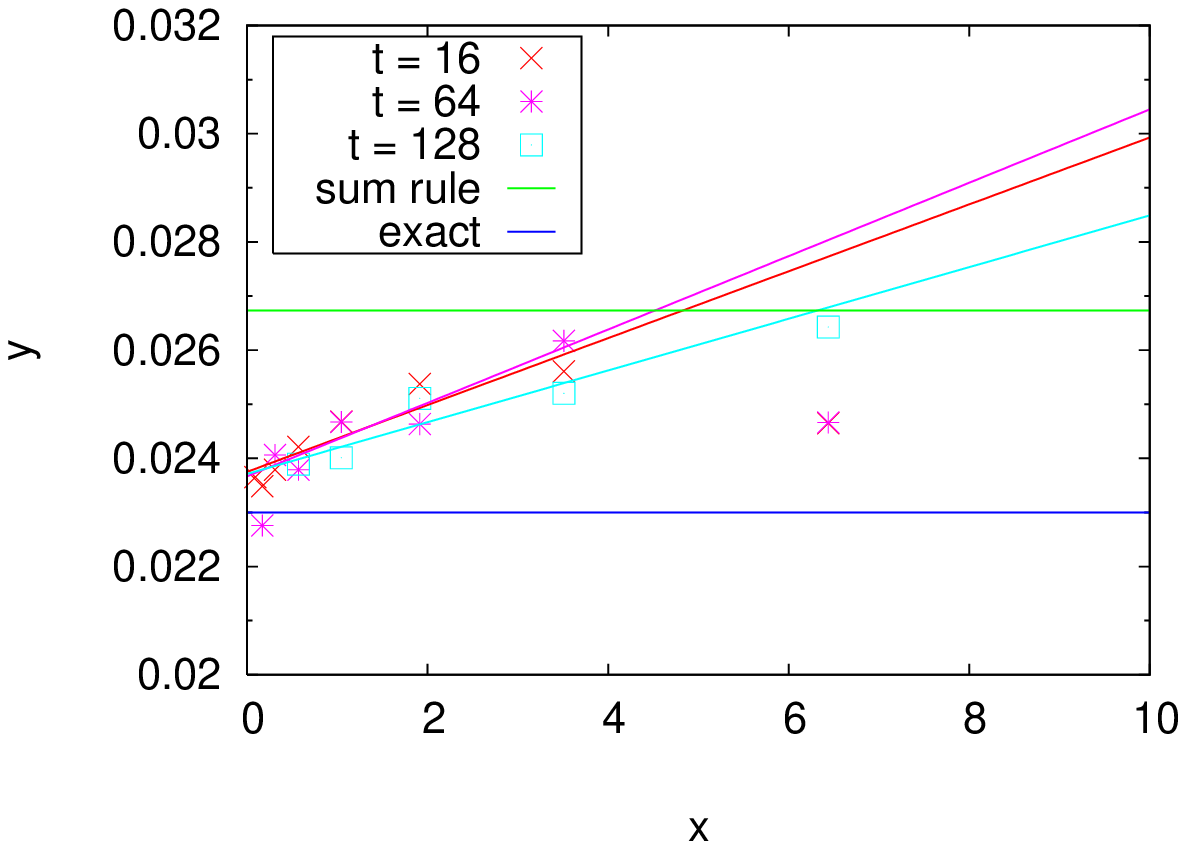}
}
\caption{(Colour online.) Study of $c_h$ using finite area scaling
implemented as in~\cite{Cardy}. The numerical data correspond to
$2AN_h(A,2A)$ in the equilibrium case and the left-hand-side of
Eq.~(\ref{eq:dyn-cd}) in the dynamic case.  Top panel: equilibrium
distribution of hull-enclosed areas at criticality.  The upper
horizontal line is the predictions for $c_h$ stemming from the use of
the sum rule.  The lower horizontal line is the analytic prediction.
The intercept of the inclined straight line with $x=0$ is the
numerical test of the analytic prediction. Bottom panel: study of $c_h$ during
coarsening after a quench from criticality to zero temperature. Note
that the prediction from the fit is slightly higher than the analytic
value, see the discussion in the text.}
\label{fig:chcd-improved}
\end{figure}

\begin{figure}[h]
\centerline{
\psfrag{x}{\large\hspace{-7mm} {$ {(A/\lambda_dt)}^{-0.875}$} }
\psfrag{y}{\large $c_h$}
\includegraphics[width=8cm]{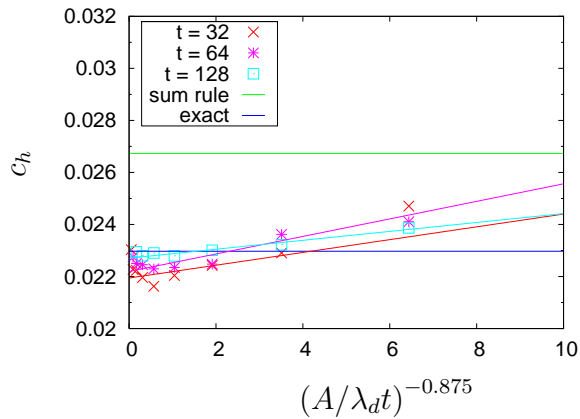}
}
\caption{(Colour online.) Study of $c_h$ during coarsening after a
quench from $T_0\to\infty$ to $T=0$ using finite area scaling
implemented as in~\cite{Cardy}.  Note that the prediction from the fit
weakly depends on time.}
\label{fig:chcd-improved2}
\end{figure}

\begin{figure}[h]
\centerline{
\psfrag{x}{\large\hspace{-7mm} {$ {(A/\lambda_dt)}^{-0.875}$} }
\psfrag{y}{\large $c_d$}
\includegraphics[width=8cm]{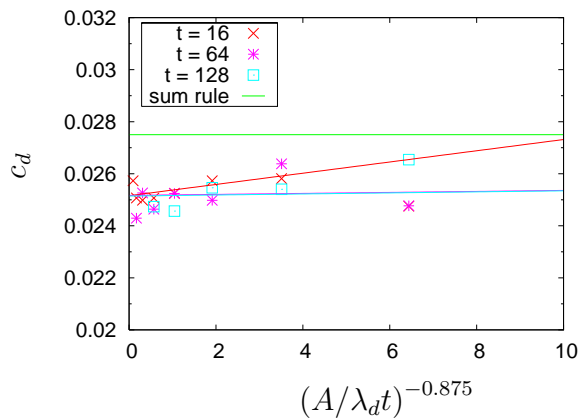}
}
\caption{(Colour online.) Study of $c_d$ using finite area scaling
implemented as in~\cite{Cardy}.  $c_d$ extracted from time-dependent
pdfs evolving at $T=0$ after a quench from criticality. The numerical
value of $c_d$ is roughly the same for all times.}
\label{fig:chcd-improved3}
\end{figure}

In the dynamic case we have
\begin{eqnarray}
N_{h,d}(A,2A;t) & \equiv & N_{h,d}(A,t) - N_{h,d}(2A,t) \nonumber \\
& = & \int_{A}^{2A} dA' \; n_{h,d}(A',t)\; . 
\end{eqnarray}
Using the predictions for $n_{h,d}$ we find
\begin{equation}
N_h(A,2A;t) = \frac{2c_h A}{(A+\lambda_h t)(2A+\lambda_h t)}
\end{equation}
and 
\begin{eqnarray}
N_d(A,2A;t) & = & \frac{2c_d (\lambda_d t)^{\tau'-2}}{1-\tau'}\; 
[(2A+\lambda_d t)^{1-\tau'} \nonumber \\ 
&& \hspace*{0.5cm} - (A+\lambda_d t)^{1-\tau'}]
\end{eqnarray}
for $T_0\to\infty$ and without the factor two and with the exponent 
$\tau'$ replaced by $\tau$ for $T_0=T_c$. To extract the values of the 
constants $c_h$ and $c_d$, we rewrite these forms as 
\begin{eqnarray}
&& \hspace*{-1.5cm} (2A)^{-1}(A+\lambda_h t)(2A+\lambda_h t)N_h(A,2A;t) 
\nonumber \\ 
&& \hspace*{2cm}= c_h +a\left(\frac{A}{\lambda_h t}\right)^{-b}
\label{eq:dyn-ch}  
\end{eqnarray}
and 
\begin{eqnarray}
&& \hspace*{-1.5cm} \frac{(\lambda_d t)^{2-\tau'} (1-\tau') N_d(A,2A;t)}
{2[(2A+\lambda_d t)^{1-\tau'}-(A+\lambda_d t)^{1-\tau'}]} \nonumber \\
&& \hspace*{2cm} = c_d + a \left(\frac{A}{\lambda_d t}\right)^{-b}
\label{eq:dyn-cd}  
\end{eqnarray}
and similarly for $T_0=T_c$. $a$ is a constant that takes different 
values for different times. 
In Figs.~\ref{fig:chcd-improved}-\ref{fig:chcd-improved3}
we show the outcome of this analysis. We use the same scale 
on the vertical axis in all plots to compare the accuracy 
of the results. 

In Fig.~\ref{fig:chcd-improved} we show results for the hull enclosed
areas. In the top panel we study equilibrium data at $T_c$. The
extrapolation of the numerical data approaches the analytic prediction
for $c_h$ within a 2\% accuracy.  We use the value of the exponent
$b$ proposed by Cardy and Ziff~\cite{Cardy}.  The estimation from the
approximate use of the sum rules is well above the analytic prediction
and numerical value.  In the bottom panel we test the value of $c_h$ in our
dynamic prediction by studying data at three instants, $t=16, \, 64,
\, 128$ MCs, after a critical Ising initial configuration.  In this
analysis we use the same exponent, $b=0.875$, as in the study of the
equilibrium data and we fit the other parameters, $c_h$ and $a$.
The fits of the time-dependent data to straight lines extrapolate to
the {\it same} value that is, however, of the order of 5\% off the
analytic prediction. Note the non-monotonic character of the
time-dependence in the slopes $a$.  One has to keep in mind that
the extrapolated value is very sensitive to the fit, in particular, to
how many data points are considered.

In the analysis of the infinite temperature initial conditions we are
forced to use dynamic results to reach, first the critical percolation
situation, and next follow the coarsening evolution. In
Fig.~\ref{fig:chcd-improved2} we display this type of data for the
instants given in the key.  The prediction from the fit is slightly
different from the analytic result at short times but it approaches the
analytic value, shown with a horizontal line, 
at sufficiently long times (look at the $t=128$ MCs
results). In conclusion we find 
\begin{equation}
c_h \sim 0.0229 \; \pm \; 0.0015
\; . 
\end{equation}

Next we study the constant $c_d$. First we use critical temperature
initial conditions using the dynamic results to avoid the ambiguity
introduced by the units restoring constant $A_0$ in equilibrium.
In short we find 
\begin{equation}
c_d \sim 0.0251 \; \pm \; 0.0015
\; . 
\end{equation}
Note that this value is slightly higher than $c_h$, in accord with the
prediction from the sum rules and the analytic argument, and the
difference between the two ($c_d-c_h\sim 0.002$) is slightly smaller
than the one that is obtained from the sum rules.

\subsection{Effect of the working temperature}
\label{sec:effectoftemperature}

Up to now we have considered quenches to a zero working temperature. 
In this Section we investigate the effect of having a finite 
temperature on the dynamics.

The arguments in Sect.~\ref{sec:analytic} rely on the $T=0$ Allen-Cahn
equation~(\ref{allen-cahn}).  Temperature fluctuations have a two-fold
effect. On the one hand they generate {\it equilibrium} thermal
domains that are not related to the coarsening process. On the other
hand they roughen the domain walls thus opposing the curvature driven
growth and slowing it down. 

Renormalization group treatments
of domain growth dynamics~\cite{Humayun} have led to the idea that
a $T=0$ fix point controls the domain growth for all $T<T_c$, i.e. that
thermal fluctuations are irrelevant to the asymptotic dynamics of the
ordering system, their contribution being limited primarily to the
renormalization of temperature-dependent prefactors.

\begin{figure}[h]
\begin{center}
\psfrag{x}{\large $r$ }
\psfrag{y}{\large\hspace{-7mm} $C(r,t)$}
\includegraphics[width=8cm]{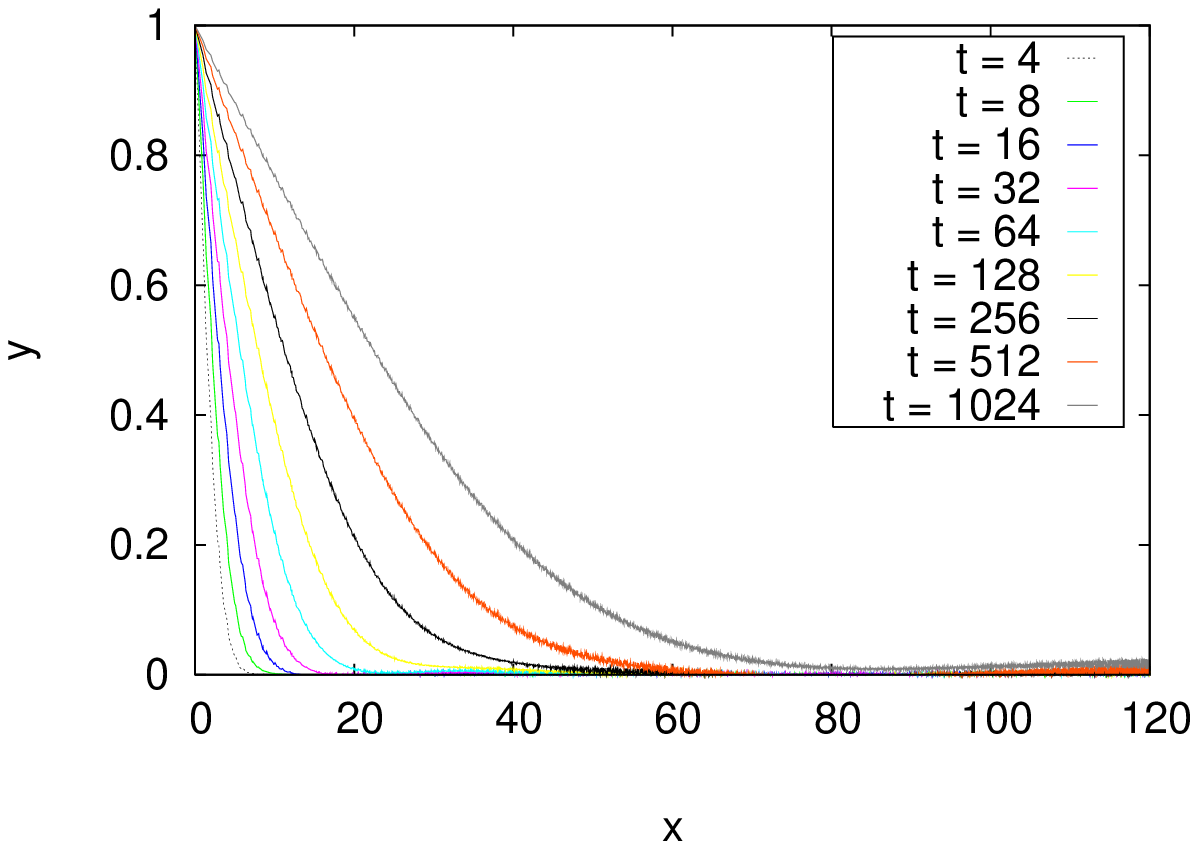}
\includegraphics[width=8cm]{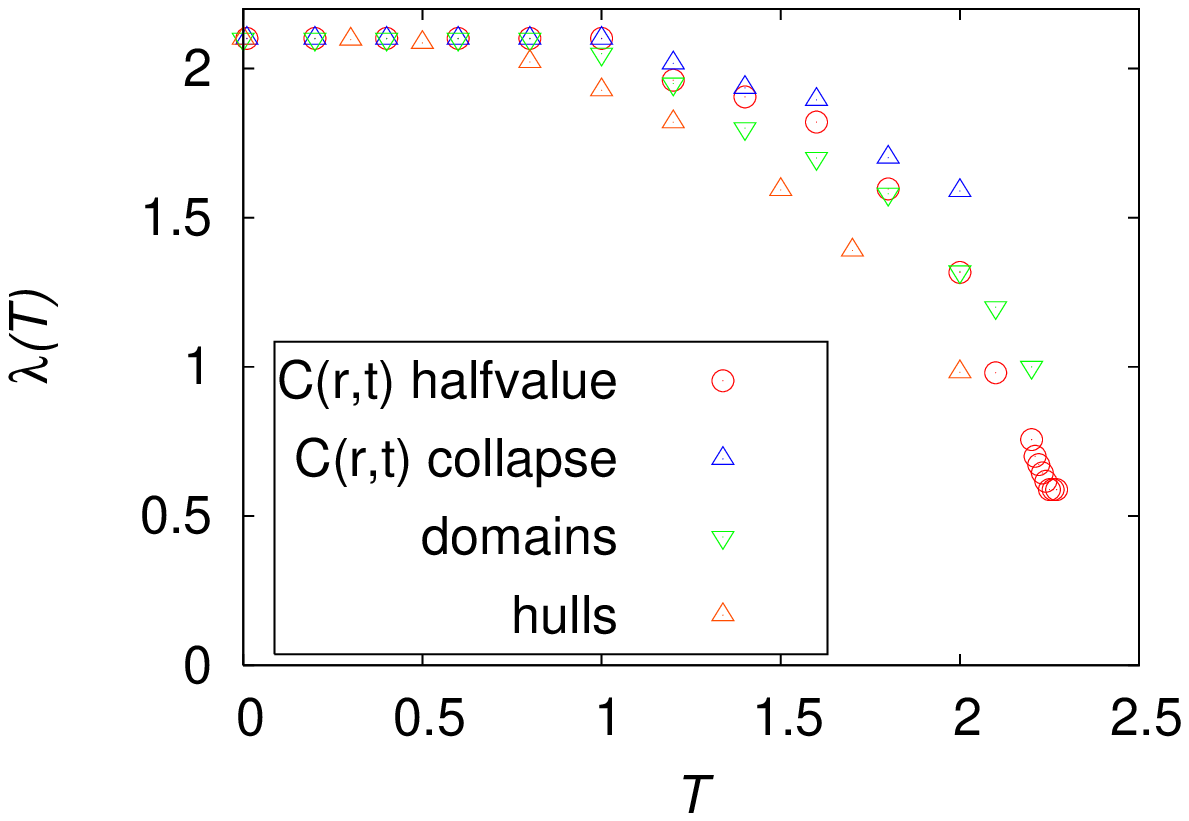}
\end{center}
\caption{(Colour online.) Top panel: spatial decay of the equal-time
correlation, Eq.~ (\ref{eq:Crt}), at fixed $T=0$ and several different
times $t$.  Bottom panel: the $T$ dependence of the parameters
$\lambda_{d,h}$.  Two sets of data points are extracted from the
analysis of the correlations shown in the top panel.  The data named
half-value are obtained from $C(r,t)=1/2$ and the data named collapse
from collapsing the curves on the range $r\gg a$.  The two other sets
are obtained from fitting $n_{d,h}(A,t)$ with $\lambda_{d,h}(T)$ as
free parameters.  }
\label{fig:Tdep-lambda}
\end{figure}

For the distribution of domain areas and hull enclosed
areas, one may expect that
once equilibrium thermal domains are
subtracted -- hulls and domains associated to the coarsening
process are correctly identified -- the full temperature dependence
enters only through the values of $\lambda_h$ and $\lambda_d$, which
set the time scale.

The first step then is to identify the temperature dependence
of the parameter $\lambda_d$. The simplest and most 
direct way to do this is to use the scaling hypothesis and 
analyze the behaviour of the spatial correlation 
\begin{eqnarray}
C(r,t) & \equiv & \frac{1}{N} \sum_{i=1}^N 
\langle \, s_i(t) s_j(t) \, \rangle|_{|\vec r_i - \vec r_j|=r} \nonumber \\
& \sim & m^2(T) \; f\left(\frac{r}{R(t)}\right)
\; ,
\label{eq:Crt}
\end{eqnarray}
where $m(T)$ is the equilibrium magnetization density and
 $a \ll r \ll L$ and $t_0 \ll t$. Using $R(t) \sim
[\lambda_d(T) t]^{1/2}$, the $T$-dependence of
$\lambda_d$ can be estimated either by collapsing all
curves or by studying the value of $r$ at which
$C(r,t)=1/2$. The resulting $\lambda_d(T)$ obtained using these
two prescriptions is shown in
Fig.~\ref{fig:Tdep-lambda}. $\lambda_d(T)$ is a monotonically decreasing
function of temperature, starting at $\lambda_d(T=0)=2.1$ and
falling-off to zero at $T_c$. These results are consistent with the
evaluation of $\lambda_{d,h}(T)$ from the analysis of $n_{d,h}(A,t)$, see
below, and it is at variance with what was previously presented
in Refs.~\cite{Kaski}.

Assuming that $\lambda_d$ vanishes at $T_c$ one can derive the way in
which it does with a simple argument~\cite{lacasse}.  We require that the coarsening
law for coarsening below $T_c$, namely $R(t) \sim [\lambda(T)
t]^{1/2}$, match critical coarsening at $T_c$, {\it viz.} $R(t) \sim
t^{1/z}$ with $z$ the dynamic exponent, for $T\to T_c$. Near (but just
below) $T_c$ the coarsening length grows as $\xi^{-a}(T) t^{1/2}$ as
long as $R(t) \gg \xi(T)$  with $\xi(T)$ the equilibrium correlation
length. For $R(t)$ comparable with $\xi(T)$, this has to be modified
by a function of $R(t)/\xi$ and, since $R(t) \sim t^{1/z}$ at $T_c$,
we can write
\begin{equation}
R(t) 
\sim \xi^{-a}(T) t^{1/2} f\left(\frac{t}{\xi^z(T)}\right)
\; . 
\end{equation} 
In the limit $\xi(T) \to \infty$, the $\xi$ - 
dependence must drop out. 
In order to cancel the time 
dependence at large times, one needs $f(x) \sim x^{1/2}$ for 
$x \to \infty$. This yields $R(t) \sim
t^{1/z}$, which fixes the exponent `$a$' as $a=(2-z)/2$, giving
$\lambda(T) \sim \xi^{-2a}(T) \sim (T_c-T)^{\nu(z-2)}$. 
Inserting the exact value $\nu=1$ and the numerical value 
$z=2.15(2)$~\cite{Leung} implies 
\begin{equation}
\lambda_d(T) \sim (T_c-T)^{0.15} \; . 
\end{equation}
Note that we are
matching two nonequilibrium growth laws -- the one below $T_c$ and the
one at $T_c$ -- not an equilibrium and a nonequlibrium one.
The data in Fig.~\ref{fig:Tdep-lambda} are still 
far from the critical region where this small power-law decay 
should show up.

\begin{figure}
\centerline{
\psfrag{x}{\large {$A$} }
\psfrag{y}{\large\hspace{-5mm} $n_h(A)$}
\includegraphics[width=8cm]{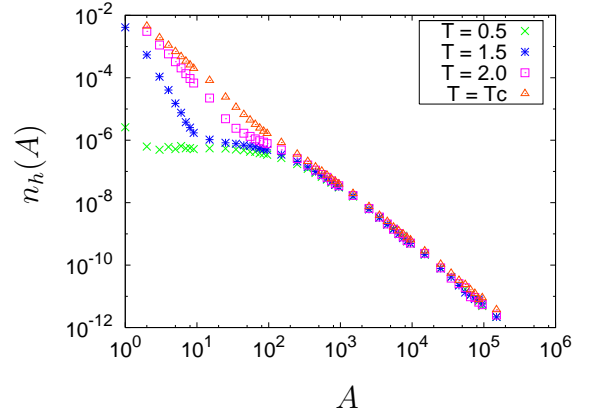}
}
\caption{(Colour online.) The number density of hull enclosed areas after
$t=128$ MCs at the working temperatures
$T=0.5,\; 1.5, \; 2$ and $T=T_c$.}
\label{fig:three-T}
\end{figure}

\begin{figure}
\begin{center}
\psfrag{x}{\large {$A$} }
\psfrag{y}{\large\hspace{-5mm} $n_h(A)$}
\includegraphics[width=8cm]{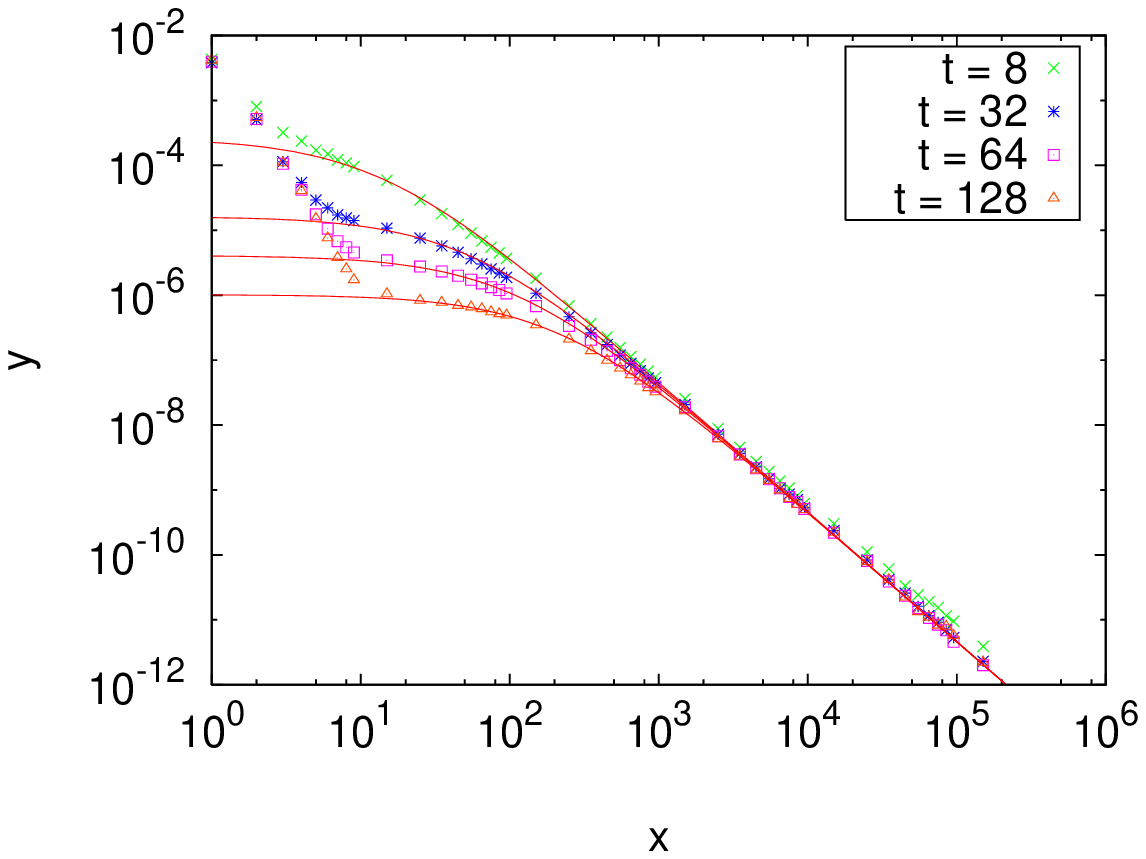}
\psfrag{x}{\large\hspace{-5mm} $A/\lambda_ht$}
\psfrag{y}{\large\hspace{-1cm} $(\lambda_ht)^2 \, n_h(A,t)$}
\includegraphics[width=8cm]{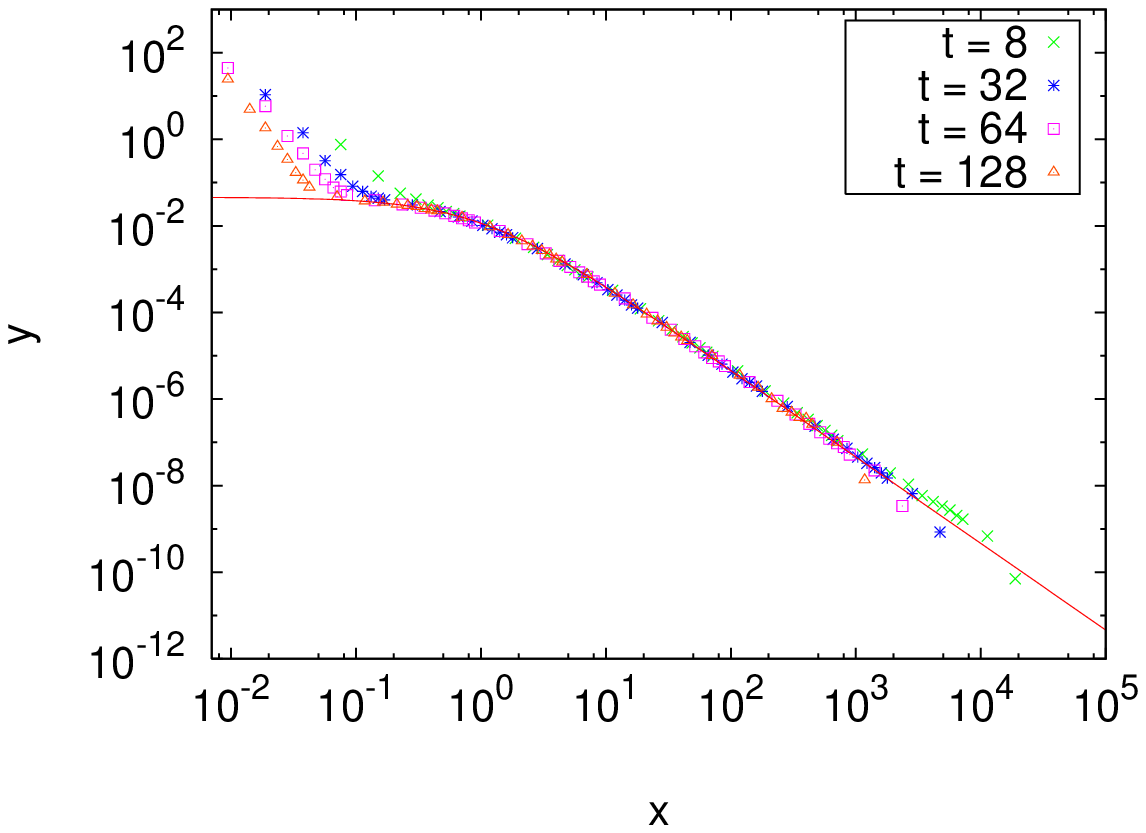}
\end{center}
\caption{(Colour online.) The number density of hulls for $T=1.5$ 
after different times (top) and the scaling of these data 
points (bottom)}
\label{fig:oneT-severaltimes}
\end{figure}

Finite working temperatures also affect the distribution of domain
areas. In Fig.~\ref{fig:three-T} the raw data at  $t=128$ MCs is
shown for four working temperatures. Upward deviations with respect to
the result of zero working temperature are prominent in the small
areas region of the figure, and increase with temperature.

\begin{figure}[h]
\centerline{
\psfrag{x}{\large {$A$} }
\psfrag{y}{\large\hspace{-5mm} $n_d(A)$}
\includegraphics[width=8cm]{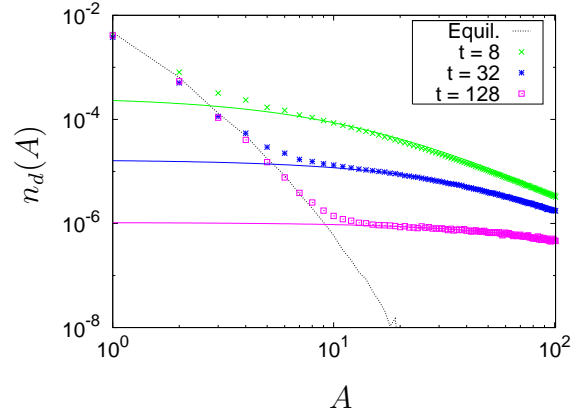}
}
\caption{(Colour online.) The contribution of
`thermal domains' obtained by simulating an equilibrated sample at the
working temperature $T=1.5$, along with the evolution of the
 distribution of domain sizes after a quench to the same temperature.}
\label{fig:Tdep-dist}
\end{figure}

In Fig.~\ref{fig:oneT-severaltimes}~(top) we display the raw data at
the working temperature $T=1.5$, for several times. Notice that
although the curves move downwards, the small areas region becomes
time independent. This region also fails to collapse (bottom) with the
proposed scaling using the temperature dependent values of
$\lambda_d(T)$. The reason is that the distribution counts thermal
equilibrium domains, that is to say fluctuations that are present in
an equilibrated sample at the working temperature, but are not due to
the coarsening process. Thus, these fluctuations should be identified
and eliminated from the statistics. We tried to apply the method
introduced by Derrida~\cite{Derrida}, and extended by Hinrichsen and
Antoni~\cite{Antoni}, to eliminate thermal domains, but the results
were not satisfactory, as not all of them could be eliminated. Thus,
instead of removing each thermal domain, we tried to directly remove
their contribution to the distributions by simulating samples in
equilibrium at the working temperature, starting with a fully
magnetized state, and computing the number density of thermal domain
areas.  These data are shown with green data points in
Fig.~\ref{fig:oneT-severaltimes} and ~\ref{fig:Tdep-dist}.
Surprisingly enough, thermal fluctuations generate areas that are
larger than one would have naively expected. Equilibrium arguments
suggest that the averaged area of thermally generated domains scale as
$A_T \sim \xi^2(T)$ with $\xi(T)/p_0\sim f^- (1-T/T_c)^{-\nu}$,
$\nu=1$ and $f^{-}=0.18$~\cite{Fisher}.  This estimate yields, for
example, $A_T\sim 4A_0$ at $T=1.5$. In equilibrium at this temperature
the average size of the domains found numerically is $\langle A_T
\rangle\sim 1.5 A_0$. However, the probability distribution of thermal
areas has a non-negligible weight -- as compared to the one of coarsening
domains -- that goes well beyond this value. For example, in
Figs.~\ref{fig:oneT-severaltimes} and \ref{fig:Tdep-dist}
we see that the crossover between the thermal area distribution
and the coarsening area distribution occurs at $A\sim 10 A_0$.

In Fig.~\ref{fig:Tdep-dist} we also present data for
the dynamic distribution at three different times, and
compared with the analytic prediction using $\lambda_d(T)$ estimated
from the analysis of the global spatial correlation, see
Fig.~\ref{fig:Tdep-lambda}. We conclude that the agreement between
analytic prediction and numerical results is very good in the region
in which the thermal domains are subdominant, {\it i.e.} when the blue
dynamic curves deviate from the green equilibrium one, indeed the
regime in which the analytic calculation is expected to apply.

\begin{figure}[h]
\centerline{
\psfrag{x}{\large {$A$} }
\psfrag{y}{\large\hspace{-5mm} $n_d(A)$}
\includegraphics[width=8cm]{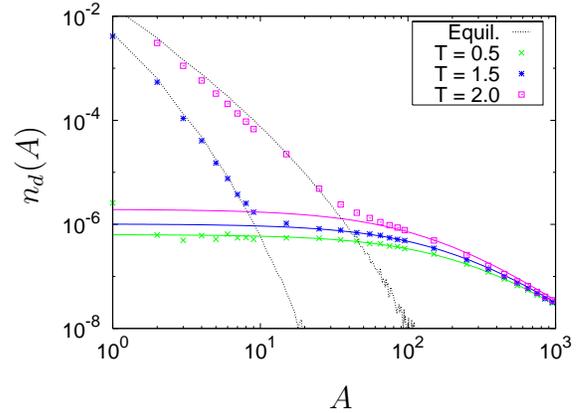}
}
\caption{(Colour online.) Zoom on the number density of domain areas
 at $t=128$ MCs at three working temperatures given in the key. The
 black lines are the equilibrium distributions at $T=1.5$ and $T=2$
 and the other lines (pink, blue and green) represent our analytic
 prediction for the coarsening areas.}
\label{fig:Tdep-dist2}
\end{figure}

One can also use the results in Fig.~\ref{fig:Tdep-dist2} to estimate
the value of $\lambda_d(T)$. Indeed, a fit of the numerical data for
areas larger than the value at which the equilibrium thermal
contribution (green points) deviates from the dynamic one, yields the
values of $\lambda_d(T)$ [and $\lambda_h(T)$] shown in
Fig.~\ref{fig:Tdep-lambda}.  This analysis allows us to extract
independent predictions for $\lambda_d(T)$ and $\lambda_h(T)$. We find
that the qualitative $T$-dependence is the same. As regards the
absolute values, the numerical data yield $\lambda_h(T) \leq
\lambda_d(T)$ on the whole range.  Note that the sum rules suggested
$\lambda_d=\lambda_h$ and the analytic prediction $\lambda_d=\lambda_h
+ {\cal O}(c_h)$.

\section{Statistics of perimeters and fractal properties} 
\label{sec:domain-walls}

The analytic argument described in Sect.~\ref{sec:analytic} can be
extended to study the distribution of domain wall lengths or
perimeters. In this Section we present the analytic prediction for this
function together with numeric results that confirm it.  We study two types 
of domain boundaries: those associated  to the hulls and those associated to 
the domains that is to say that 
include external and internal perimeters.  In
the simulations we define the length of the 
boundary as the number of broken bonds.

\subsection{Initial conditions}

\subsubsection{Equilibrium at $T_0=T_c$}

\begin{figure}[h]
\centerline{
\psfrag{x}{\large $p$}
\psfrag{y}{\large $A$}
\includegraphics[width=8cm]{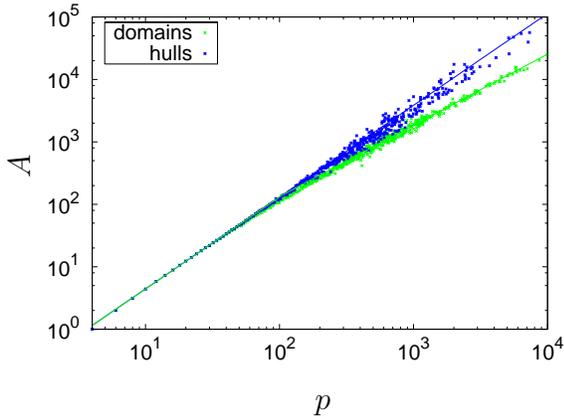}
}
\caption{(Colour online.) Relation between areas and perimeters in equilibrium at $T_0=T_c$.}
\label{fig:new1}
\end{figure}

In equilibrium we find numerically that the domain areas and their
corresponding boundaries are related by (see Fig.~\ref{fig:new1}
where the scatter plots have been averaged to make the trend clearer)
\begin{eqnarray*}
A_h &\sim& c^{(i)}_h \; p^{\alpha^{(i)}_h}, \;\; \mbox{with} \;\; \alpha^{(i)}_h \sim 1.47 \pm 0.1,
\\
\\
A_d &\sim& 
\left\{
\begin{array}{lll}
c^{(i>)}_d \; p^{\alpha^{(i>)}_d},
 \; &  \alpha^{(i>)}_d \sim
1.14\pm 0.1 & {\mbox{for}} \;\; p \stackrel{>}{\sim} 50 
\; ,
\\
\\
c^{(i<)}_d \; p^{\alpha^{(i<)}_d},
 \; & \alpha^{(i<)}_d \sim
1.47\pm 0.1 & {\mbox{for}} \;\; p \stackrel{<}{\sim} 50 
\; ,
\end{array}
\right.
\label{eq:initial-powers-domains-Tc}
\end{eqnarray*}
in the whole range of variation. Note that the longest lengths,
$p\approx 10^3-10^4$ may be affected by finite size effects given that
the linear size of the simulating box is $L=10^3$.  The spanning
clusters are not counted (note that their perimeters would be severely
under estimated due to the periodic boundary conditions).
The exponent $\alpha^{(i<)}_d \sim 1.47\pm 0.1$ is consistent with the result in~\cite{Cambier}
mentioned in Sect.~\ref{sec:Cambier}. The constants take the values
$c^{(i)}_h = 0.15$, $c^{(i<)}_d = 0.15$ and $c^{(i>)}_d = 0.70$.
The difference between the small and large $p$ regimes in the relation
between areas and perimeters for the domains is due to the existance of
holes in the large structures. The small domains and hulls are just the same
objects because the former do not have holes within.

\begin{figure}[h]
\centerline{
\psfrag{x}{\large $p$ }
\psfrag{y}{\large\hspace{-5mm} $n_{h,d}(p)$}
\includegraphics[width=8cm]{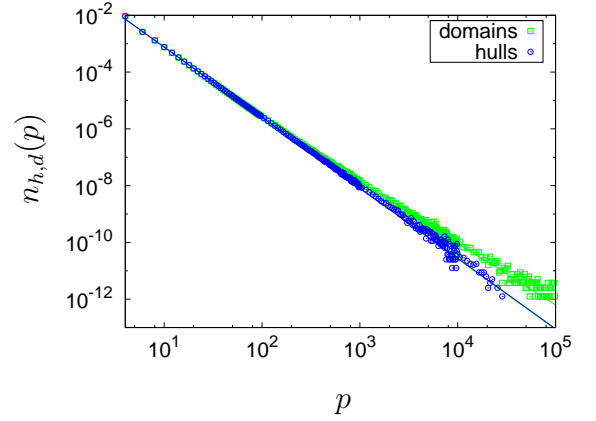}
}
\caption{(Colour online.) Distribution of domain and hull lengths in equilibrium at $T_0=T_c$.}
\label{fig:new2}
\end{figure}

Numerically, we find that 
the number densities of hull and domain lengths
at critical Ising conditions are (see Fig.~\ref{fig:new2})
\begin{eqnarray*}
n_h(p,0) &\sim& p^{-\zeta_h}\;\; \mbox{with} \;\; \zeta_h \sim 2.48\pm 0.05.
\\
\\
n_d(p,0) &\sim& \left\{
\begin{array}{lll}
p^{-\zeta^>_d}, 
\; & \zeta^>_d  \sim
2.17\pm 0.05 & {\mbox{for}} \;\; p \stackrel{>}{\sim} 50 
\; ,
\\ \\
p^{-\zeta^<_d},
\; & \zeta^<_d  \sim
2.48\pm 0.05 & {\mbox{for}} \;\; p \stackrel{<}{\sim} 50
\; ,
\end{array}
\right.
\label{eq:initial-powers-domains-Tc}
\end{eqnarray*}
The value of $\zeta_h$ is to be
compared to the analytic result $\zeta_h =27/11\approx
2.454$~\cite{Vander}. It is interesting to notice that
the distribution of domain lengths is not a single power law
in constrast to the distribution of domain areas.

\subsubsection{Equilibrium at $T_0=\infty$}

After a few time-steps evolving at $T=0$ from the
infinite temperature initial condition, we reach critical
percolation conditions. 

\begin{figure}[h]
\centerline{
\psfrag{x}{\large $p$ }
\psfrag{y}{\large $A_h$}
\includegraphics[width=8cm]{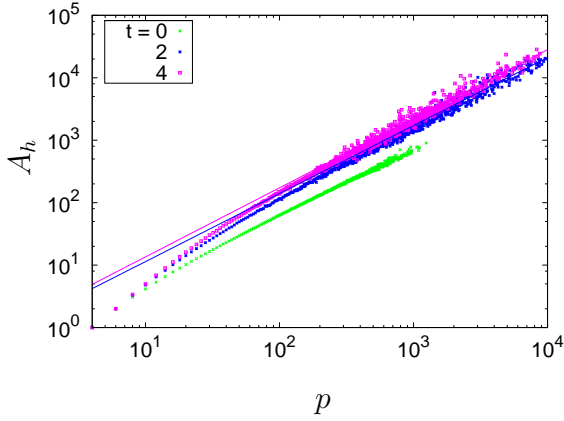}
}
\caption{(Colour online.) Relation between areas and perimeters in equilibrium at $T_0\to\infty$
and after a few time steps when critical percolation is approximatively reached.}
\label{fig:new3}
\end{figure}

In Fig.~\ref{fig:new3}, we show the area perimeter relation
for hulls, in equilibrium at $T_0 \to \infty$ and after a few
time steps. The analysis of this figure and the corresponding
one for domains yields
\begin{eqnarray*}
A_h &\sim& c^{(i')}_h \; p^{\alpha^{(i')}_h}, \;\; \mbox{with} \;\; \alpha^{(i')}_h \sim 1.12 \pm 0.1
\\
\\
A_d &\sim& 
\left\{
\begin{array}{lll}
c^{(i'>)}_d \; p^{\alpha^{(i'>)}_d},
 \; &  \alpha^{(i'>)}_d \sim
1.01\pm 0.1 & {\mbox{for}} \;\; p \stackrel{>}{\sim} 50 
\; ,
\\ \\
c^{(i'<)}_d \; p^{\alpha^{(i'<)}_d},
 \; & \alpha^{(i'<)}_d & {\mbox{for}} \;\; p \stackrel{<}{\sim} 50 
\; ,
\end{array}
\right.
\label{eq:initial-powers-domains-Tc}
\end{eqnarray*}
The constants take the values
$c^{(i')}_h = 0.96$, and $c^{(i'>)}_d = 1.50$.
The exponent $\alpha^{(i'<)}_d$ cannot be determined numerically since
critical percolation is not accessible exactly.

\begin{figure}
\centerline{
\psfrag{x}{\large $p$ }
\psfrag{y}{\large\hspace{-5mm} $n_h(p)$}
\includegraphics[width=8cm]{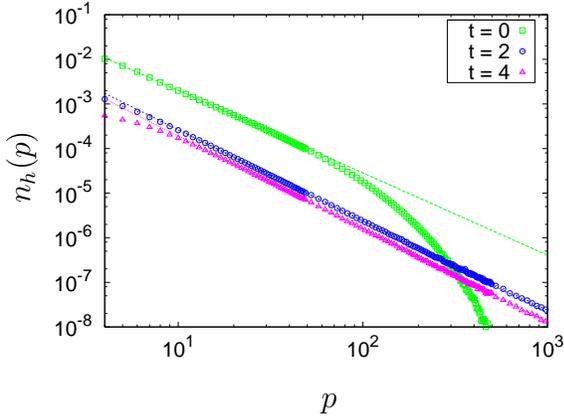}
}
\caption{(Colour online.) The number density of hull lengths 
in equilibrium at $T_0\to\infty$ labelled as $t=0$ and 
the same quantity evaluated at two later times after evolution
at zero working temperature. The lines are 
$n_h \sim p^{-2.05}$; see the text for a discussion.}
\label{fig:pdf-perim-equil-Tinf}
\end{figure}

As we can see in Fig.~\ref{fig:pdf-perim-equil-Tinf}, the
initial weight of the number density at large values of the 
perimeter is lower than expected at critical percolation. In a few
time-steps long perimeters develop and the weight reaches 
the asymptotic power law at large values of $p$ while it 
looses weight at small values of $p$. This effect is the same 
as the one observed in the study of the initial and early times
number density of areas, see the discussion in Sect.~\ref{sec:infTinit} 
and Fig.~\ref{fig:evolinitcondT05fromTinfhull}. For the hull and domain
length distributions one finds
\begin{eqnarray*}
n_h(p,0) &\sim& p^{-\zeta^{'}_h} \;\; \mbox{with} \;\; \zeta^{'}_h \sim 2.12\pm 0.05
\\
\\
n_d(p,0) &\sim& \left\{
\begin{array}{lll}
p^{-\zeta^{'>}_d}, 
\; & \zeta^{'>}_d  \sim
2.01\pm 0.05 & {\mbox{for}} \;\; p \stackrel{>}{\sim} 50 
\; ,
\\ \\
p^{-\zeta^<_d},
\; & \zeta^<_d & {\mbox{for}} \;\; p \stackrel{<}{\sim} 50
\; ,
\end{array}
\right.
\label{eq:initial-powers-domains-Tc}
\end{eqnarray*}
The analytical result for the hull exponent in critical percolation is 
$\zeta'_h=15/7 \approx 2.14$~\cite{Saleur-Duplantier}.

\subsubsection{General comments on both initial cases}

It is interesting to note that the exponents characterising the number
density of perimeter lengths at the two initial conditions are
significantly different. They are approximately equal to 2.5 at $T_c$ and 2
at $T_0\to\infty$. This is to be contrasted with the behaviour of the
area number densities for which the exponents were identical for hull
enclosed areas and very close indeed for domains.

The exponents $\alpha$ and $\zeta$ 
are linked by the fact that 
each hull-enclosed area or domain area is in one-to-one relation
to its own boundary. Thus, $n_h(A,0) dA = n_h(p,0) dp$
and one finds
\begin{equation}
n_h(p,0) \sim \frac{c_h \alpha^{(i)}_h}{c^{(i)}_h} 
\; p^{-1-\alpha^{(i)}_h} \; ,
\end{equation}
which implies 
\begin{equation}
\zeta_h = 1+\alpha_h^{(i)} \; ,
\label{eq:zetah} 
\end{equation}
These conditions are also satisfied for the primed ($T_0 \rightarrow \infty$) quantities.
Within our numerical accuracy these relations are respected,
for instance
\begin{eqnarray}
\begin{array}{lll}
\zeta'_h \sim 2.12 \; , & \qquad {\alpha'}_h^{(i)} \sim 1.12 \;, 
& \qquad T_0\to\infty
\; , 
\nonumber\\
\zeta_h\sim 2.48 \; , & \qquad \alpha^{(i)}_h \sim 1.47 \;, & \qquad T_0=T_c
\; 
\end{array}
\end{eqnarray}

Similarly, for domain areas and domain boundaries one obtains
\begin{equation}
n_d(p,0) \sim \frac{c_d\alpha^{(i)}_d}{{c_d^{(i)}}^{\tau-1}} \; p^{-1-(\tau-1) 
\alpha^{(i)}_d} \; ,
\end{equation}
therefore,
\begin{equation}
\zeta_d = 1+(\tau-1) \alpha_d^{(i)} \; ,
\label{eq:zetad} 
\end{equation}
These relations are satisfied for both $\alpha^{(i>)}_d$
and $\alpha^{(i<)}_d$
as well as for primed ($T_0 \rightarrow \infty$) quantities.
They are respected by our measures.

The main sources of error in the determination of the exponents and
the constants in the study of the initial conditions are the following: 
(i) statistical errors, although we have a rather good sampling; 
(ii) the choice of the large area-perimeter limit that is not perturbed 
by finite size effects, and (iii) the fact that the $T_0\to\infty$ initial 
condition is not exactly at critical percolation. We estimated the magnitude 
of the error to be $\pm 0.1$ in the $\alpha$ exponents, and $\pm 0.05$ in the
$\zeta$ exponents, which correspond to, roughly, less than $10\%$ in
both cases. Within this level of accuracy, the relations between 
exponents (\ref{eq:zetad}) and (\ref{eq:zetah}) are satisfied.

\subsection{Time evolution at zero temperature}

\subsubsection{Hulls}

\begin{figure}[h]
\begin{center}
{
\psfrag{x}{\large $p$ }
\psfrag{y}{\large $A_h$}
\includegraphics[width=8cm]{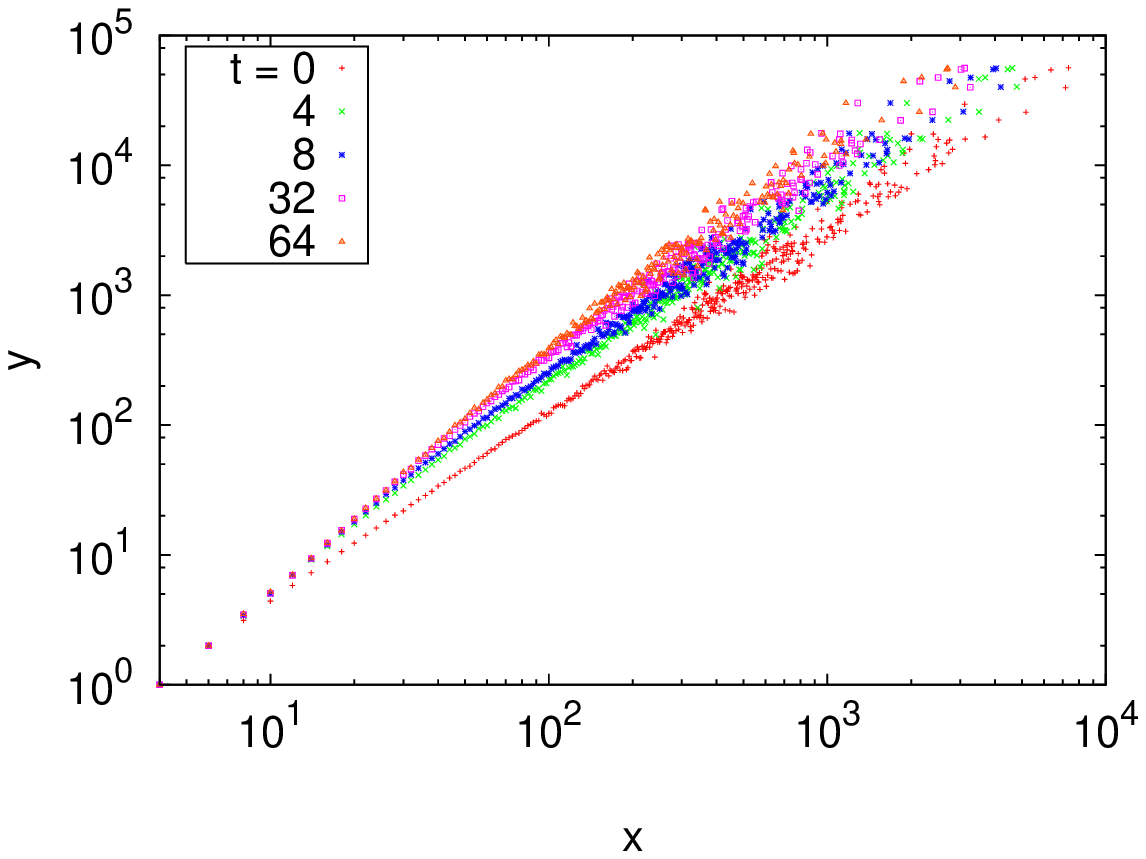}
\psfrag{x}{\large\hspace{-10mm} $p/\sqrt{\lambda_h t}$ }
\psfrag{y}{\large\hspace{-5mm} $A_h/\lambda_h t$}
\includegraphics[width=8cm]{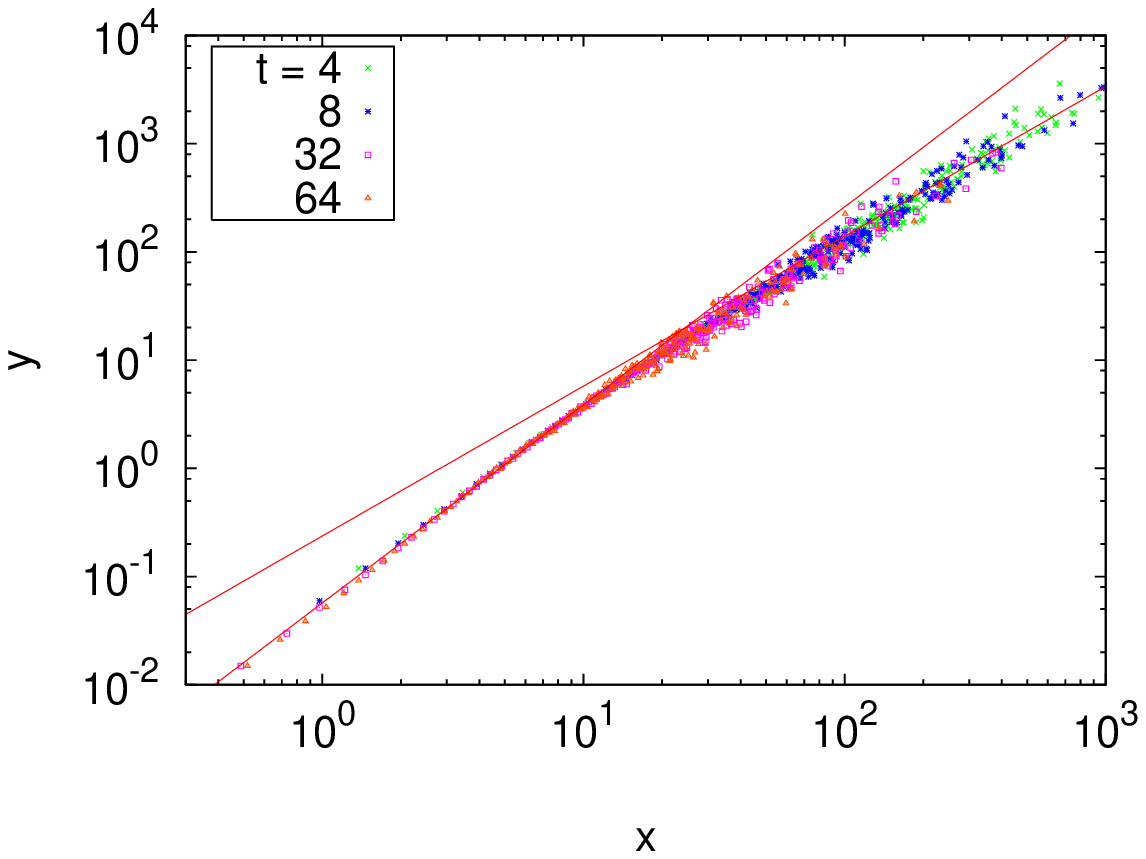}
}
\end{center}
\caption{(Colour online.) Time evolution of the hull enclosed area
vs. perimeter relation for $T_0=T_c$ and different
times indicated in the legend.}
\label{fig:new4-5}
\end{figure}

After a quench from $T_0=T_c$, the hull enclosed areas and their corresponding
perimeters, during coarsening at zero temperature, obey the scaling 
relations (see Fig.~\ref{fig:new4-5})

\begin{figure}[h]
\begin{center}
{
\psfrag{x}{\large\hspace{-10mm} $p/\sqrt{\lambda_h t}$ }
\psfrag{y}{\large\hspace{-5mm} $A_h/\lambda_h t$}
\psfrag{zuluzulu}{$T_0 = T_c \;\;\;\;$}
\psfrag{wuluzulu}{$T_0 = \infty$}
\includegraphics[width=8cm]{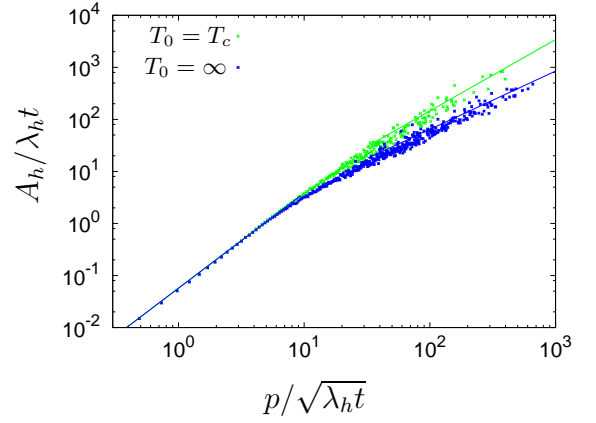}
}
\end{center}
\caption{(Colour online). Hull enclosed area vs. perimeter relation
at time $t=32$ MCs from two different initial conditions.}
\label{fig:new6}
\end{figure}

\begin{equation} 
\frac{A}{\lambda_h t} 
\sim \eta_h 
\left(\frac{p}{\sqrt{\lambda_h t}}\right)^{\alpha_h},
\label{eq:t-dep-ApTc}
\end{equation}
with
\begin{equation}
\left.
\begin{array}{l}
\alpha^>_h \sim 1.37 \pm 0.2
\\
\eta_h^> \sim 0.26
\end{array}
\right\}
\;\; \mbox{for} \;\; 
 \frac{A}{\lambda_h t} \; \stackrel{>}{\sim} 50 \; , 
\end{equation}
and
\begin{equation}
\left.
\begin{array}{l}
\alpha^<_h \sim 1.83 \pm 0.2
\\
\eta_h^< \sim 0.06
\end{array}
\right\}
\;\; \mbox{for} \;\; 
 \frac{A}{\lambda_h t} \; \stackrel{<}{\sim} 10 \; , 
\end{equation}
and, after a quench from $T_0\to\infty$,
\begin{equation}
\frac{A}{\lambda_h t} 
\sim \eta'_h 
\left(\frac{p}{\sqrt{\lambda_h t}}\right)^{\alpha'_h},
\label{eq:t-dep-ApTinfty}
\end{equation}
with
\begin{equation}
\left.
\begin{array}{l}
{\alpha'}^>_h \sim 1.12 \pm 0.2
\\
{\eta'}_h^> \sim 0.38
\end{array}
\right\}
\;\; \mbox{for} \;\;
 \frac{A}{\lambda_h t} \; \stackrel{>}{\sim} 50 \; , 
\end{equation}
and
\begin{equation}
\left.
\begin{array}{l}
{\alpha'}^<_h \sim 1.83  \pm 0.2
\\
{\eta'}_h^< \sim 0.057
\end{array}
\right\}
\;\; \mbox{for} \;\; 
 \frac{A}{\lambda_h t} \; \stackrel{<}{\sim} 10 \; .
\end{equation}
We note that the relation between area and perimeter exhibits two 
distinct regimes.  During the coarsening process a characteristic scale
$A^*(t) \sim \lambda_h t$ develops such that domains with area $A>A^*$
have the same exponent as in the initial condition (structures that
are highly ramified with $\alpha$ smaller than two) and domains with
$A<A^*$ are regular ($\alpha \sim 2$) (as shown in Fig.~\ref{fig:new6} the
structure of these small domains does not depend on the initial condition).
This phenomenon is reminiscent
of an {\it unroughening transition} occurring at a velocity
$\lambda_h$.  The same features were observed by Grest and
Srolovitz~\cite{Grest} and Fialkowski and Holyst~\cite{Holyst} 
in the study of the {\it domain} fractal dimension
during coarsening.

The hull structures of any size do not have holes, therefore
the crossover we see is of pure dynamical origin, contrasting
the idea presented in~\cite{Jacobs}, where the crossover in the domains
where explained by only geometrical reasons.

Note that we estimated the error in the exponents $\alpha$ to be $\pm
0.2$ and thus more important than in the analysis of the initial
conditions. The reason is that the crossover from the small area to
the large area regime is not sufficiently sharp and the choice of the
fitting interval introduces an additional source of error. Indeed, 
note that in Eqs.~(\ref{eq:t-dep-ApTc}) and (\ref{eq:t-dep-ApTinfty})
we did not use the intermediate regime $10 \leq \frac{A}{\lambda_h t} \leq 50$ 
to fit the power laws.

In analogy with the derivation in Sect.~\ref{sec:analytic} for the
time-dependent number density of domain areas, the time-dependent
number densities of hull and domain wall lengths are given by
\begin{equation}
n_{h,d}(p,t) = \int dp_i \; \delta(p-p(t,p_i)) \, n_{h,d}(p_i,t_i)
\end{equation}
with $n_{h,d}(p_i,t_i)$ the initial condition and $p(t,p_i)$ the
perimeter length of a boundary at time $t$ that had initial length
$p_i$ at time $t_i$.

Let us here discuss the hull lengths. In this case one can simply use
the exact number density of hull enclosed areas, $n_h(A,t) \sim
c_h/(A+\lambda_h t)^2$ for, say, $T_0=T_c$ and
Eq.~(\ref{eq:t-dep-ApTc}) to relate time-dependent areas to their
perimeters on the two regimes of large and small areas.
After a little algebra one derives
\begin{equation}
(\lambda_h t)^{3/2}
\; 
n_h(p,t) \sim 
\frac{
\alpha^<_h \eta^<_h c_h \left(
\frac{p}{\sqrt{\lambda_h t}}\right)^{\alpha^<_h-1}} 
{\left[ 
1+\eta^<_h \left(\frac{p}{\sqrt{\lambda_h t}}\right)^{\alpha^<_h} 
\right]^{2}
}
\label{eq:analytic-np-small}
\end{equation}
for small areas, $A/\lambda_h t<10$, and 
\begin{equation}
(\lambda_h t)^{3/2}
\; 
n_h(p,t) \sim 
\frac{
\alpha^>_h \eta^>_h c_h \left(
\frac{p}{\sqrt{\lambda_h t}}\right)^{\alpha^>_h-1}} 
{\left[ 
1+\eta^>_h \left(\frac{p}{\sqrt{\lambda_h t}}\right)^{\alpha^>_h} 
\right]^{2}
}
\label{eq:analytic-np-large}
\end{equation}
for large areas $A/\lambda_h t>50$. 
Note that these expressions satisfy scaling -- see Eq.~(\ref{eq:scaling-np}).
Interestingly, the scaling
function, $f_<(x)=x^{\alpha_h^<-1}/(1+\eta_h^< x^{\alpha_h^<})^2$ with
$x=p/\sqrt{\lambda_h t}$ reaches a maximum at 
\begin{equation}
x_{max} = \left(\frac{\alpha_h^<-1}{\eta^<_h (\alpha_h^<+1)}\right)
^{1/\alpha_h^<}
\label{eq:maximum}
\end{equation} 
and then falls-off to zero as another power-law.  There is then a
maximum at a finite and positive value of $p$ as long as
$\alpha_h^<>1$, that is to say, in the regime of not too large
areas. The numeric evaluation of the right-hand-side yields $x_{max} =
p_{max}/(\sqrt{\lambda_h t}) \sim 3$ which is in the range of validity
of the scaling function $f_<$. The behaviour of the time-dependent
perimeter number density for long perimeters is controlled by
Eq.~(\ref{eq:analytic-np-large}) that falls-off as a power law $f_>(x)
\sim x^{-(1+\alpha_h^>)}$. Although the function $f_>$ also has 
a maximum, this one falls out of its range of validity.

Above we used the critical Ising parameters. The results 
after a quench from $T_0\to \infty$ follow the same functional 
form with the corresponding primed values of $\alpha$ and $\eta$ 
and $c_h \to 2c_h$.

The power law describing the tail of the number density of long
perimeters is the same as the one characterising the initial
distribution, since $\alpha_h^>=\alpha_h^{(i)}$ and then
$1+\alpha_h^>=\zeta_h$. Therefore, the decay of the time-dependent
number density at long perimeters after a quench from $T_0=T_c$ and
$T_0\to\infty$ are distinguishably different with $\zeta_h\approx 2.5$
and $\zeta'_h\approx 2$. This is to be contrasted with the small
difference in the area number densities that fall with two power laws
that are so close (powers of 2 and 2.05) that are impossible to
distinguish numerically.

\begin{figure}[h]
\begin{center}
\psfrag{x}{\large $p$ }
\psfrag{y}{\large\hspace{-5mm} $n_h(p)$}
\includegraphics[width=8cm]{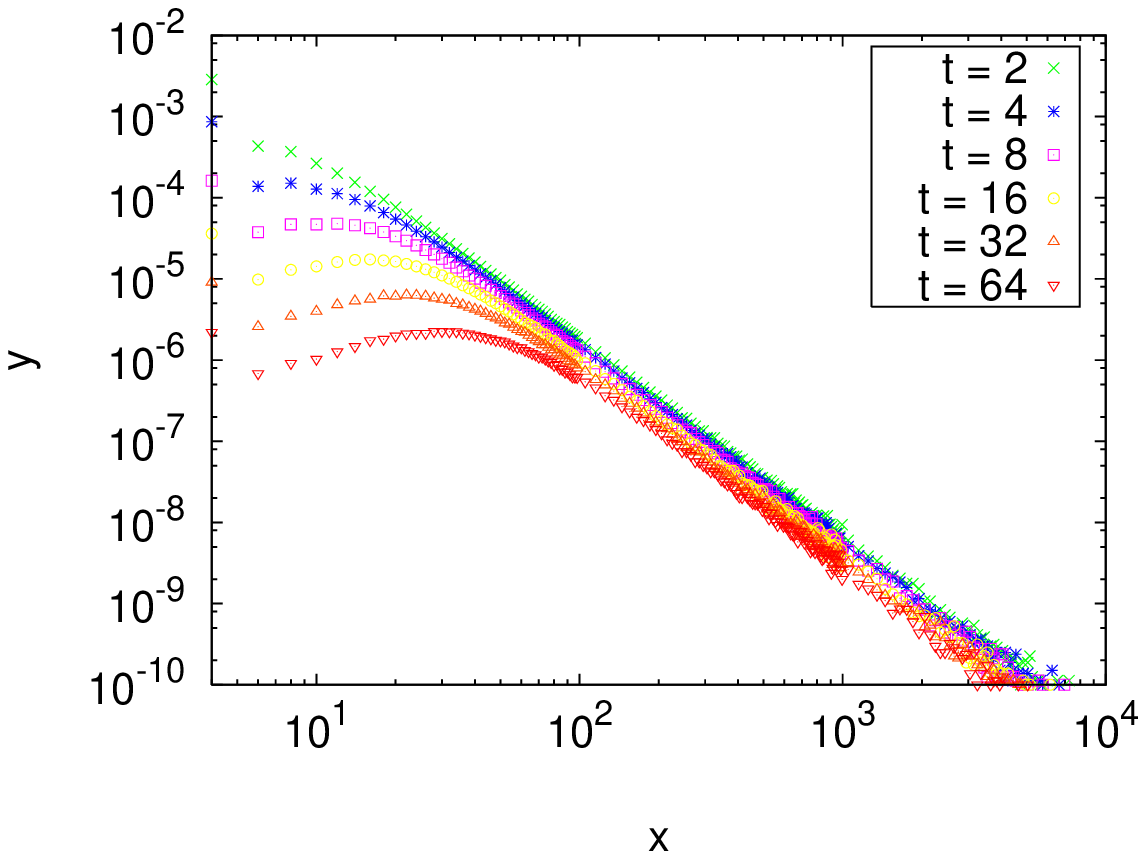}
\includegraphics[width=8cm]{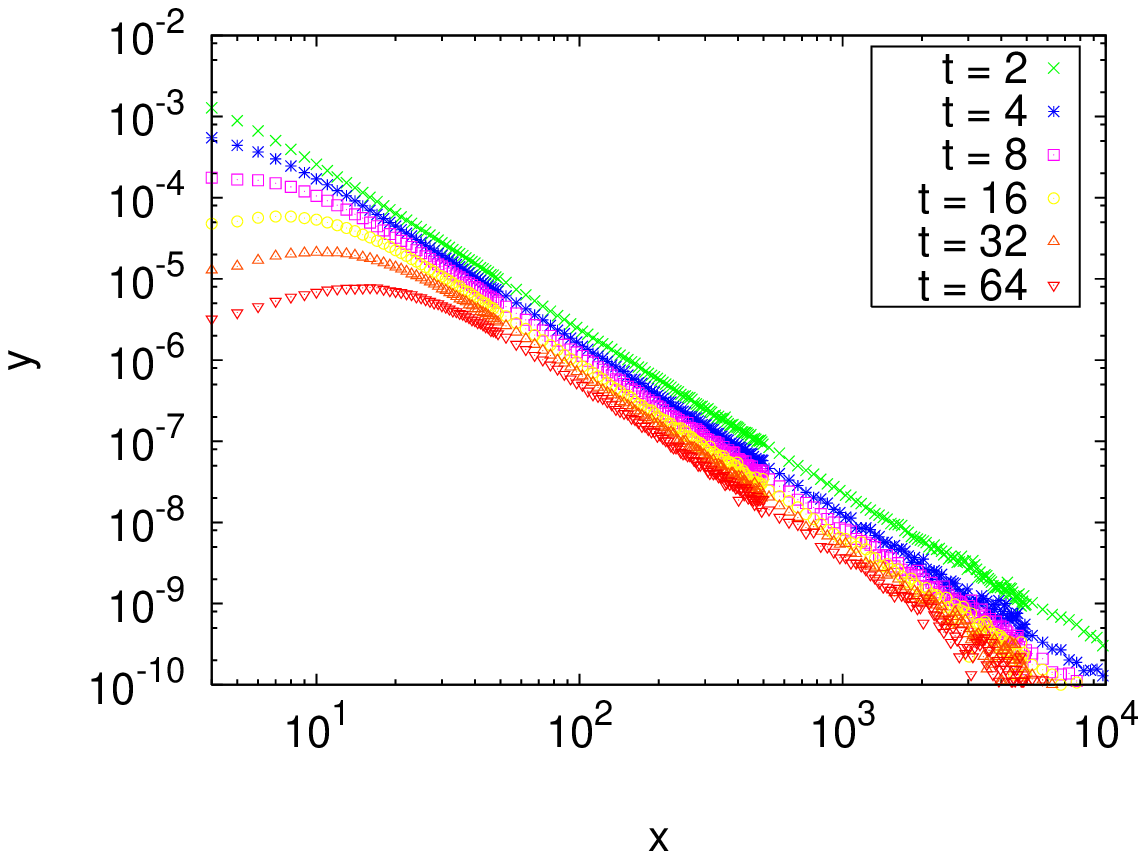}
\end{center}
\caption{(Colour online.) The time-dependent number density of
perimeters evolving at $T=0$ from an initial condition at $T_0=T_c$
(top) and $T_0\to\infty$ (bottom). Note that the time-dependence is
visible in the whole range of values of $p$ (while in the area number 
densities the large area tails were very weakly dependent on time, see
Fig.~\ref{fig:hulls}).}
\label{fig:perimeters-hulls}
\end{figure}

 In Fig.~\ref{fig:perimeters-hulls}, top and bottom, we display the
time-dependent perimeter number densities for a system evolving at
zero temperature after a quench from $T_0=T_c$ and $T_0\to\infty$,
respectively.  Notice that the perimeter length definition we are
using on the lattice can only take even values and thus when
constructing the histogram we have to take into account the extra
factor of $2$ in the binning.

\begin{figure}[h]
\begin{center}
\psfrag{y}{\large\hspace{-10mm} ${(\lambda_h t)}^{3/2} \, n_h(p)$ }
\psfrag{x}{\large\hspace{-5mm} $p/ \sqrt{\lambda_h t}$}
\includegraphics[width=8cm]{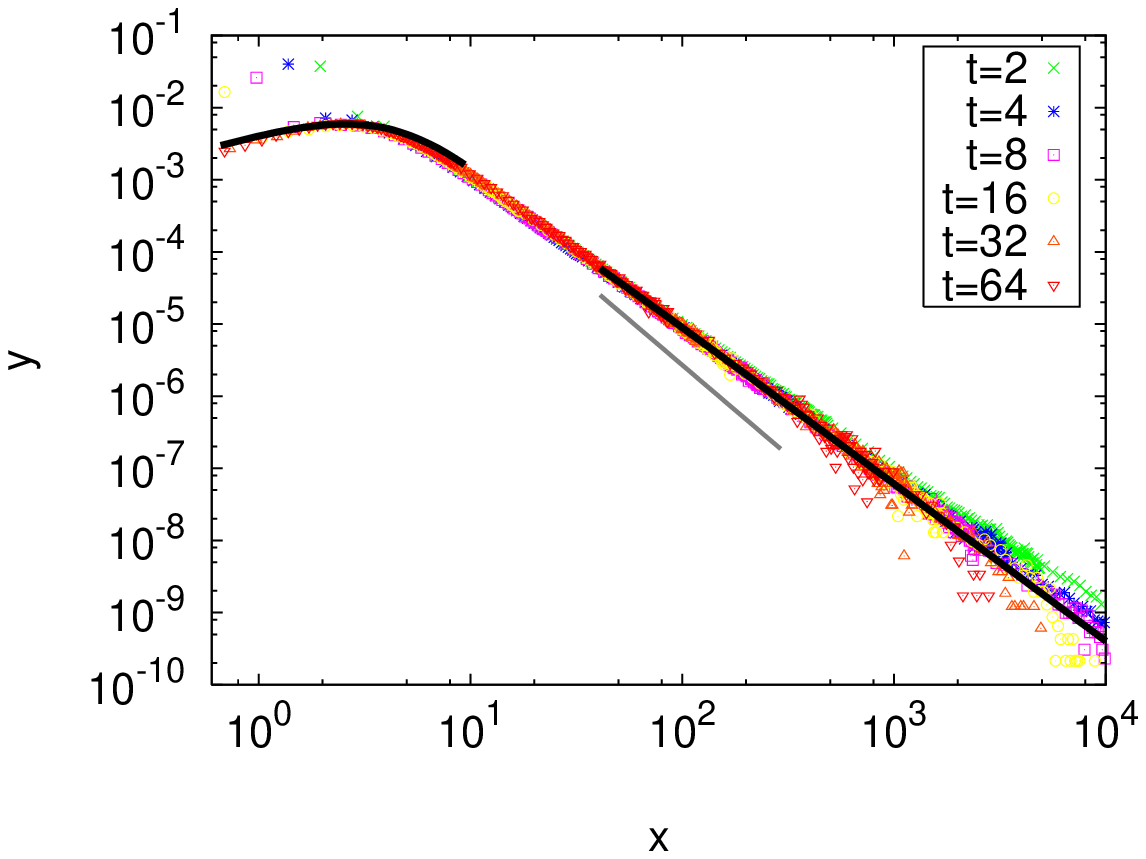}
\includegraphics[width=8cm]{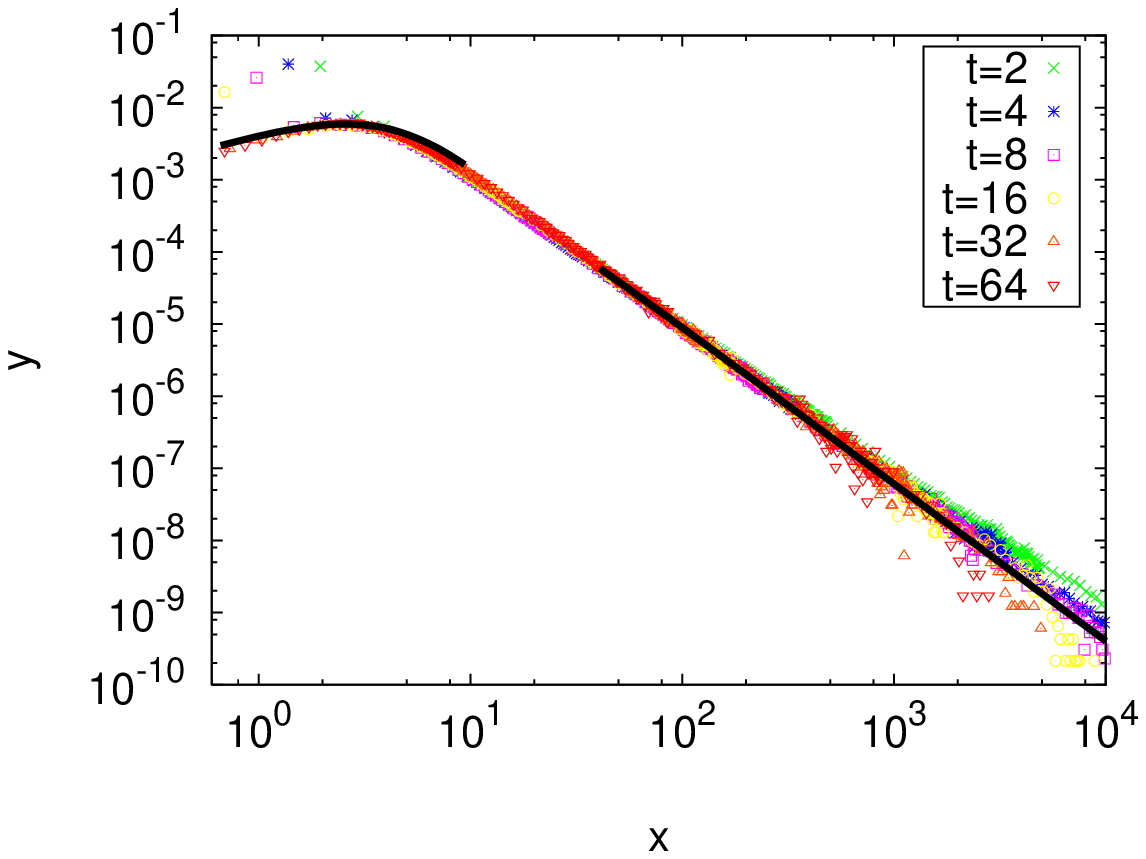}
\end{center}
\caption{(Colour online.) Scaling of the time-dependent number density
of hull lengths evolving at $T=0$ from an initial condition at
$T_0=T_c$ (top) and $T_0\to\infty$ (bottom). The solid black lines
represent the theoretical prediction valid for $A/\lambda_h t<10$
and for $A/\lambda_h t>50$. The
agreement between theory and numerical data is again very
impressive. The small grey line in the top plot represents the slope in
the bottom plot, showing that in contrast with domain size distribution,
perimeter distribution is very sensitive to initial conditions.
The isolated data points that lie above the scaling
function correspond to reversed, isolated spins within a
bulk of the opposite sign that give rise to a perimeter length $p=4$
(four broken bonds). The area number densities also showed this
anomalous behaviour for $A=1$.}
\label{fig:perimeters-hulls-scaling}
\end{figure}
In Fig.~\ref{fig:perimeters-hulls-scaling} we display the scaling plot
of the number density of hull lengths and we compare it to the
analytic prediction~(\ref{eq:analytic-np-small}) and
(\ref{eq:analytic-np-large}).  The data are in remarkably good
agreement with the analytic prediction; the lines represent the
theoretical functional forms for long and short lengths, and describe
very well the two limiting wings of the number density. The maximum
is located at a value that is in agreement with the prediction,
Eq.~(\ref{eq:maximum}).

\subsubsection{Domains}

We studied the relation between domain areas and their corresponding
perimeters during coarsening at zero temperature finding that the
scaling forms
\begin{equation} 
\frac{A}{\lambda_d t} 
\sim \eta_d 
\left(\frac{p}{\sqrt{\lambda_d t}}\right)^{\alpha_d},
\label{eq:t-dep-ApTc-domains}\\
\end{equation}
with
\begin{equation}
\left.
\begin{array}{l}
\alpha^>_d \sim 1.16 
\\
\eta_d^> \sim 0.63
\end{array}
\right\}
\;\; \mbox{for} \;\;
 \frac{A}{\lambda_d t} \; \stackrel{>}{\sim} 50 \; , 
\end{equation}
and
\begin{equation}
\left.
\begin{array}{l}
\alpha^<_d \sim 1.83 
\\
\eta_d^< \sim 0.057
\end{array}
\right\}
\;\; \mbox{for} \;\;
 \frac{A}{\lambda_d t} \; \stackrel{<}{\sim} 10 \; , 
\end{equation}
after a quench from $T_0=T_c$ and
\begin{equation}
\frac{A}{\lambda_d t} 
\sim \eta'_d 
\left(\frac{p}{\sqrt{\lambda_d t}}\right)^{\alpha'_d},
\label{eq:t-dep-ApTinf-domains}
\end{equation}
with
\begin{equation}
\left.
\begin{array}{l}
{\alpha'_d}^> \sim 1.01 
\\
{\eta'_d}^> \sim 0.52
\end{array}
\right\}
\;\; \mbox{for} \;\;
 \frac{A}{\lambda_d t} \; \stackrel{>}{\sim} 50 \; , 
\end{equation}
and
\begin{equation}
\left.
\begin{array}{l}
{\alpha'_d}^< \sim 1.83 
\\
{\eta'_d}^< \sim 0.06
\end{array}
\right\}
\;\; \mbox{for} \;\;
 \frac{A}{\lambda_d t} \; \stackrel{<}{\sim} 10 \; , 
\end{equation}
after a quench from $T_0\to\infty$. 

\begin{figure}[h]
\begin{center}
\psfrag{x}{\large\hspace{-10mm} $p_d/\sqrt{\lambda_d t}$ }
\psfrag{y}{\large\hspace{-5mm} $A_d/\lambda_d t$}
\includegraphics[width=8cm]{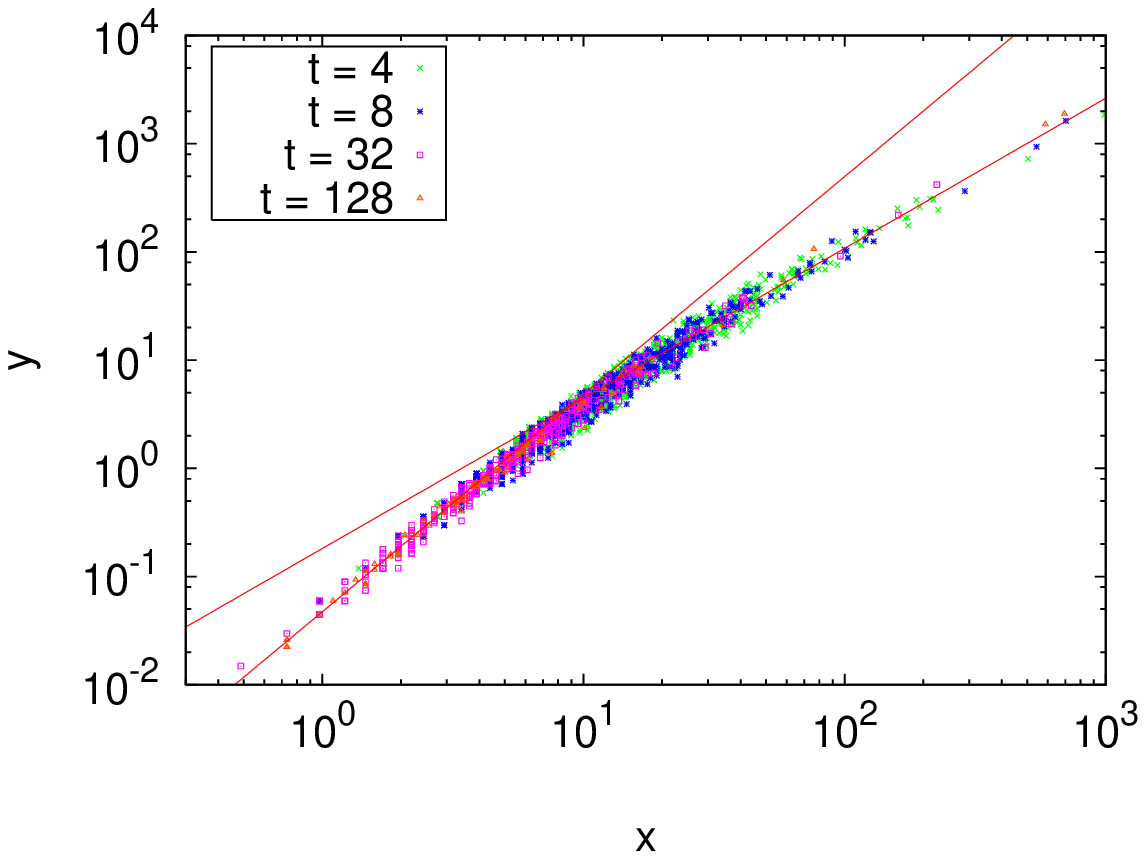} 
\includegraphics[width=8cm]{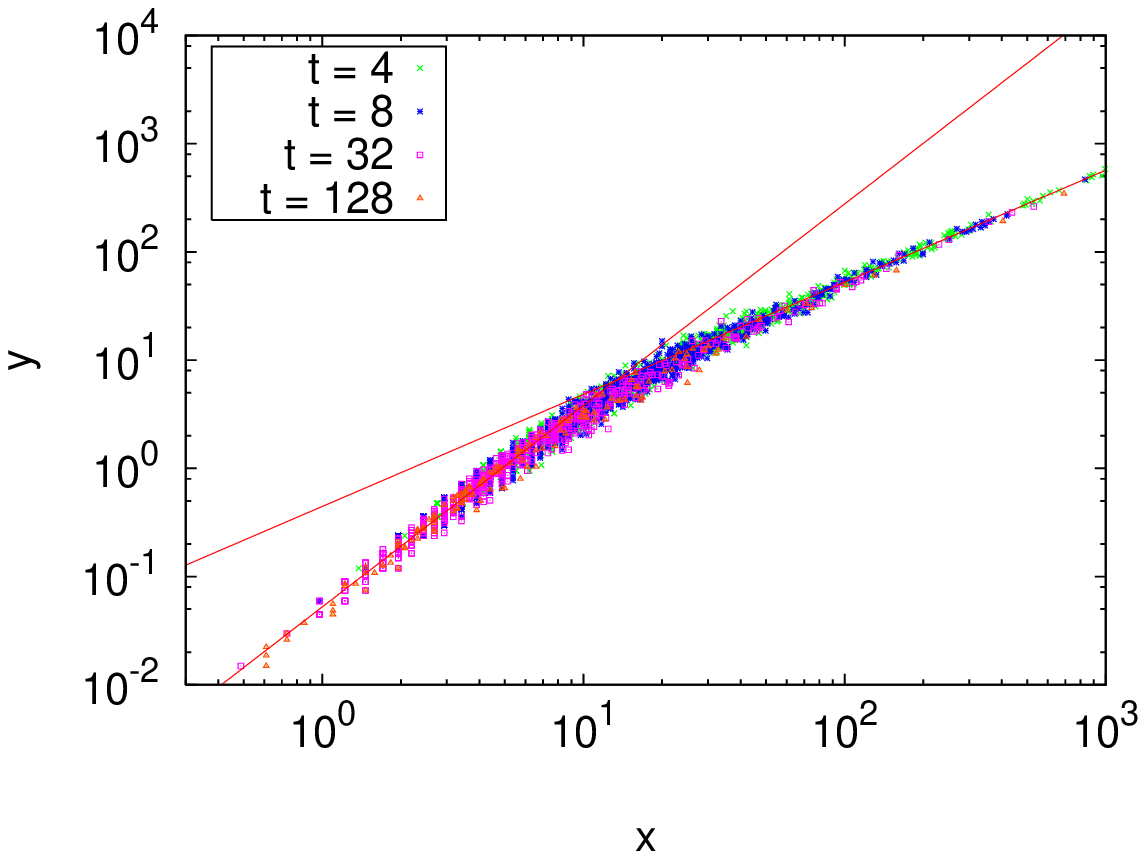}
\end{center}
\caption{(Colour online.) The time-dependent relation between the area and the
perimeter of a cluster of aligned spins (domain) evolving at $T=0$
after a quench from $T_0=T_c$ (top) and
$T_0\to\infty$ (bottom). In both cases $\lambda_d=2.1$.}
\label{fig:t-dep-Ap1}
\end{figure}

These results are shown in Fig.~\ref{fig:t-dep-Ap1}. Note
that the large area results match the behaviour of the initial conditions
in both cases, and small domains are much more compact that the initial ones.

\begin{figure}[h]
\begin{center}
\psfrag{y}{\large\hspace{-10mm} ${(\lambda_d t)}^{3/2} \, n_d(p)$ }
\psfrag{x}{\large\hspace{-5mm} $p/ \sqrt{\lambda_d t}$}
\includegraphics[width=8cm]{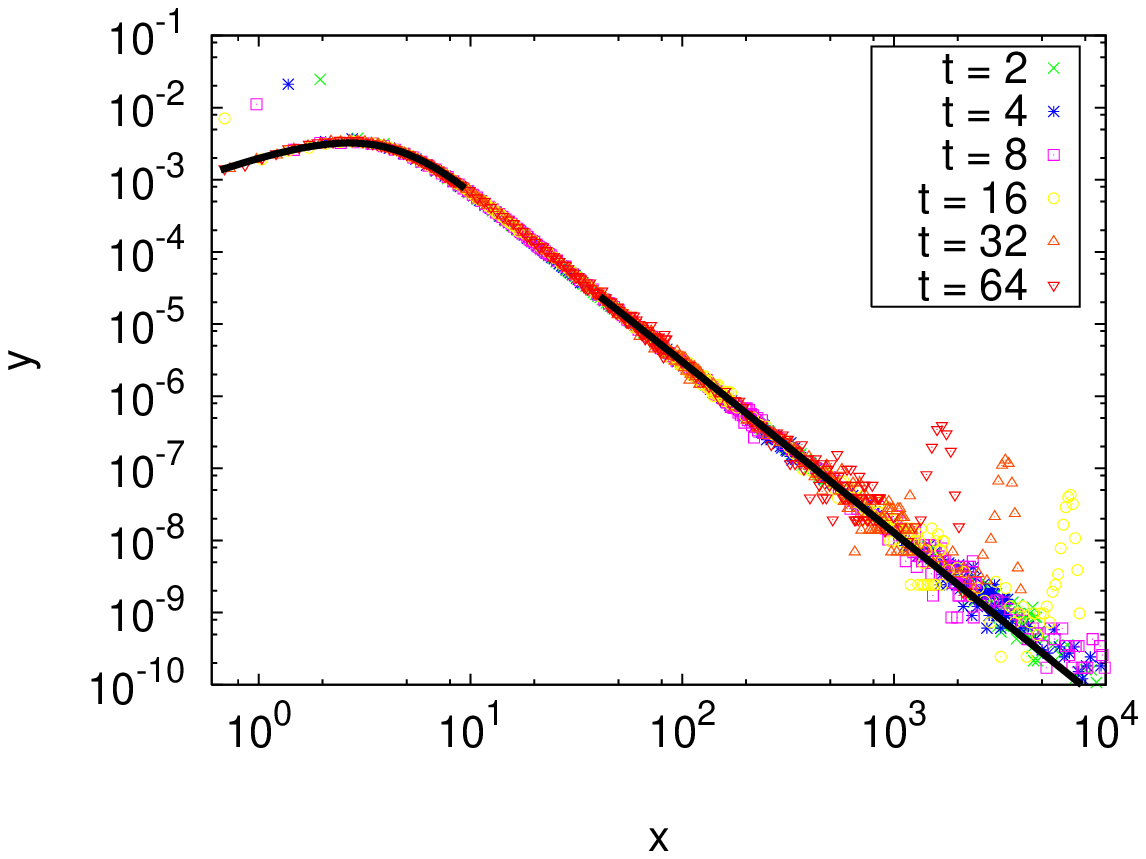}
\includegraphics[width=8cm]{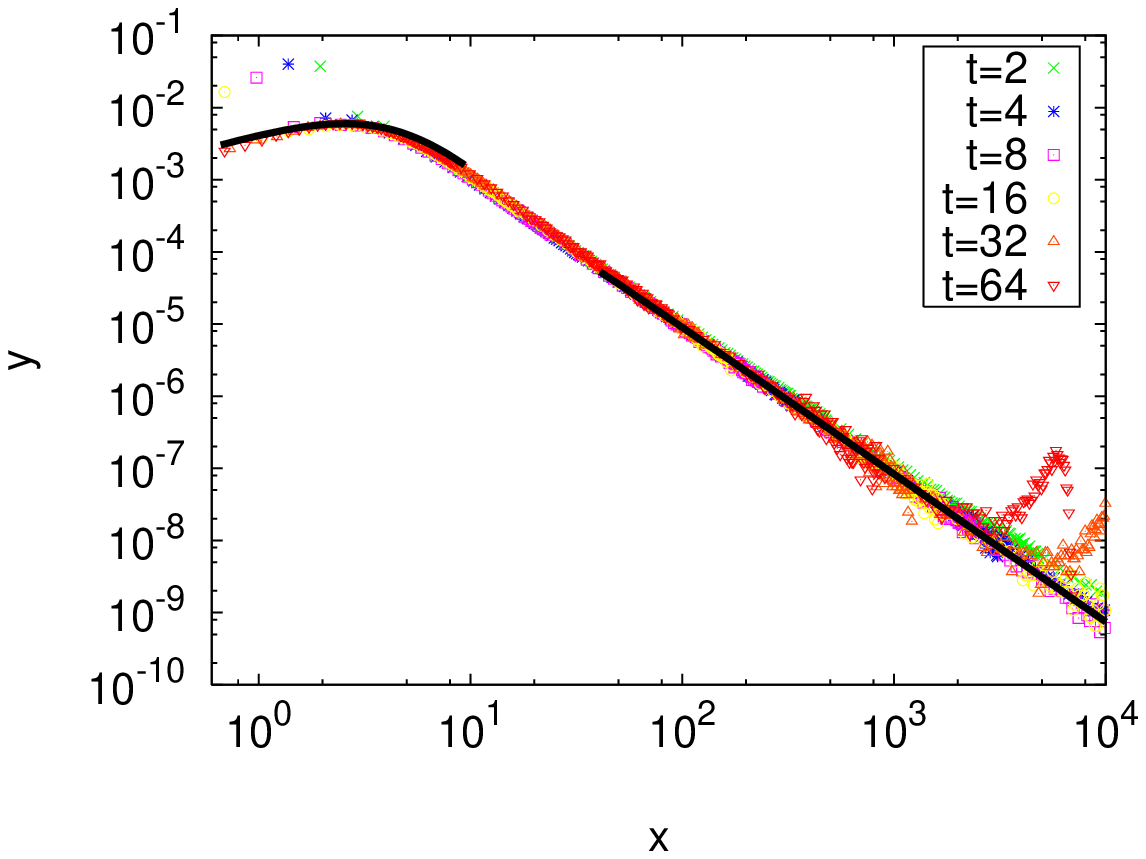}
\end{center}
\caption{(Colour online.) (Colour online.) Scaling of the time-dependent number density
of domain wall lengths evolving at $T=0$ from an initial condition at
$T_0=T_c$ (top) and $T_0\to\infty$ (bottom). The solid black lines
represent the theoretical prediction valid for $A/\lambda_h t<10$
and for $A/\lambda_h t>50$. The
origin of the isolated data points is the same as in
Fig.~\ref{fig:perimeters-hulls-scaling}.}
\label{fig:perimeters-domain-scaling}
\end{figure}

With the same line of argument exposed above we can analyse the statistics of 
the domain walls, that is to say, including external and internal 
perimeters. One finds basically the same results as for the hulls;
for critical Ising initial conditions:
\begin{equation}
(\lambda_d t)^{3/2}  \; n_d(p,t) 
\sim 
\frac{\alpha_d^< \eta_d^< c_d 
\left( \frac{p}{\sqrt{\lambda_d t}}\right)^{\alpha_d^<-1} 
}
{
\left[
1+ \eta_d^< \left( \frac{p}{\sqrt{\lambda_d t}}\right)^{\alpha_d^<} 
\right]^{\tau}}
\end{equation}
for small areas and its obvious modification for large areas. For
$T_0\to\infty$ one replaces $\eta_d$ and $\alpha_d$ by the primed
quantities and $c_d \to 2c_d$. The scaling analysis of the number
density of domain wall lengths is displayed in
Fig.~\ref{fig:perimeters-domain-scaling} for both initial conditions.
Once again we find a very good agreement between the analytic predictions 
and the numerical data.

\subsection{Finite temperature evolution}

Once we analysed the statistics of perimeters in the 
zero temperature dynamics we focus on the effects of a 
finite working temperature. We briefly list 
the results below without presenting the 
data. 

\subsubsection{The area-perimeter relations}

 For large areas we find the same exponent as for zero
temperature coarsening that is also the initial condition exponent
($T_0\to\infty$ or $T_0=T_c$).  This is reasonable since the large
structures are still `unaware' of the coarsening process and thus
retain the form they had in the initial configuration. For small
areas, instead, we see domain walls roughening due to thermal agitation
but it is hard to extract the value of the exponent $\alpha^<$ 
with sufficient accuracy. 


\subsubsection{Perimeter number densities}

The scaling of the perimeter number densities and the functional form
for the scaling function predicted analytically describe the numerical
data with high precision once the values of the exponents $\alpha$,
the prefactors $\eta$, and the parameter $\lambda$ are modified to
take into account thermal agitation (for absolute area values larger
than $A\sim 10 A_0$, this value limits the range of areas where the
effect of thermal fluctuations is larger than the ones of coarsening).
The analytic prediction is very accurate in the region of 
small coarsening domains,
$A/\lambda_{d,h} t < 10$ and $A>10A_0$ where the maximum is located,
and in the region of large coarsening domains, $A/\lambda_{d,h} t > 10$ and
$A>10A_0$, for both domains and hulls and the two initial conditions.

\section{Conclusions}

In this paper we studied the statistics and geometry of hull enclosed
and domain areas and interfaces during the non-equilibrium dynamics of
curvature driven pure coarsening in two dimensions. The analytical
part of our work relies on the Allen-Cahn equation derived from the
continuous Ginzburg-Landau field-theory in two-dimensions while the
numerical part of it dealts with Monte Carlo simulations of the
$2d$IM. Our main results are: 
\newline 
(i) We proved scaling of the various number densities studied.
\newline
(ii) We derived the exact number density of hull enclosed areas
and hull lengths; we obtained approximate expressions for the 
number density of domain areas and domain wall lengths.
\newline
(iii) The geometrical properties and distribution of the time
dependent large structures (by large we mean much larger than the
average ones) are the ones of critical continuous percolation (for
all initial conditions equilibrated at $T_0>T_c$) and critical
Ising (for $T_0=T_c$).  The long interfaces retain the fractal
geometry imposed by the equilibrium initial condition and the scaling
function of all number densities decay as power laws.
\newline
(iv) Instead, small structures progressively become regular 
 and the area-perimeter relation is $A\sim p^2$.
\newline
(v) We took into account the effects of a finite working temperature
by correctly eliminating purely thermal fluctuations and thus
correctly identifying the coarsening structures.  The temperature
effect thus amounts to introducing the temperature dependence in the
prefactor in the growth law, $R(t) \sim [\lambda(T) t]^{1/2}$.
$\lambda(T)$ is a monotonically decreasing function of $T$ that
vanishes at $T_c$.

It is important to stress that our analytic results rely on the use of
the Allen-Cahn result for the velocity of an almost flat interface.
Thus, they would be expected to hold only in a statistical sense and
for large structures in the lattice model.  Surprisingly, we
found with numerical simulations that the number density area
distributions in the $2d$IM match the analytic predictions for very
small structures, and even after a few MC steps evolution of a
critical Ising initial condition for which rather rough interfaces
exist.

Using the Allen-Cahn result and a variety of numerical measurements we
verified the well-known result for pure coarsening with non-conserved
order parameter: there is a characteristic growing length that
increases in time as $t^{1/2}$.  The mesoscopic analysis presented
here allows us to demonstrate that the reason for the growth of the
characteristic length is the disappearance of small structures.

Our analytic results hold only in two dimensions. In Appendix
\ref{sec:oned} we summarized the behaviour of the distribution of
domain lengths in the Ising chain. As expected from scaling arguments,
this quantity scales with the typical domain length, $R(t) \sim
t^{1/2}$ but the form of the scaling function is very different from
the one in two dimensions: the pdf vanishes at zero scaling argument
($x=0$), it then increases linearly to reach a maximum and then
falls-off to zero exponentially.  On the other hand, we cannot extend
the analytic argument to dimensions higher than two since the hull
enclosed volumes no longer decrease with time in a manner that does
not depend on their own volume~\cite{Comment}.

This paper and \cite{us} open the way to a number of related
studies. For instance, it would be interesting to extend the analysis
presented here to dynamic clusters of correlated spins (droplets),
that are known to describe thermal fluctuations close to the
transition (see ~\cite{coniglio} and ~\cite{Sator} for a review). These droplets are smaller
than the geometric clusters in which they are embedded because some of
the neighboring parallel spins are discarded (by a temperature
dependent criteria) for not being correlated (remember that even at
infinite temperature, where all spins are uncorrelated, there are
domains of parallel spins).

Two-dimensional coarsening with conserved order parameter is another 
problem that deserves a careful study along these lines. 

In~\cite{us} we include a preliminary analysis of the hull and domain
structure in the finite temperature dynamics of the bi-dimensional
random bond ferromagnetic Ising model after a quench from infinite
temperature.  In such a disordered case a finite working temperature
is necessary to help the interfaces depin from pinning centers in the
quenched disordered potential through thermal activation. We first
computed the typical domain radius that scales the time-dependent
spatial correlation, $C(r,t) \sim f(r/R(t))$, in the scaling
regime. Due to the presence of quenched disorder $R(t)$ strongly
depends on temperature and the strength of randomness and it is slower
than the simple square root behaviour of the pure Ising case.  We then
showed numerically that the number density of hull enclosed and domain
areas scale as $R^4(t) n_{h,d}(A,t) \sim g(A/R^2(t))$ for areas
satisfying $10^{-1} \stackrel{<}{\sim} A/R^2(t)$ and that are smaller
than the cut-off set by finite size effects. The effect of a
non-trivial typical radius $R(t)$ determined by the quenched disorder
can be tested in the intermediate regime, say $10^{-1}
\stackrel{<}{\sim} A/R^2(t)\stackrel{<}{\sim} 10^1$, where the quality
of the scaling plot is excellent.  The scaling function $g(x)$ does
not depend on the disorder strength satisfying the hyper-scaling
hypothesis~\cite{BrayReview}.  For smaller areas, say $A/R^2(t)
\stackrel{<}{\sim} 10^{-1}$, the contribution of thermal domains with
domain walls roughened by disorder is important.  We shall give more
details on the domain morphology of the quenched disordered coarsening
problem in a separate publication~\cite{us-EPL}.

These results give an idea of the richness and complexity of
coarsening phenomena even in the absence of quenched randomness. We
expect them to be of help in understanding the fluctuating dynamics of
even more complex situations, like spin-glasses and 
glassy problems~\cite{Chamon}, in
which the mere existence of a domain growth of two competing 
equilibrium phases is not even established. 

\appendix
\section{Algorithm used to identify hull enclosed areas} 
\label{app:algorithm}

In order to obtain the size of each hull in an $L\times L$ system, 
a biased walk along the interior border of each 
domain is performed, with the hull enclosed area
being updated at each step. This algorithm is related to, but different from the
Grossman-Aharony proposal.

{\em Labelling}. The $N$ sites are initially indexed from 0 
to $N-1$ (top-left=0) while the domains (geometric, Coniglio-Klein, etc)
are identified and labelled by the Hoshen-Kopelman algorithm~\cite{Hoshen}.  
By construction, all sites in each domain receive the (unique) label 
corresponding to the smaller index among its spins.

{\em Starting point.} The putative starting site for the walk is
the spin whose index identifies the cluster. In some cases (for example, 
when the cluster crosses a border), it may not be the leftmost/topmost 
site, as it should in order to be counted correctly by the implementation
of the algorithm below. Although such domains may be excluded from the 
statistics, for finite sizes the introduced bias is unacceptable. Thus, 
before starting, we try to find another site in the same cluster above or
to the left of the original spin, where a new starting site may be found. 
Once this first site 
is correctly identified, we assign a height $y_0$. 

{\em The walk.} From the starting point, we try to turn clockwise
around the domain border. Viewed from the incoming direction,
the attempted move is performed in the sequence: left, front, right and
backwards. For the sake of notation, we label the four directions with the 
indices shown in the table \ref{tab.deltaA}.
Thus, for example, if the previous step was 0, the attempted
order will be 1, 0, 3 and 2. 

{\em The area.} The first step sets the area to the value
$A_1=y_0+1$. As the walk proceeds, both and area and the height
are updated, $A_{i+1}=A_i+\Delta A$ and $y_{i+1}=y_i+\Delta y$
(in this order), where $\Delta A$ and $\Delta y$ depend also on the
former direction (see table~\ref{tab.deltaA}). At the end of the walk, 
when the 
departure site in reached, if the last direction is 1, we increase $A$ 
by $-y$. Care should be taken when the starting point has 
right and bottom neighbours belonging to the same cluster (but not the bottom
right diagonal), because in this case the walk should only be
finished after the second time the starting point is visited.

\begin{table}
\begin{center}
\begin{tabular}{||c|c|c|c||} \hline
$t-1$ & $t$ & $\Delta A$ & $\Delta y$ \\ \hline
0&0&0 & 1 \\
0&1&0&0 \\
0&2&$y+1$&$-1$ \\
0&3&$y+1$&0 \\
1&0&$-y$&1 \\
1&1&$-y$&0 \\
1&2&0&$-1$ \\
1&3&1&0 \\
2&0&$-y$&1 \\
2&1&$-y$&0 \\
2&2&0&$-1$ \\
2&3&0&0 \\
3&0&0&1 \\
3&1&1&0 \\
3&2&$y+1$&$-1$ \\
3&3&$y+1$&0 \\ \hline
\end{tabular}
\includegraphics[width=2.5cm]{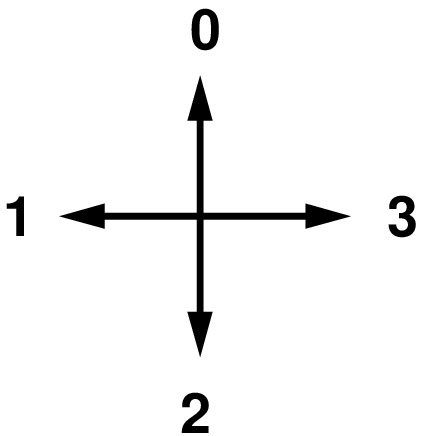}
\end{center}
\label{tab.deltaA}
\caption{(Colour online.) Incremental area contribution for each step
during the oriented walk along the domain border. The values depend
both on the present ($t$) and former steps ($t-1$). On the right we
show the labelling of the four possible directions.}
\end{table}

\section{The one dimensional case}
\label{sec:oned}

For the sake of comparison we present in this Appendix results for the
domain size distribution in the one-dimensional Ising chain.  The 1d
ferromagnet with nearest neighbours interactions is given by the
Hamiltonian
\begin{equation}
{\cal H} = -J \sum_{i} s_is_{i+1}
\label{Eq.H}
\end{equation}
where the spins are $s_i=\pm 1$ and $J>0$. 
The system evolves through a Heat Bath dynamics, where each
spin tries to align with its local field (with a temperature
dependent probability).

\begin{figure}
\psfrag{t}{\large $t$}
\psfrag{L}{\large $\ell(T)$}
\includegraphics[width=5.5cm,angle=-90]{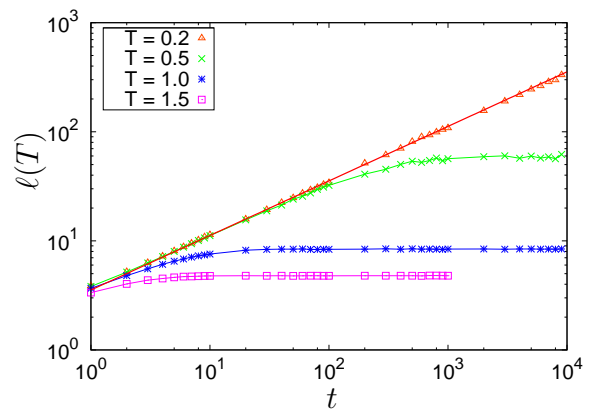}
\caption{(Colour online.) The average domain length at finite temperature, 
$\ell(t) = 2\sqrt{\pi t}$ independently of temperature 
in the domain-growth regime before saturation.}
\label{fig:domain_t_1d}
\end{figure}

We measure the domain length distribution at time $t$, $n(\ell,t)$
($1\leq \ell \leq L$) after quenching the system from $T_0\to\infty$ to
a final, working temperature, $T$. There is no finite
static critical temperature and the system orders ferromagnetically only 
at $T=0$. At all finite temperatures a completely disordered
initial condition starts evolving in a coarsening regime in 
which regions of finite 
length order and subsequently crosses over to equilibrium in the paramagnetic 
phase.
Differently from the higher
dimensional case where domain growth is driven by interfacial tension,
in $d=1$ coarsening is driven by the diffusion of domain walls and
annihilation when they meet. 

\begin{figure}
\psfrag{l}{\hspace{-8mm}\large $\ell/\langle \ell(t)\rangle$}
\psfrag{P}{\hspace{-1cm}\large $\langle \ell(t)\rangle n(\ell)$}
\psfrag{x1}{$t$}
\psfrag{y1}{$\langle \ell\rangle$}
\includegraphics[width=8cm]{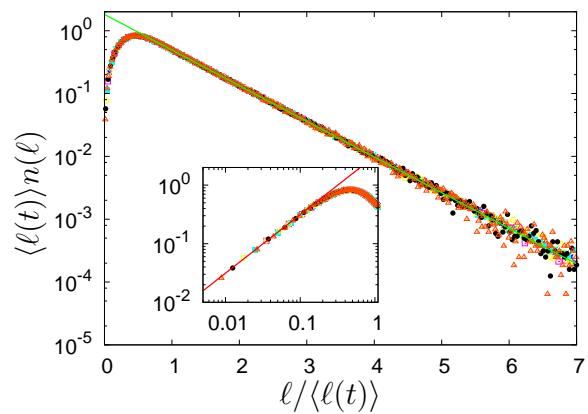}
\caption{(Colour online.) Collapsed distributions of the (rescaled)
probability distribution of domain lengths after a quench to $T=0$ for
$L=10^5$ and times $t=2^3,\ldots,2^{9}$.  All data collapse onto an
universal curve whose tail is exponential~\cite{DeZe96}.  Inset: the
distribution for short length scales where the universal function is
$f(x)\simeq \pi x$. Notice that the average domain length is different
from the typical one.}
\label{fig.dist1d}
\end{figure}

In the infinite temperature initial condition the spins are uncorrelated and
$n(\ell,0)=2^{-\ell}$, while in equilibrium, the normalized distribution is
also exponential
as it corresponds to the
distribution of domains of a paramagnet in a field: $n(\ell,\infty)=
r(1-r)^{\ell-1}$, with $r=[1+\exp(2\beta)]^{-1}$~\cite{DeHa05}.  

During the coarsening
regime, the distribution of domain sizes obeys the scaling behavior
\begin{equation}
n(\ell,t) = \langle \ell(t)\rangle^{-1} f\left( \frac{\ell}{\langle
  \ell(t)\rangle} \right)
\end{equation}
where the time dependent average length is $\langle
\ell(t)\rangle=2\sqrt{\pi t}$~\cite{Amar-Family,DeZe96}, independently
of temperature
and saturates at
$\langle \ell(\infty)\rangle=1/r$ after a time that roughly grows as
$\exp(4/T)$, see Fig.~\ref{fig:domain_t_1d}.  At zero working temperature 
the universal function $f(x)$
is given by~\cite{Amar-Family,AlAv95,DeZe96,KrNa97}
\begin{equation}
f(x) = \left\{ 
\begin{array}{ll}
\pi x & \;\; x\ll 1  \; , 
\\
\exp(-Ax+B) & \;\; x\apprge 1 \; , 
\end{array} \right.
\label{Eq.Derrida}
\end{equation}
where the constants are known exactly:
$A=1.3062$ and $B=0.597$~\cite{DeZe96}, see
Fig.~\ref{fig.dist1d}. Differently from higher dimensions, here as
large clusters coalesce, there is no formation of small clusters
(compared with the typical cluster). This fact is at the origin of the
dip in the pdf close to $\ell/\langle \, \ell\, \rangle=0$.

\begin{figure}
\psfrag{l}{\large $\ell$}
\psfrag{P}{\large $n(\ell)$}
\includegraphics[width=8cm]{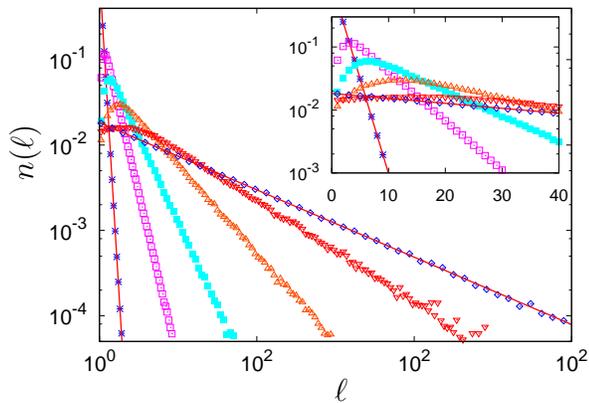}
\caption{(Colour online.) Domain length distribution at $T=0.5$ for
increasing values of time (from left to right: $0$, $4$, $16$, $64$,
$256$ and $\infty$.  The initial, $n(\ell,0)$, and final
distributions, $n(\ell,\infty)$, are also indicated with solid lines
(see text). The small $\ell$ region is shown in the inset.  }
\label{fig:dist_1d_short}
\end{figure}

\begin{figure}
\psfrag{P}{\large $n(\ell)$}
\psfrag{l}{\large $\ell$}
\includegraphics[width=8cm]{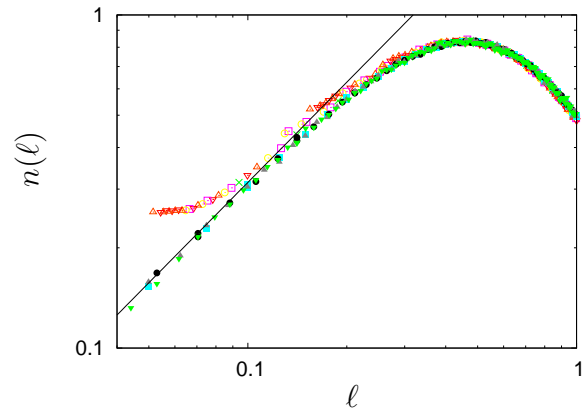}
\caption{(Colour online.) Temperature dependence of the domain length
distribution at small lengths. Filled symbols are for $T=0$ while
hollow ones are for $T=0.5$. The line is Eq.~(\ref{Eq.Derrida}) for
small $x$.}
\label{fig:temp-small-x}
\end{figure}

At finite working temperatures, because of the thermal fluctuations,
small clusters are created at a constant temperature dependent
rate. As say one spin flips within a domain not only there is a
new domain with length $\ell=1$ but the host domain has been
cut in two pieces of relative much shorter length.  
Thermal agitation thus decreases the depth of the dip. 
When $x\apprge 1$, $f(x)$ is still well fitted by 
Eq.~(\ref{Eq.Derrida}). In the opposite regime, 
$x\ll 1$, instead thermal fluctuations diminish the dip and the
behavior deviates from the linear scaling function, as can be
seen in Fig.~\ref{fig:temp-small-x}.

\vspace{1cm}
\noindent\underline{Acknowledgements} LFC is a member of Institut
Universitaire de France. AJB and LFC thank the Newton Institute at the
University of Cambridge, UK, JJA the LPTHE Jussieu, and AS and LFC the
Universidad Nacional de Mar del Plata, Argentina, for hospitality
during the preparation of this work. AS and LFC acknowledge financial
support from Secyt-ECOS P. A01E01 and PICS 3172, AS, JJA and LFC
acknowledge financial support from CAPES-Cofecub research grant
448/04. JJA is also partially supported by the Brazilian agencies CNPq
and FAPERGS. We thank S. Bustingorry and A. B.  Kolton for useful
discussions.


\begin{thebibliography}{99}
\bibitem{Comment} In three dimensions the time variation of the hull
enclosed volume is not independent of its size. This can be easily
seen by considering a spherical bubble for which $\frac{dV}{dt} =
4 \pi R^2 \frac{dR}{dt}$. In accordance with Eq.~(\ref{allen-cahn}), $\frac{dR}{dt} = 
- \frac{\lambda_h}{2 \pi}  \kappa$, and $\kappa=2/R$ for a sphere, so,
$\frac{dV}{dt} = - 4  \lambda_h  R$.
This is the reason why our arguments cannot be extended to
higher dimensions.
\end{thebibliography}
\end{document}